\documentclass[aps,prd,notitlepage,nofootinbib,superscriptaddress,groupedaddress]{revtex4-2}

\usepackage[latin9]{inputenc}
\usepackage{mathrsfs}
\usepackage{bm}
\usepackage{amsmath}
\usepackage{amsthm}
\usepackage{amssymb}
\usepackage{graphicx}
\usepackage{esint}
\usepackage{hyperref}
\usepackage{tabularx}

\theoremstyle{plain}

\theoremstyle{plain}

\ifx\proof\undefined
\newenvironment{proof}[1][\protect\proofname]{\par
	\normalfont\topsep6\p@\@plus6\p@\relax
	\trivlist
	\itemindent\parindent
	\item[\hskip\labelsep\scshape #1]\ignorespaces
}{%
	\endtrivlist\@endpefalse
}
\providecommand{\proofname}{Proof}
\fi

\usepackage[table ]{ xcolor}

\usepackage{amssymb}
\usepackage{graphicx,color}

\usepackage{amsmath,amsthm}

\usepackage{bm}

\def\beq{\begin{equation}}
\def\eeq{\end{equation}}
\def\bi{\begin{itemize}}
\def\ei{\end{itemize}}
	\def\ba{\begin{array}}
	\def\ea{\end{array}}
	\def\bfig{\begin{figure}}
	\def\efig{\end{figure}}

	\def\C{\mathbb{C}}
	\def\R{\mathbb{R}}
	\def\Z{\mathbb{Z}}

	\newtheorem{theorem}{Theorem}[section]
	
	\newtheorem{corollary}{Corollary}[theorem]
	\newtheorem{lemma}[theorem]{Lemma}

	\newcommand{\SLDC}{{\mathrm{SL}(2,\mathbb{C})}}

	\newcommand{\Slc}{\mathrm{SL}(2,\mathbb{C})}

	\newcommand{\Su}{\mathrm{SU}(2)}

	\def\be{\begin{eqnarray}}
	\def\ee{\end{eqnarray}}

	\newcommand{\cb}{\mathcal B}
	
	\newcommand{\cd}{\mathcal D}
	\newcommand{\ce}{\mathcal E}

	\newcommand{\ch}{\mathcal H}
	\newcommand{\ci}{\mathcal I}
	\newcommand{\cj}{\mathcal J}
	\newcommand{\ck}{\mathcal K}

	\newcommand{\cn}{\mathcal N}
	\newcommand{\co}{\mathcal O}

	\newcommand{\calr}{\mathcal R}
	\newcommand{\cs}{\mathcal S}
	
	\newcommand{\cu}{\mathcal U}
	
	\newcommand{\cw}{\mathcal W}
	\newcommand{\cx}{\mathcal X}
	\newcommand{\cy}{\mathcal Y}

	\newcommand{\sa}{\mathscr{A}}
	\newcommand{\sm}{\mathscr{M}}

	\newcommand{\fa}{\mathfrak{a}}

	  \newcommand{\Fg}{\mathfrak{G}}
	\newcommand{\fh}{\mathfrak{h}}

	\renewcommand{\a}{\alpha}
	\renewcommand{\b}{\beta}
	\newcommand{\g}{\gamma}

	\newcommand{\eps}{\varepsilon}

	\newcommand{\sig}{\sigma}
	
	\renewcommand{\l}{\lambda}
	\renewcommand{\L }{\Lambda}
	\renewcommand{\o}{\omega}

	\newcommand{\rmd}{\mathrm d}

	\newcommand{\lt}{\left}
	\newcommand{\rt}{\right}

	\newcommand{\lag}{\left\langle}
	\newcommand{\rag}{\right\rangle}
	
	\newcommand{\bbc}{\mathbb{C}}

	\newcommand{\sn}{\mathscr{N}}

	\newcommand{\re}{\mathrm{Re}}

\newcommand{\PT}{\mathscr{P}}

\begin{document}

\title{Pure integral formula of spinfoam amplitude and curvature corrections in effective theory}

\author{Muxin Han}
\email{hanm(At)fau.edu}
\affiliation{Department of Physics, Florida Atlantic University, 777 Glades Road, Boca Raton, FL 33431, USA}
\affiliation{Institut f\"ur Quantengravitation, Universit\"at Erlangen-N\"urnberg, Staudtstr. 7/B2, 91058 Erlangen, Germany}


\begin{abstract}
We develop a new analytic method for the semiclassical analysis of the Lorentzian EPRL spinfoam amplitude on a fixed triangulation. We derive a pure integral formula for the spinfoam amplitude that trades the discrete sum over internal spins for integrals over continuous variables. This reformulation is free of discrete sums and provides a well-behaved action that is well-adapted to the stationary phase approximation in the large-spin regime. Integrating out the non-geometric variables yields an effective action for the edge lengths of the Regge geometry. Applying the stationary phase approximation and expanding this effective action to second order in perturbations around a flat background, we obtain the classical Regge action at leading order, supplemented by a quadratic correction in the deficit angles $\varepsilon$. This curvature correction originates from two distinct sources: a holonomy correction resulting from the new action and the fluctuations of non-Regge areas. We show that the quadratic $\varepsilon$ correction appears at a higher order in the Barbero-Immirzi parameter $\gamma$ relative to the Regge action. Together with the large-spin expansion, this yields a two-parameter expansion for the spinfoam effective action, whose leading order is the quadratic Regge action. In the long wavelength regime, the latter approaches the action of a massless spin-2 graviton.
\end{abstract}

\maketitle

\tableofcontents

\section{Introduction}

Spinfoam theory provides a covariant, path-integral formulation of Loop Quantum Gravity (LQG) in 4-dimensions \cite{Perez2012,rovelli2014covariant,EPRL,FK}. The spinfoam amplitude defined on a 2-complex $\ck$ is a sum over $\Su$ spins labelling the faces of $\ck$ and an integral over $\Slc$ holonomies and spinors, and the spins are interpreted as areas of triangles. A central question since the construction of these models has been whether the amplitude reproduces classical general relativity in the semiclassical regime, and this question has mostly been addressed in the large-spin regime by the stationary phase method. 
The asymptotic analysis of the vertex amplitude identified the critical points of the integral that correspond nondegenerate Lorentzian 4-simplex geometries, and the amplitude oscillates with the Regge action \cite{semiclassicalEu,semiclassical,Kaminski:2017eew}. This analysis was extended to simplicial complexes with many 4-simplices \cite{CFsemiclassical,HZ1,HZ,hanPI}, where the critical points are identified to correspond to Regge geometries on the entire complex. 
When the sum over internal spins is included, however, the real critical points (within the integration cycle) are subject to the additional constraint that the deficit angles $\eps_h$ of the corresponding geometry satisfy $\g\eps_h\in2\pi\Z$, where $\g$ is the Barbero-Immirzi parameter; this is known as the ``flatness'' of spinfoams \cite{flatness,frankflat,LowE1,Oliveira:2017osu}. Curved geometries were argued to survive in a regime where the Barbero-Immirzi parameter $\g$ is small and the spins are large with the areas kept fixed \cite{claudio,claudio1,lowE2,propagator2,propagator3}. The introduction of complex critical points (away from the original integration cycle) \cite{Han:2021kll,Han:2023cen} made this statement precise: curved Regge geometries arise from complex critical points of the spinfoam amplitude, and the resulting effective dynamics in the large spin regime has been analyzed numerically on a variety of complexes \cite{Han:2023cen,Han:2024ydv,Han:2024lti}. 
The results from these numerical experiments suggest that, to leading order, the effective dynamics agree with classical Regge calculus for curved geometries, with corrections appearing only at higher orders in $\g$.


A complementary line of work has studied the effective dynamics of spinfoams through area-Regge calculus \cite{Dittrich:2021kzs,Dittrich:2022yoo,Borissova:2022clg}, where the fluctuations of areas that are not induced by lengths modify the graviton dynamics. 
At leading order, the effective dynamics again reproduce classical Regge calculus, while corrections arise at higher orders in curvature. In the continuum limit, these corrections yield a Weyl-squared $\cw^2$ term.


Despite these advances, a systematic analytic derivation of the effective dynamics of the full spinfoam amplitude, including the sum over internal spins and the corrections to the Regge action, has remained an open problem.
Moreover, these results suggest that the corrections are expected to arise at higher orders in the Barbero-Immirzi parameter $\gamma$, and are associated with high-curvature effects. A key objective of this paper is to confirm these features.

In this paper we develop a new analytic method for the semiclassical analysis of the Lorentzian EPRL spinfoam amplitude on a fixed triangulation. We apply this method to derive the effective dynamics of perturbative gravity around a flat geometry. Our main results are the following: 

First, we derive a \emph{pure integral formula} for the spinfoam amplitude: the sum over internal spins is traded for integrals over continuous real variables $\eta$, so that the amplitude is written as a pure integral $\int e^{S}$ with no discrete sums, with the new spinfoam action $S$. The variables $\eta$ still relate to areas semiclassically.

This pure integral formula is important because it is well adapted to the stationary phase method for the large spin analysis. 
For a comparison, in earlier works e.g. \cite{LowE1,Han:2021kll,Han:2023cen}, the sum over spins in the amplitude is handled by the Poisson summation formula, which produces an integral over continuous areas with the standard spinfoam action $S_{\rm SF}$. 
However, this approach comes with the drawback of leaving a residual sum over integers.
The stationary phase analysis can then only be carried out term by term in the residual sum, and the sum has been difficult to compute. 
In our new approach, the spinfoam amplitude is reformulated as a pure integral expression without discrete sums, which greatly simplifies the large spin analysis.
An additional advantage is that the new action $S$ exhibits better behavior than $S_{\rm SF}$: specifically, the bulk part of $S$ is a single-valued function of the $\Slc$ holonomies, unlike the Poisson summation approach where $S_{\rm SF}$ is multi-valued.
This difference is responsible for a part of the curvature corrections in the effective theory, which we call the holonomy correction below.

We integrate out the non-geometric spinfoam variables, including the group elements, spinors, and the non-Regge areas that are not determined by lengths of Regge geometry. 
This yields an effective action $\ci_{\rm eff}(\ell)$ depending on the edge lengths $\ell$ of Regge geometry. 
Schematically, letting $\cx$ represent the collection of non-geometric spinfoam variables, the amplitude takes the form
\be
\int\, e^{S}=\int\rmd\rho(\ell) \int \rmd\mu(\cx)\, e^{S}=\int\rmd\rho(\ell)\, e^{\ci_{\rm eff}(\ell)}.
\ee
The effective dynamics of Regge geometry is encoded in the action $\ci_{\rm eff}(\ell)$.
Applying the stationary phase approximation in the regime of large boundary spins, and expanding the effective action $\ci_{\rm eff}(\ell)$ to second order in the perturbation $\delta\ell$ on the flat background, we obtain the Regge action together with a quadratic correction in the deficit angles,
\be
\ci_{\rm eff}(\ell) = iS_{\rm Regge}-\frac{\g^2}{8}\sum_{h_{1}h_{2}}K_{h_1,h_2}\eps_{h_1}\eps_{h_2}+\co\lt(\delta \ell^3\rt)\ ,\label{ciandKIntoduction}
\ee
where the correction matrix $K$ is expressed in terms of the spinfoam data at the flat background.

The $\eps^2$-correction in \eqref{ciandKIntoduction} has two distinct sources: (1) the \emph{holonomy correction}, arising because the bulk part of the new action $S$ is single-valued on holonomies rather than on their logarithms, and thus semiclassically, it is single-valued on $e^{i\g\eps_h/2}$ rather than on $\eps_h$, and (2) the fluctuations of the \emph{non-Regge areas}, i.e. the area degrees of freedom that are not determined by edge lengths. The latter has the same origin as the correction found in area-Regge calculus \cite{Dittrich:2021kzs,Dittrich:2022yoo}.  

Perturbatively at quadratic order in $\delta\ell$, the holonomy correction turns out to be quadratic in the equations of motion, and it can be removed via a field redefinition. 
It resembles $\calr^2$ and $\calr_{\mu\nu}\calr^{\mu\nu}$ terms in the effective field theory of pure gravity \cite{tHooft:1974toh,Burgess:2003jk}. Consequently, at the perturbative level, the holonomy correction only affects off-shell observables, although it is anticipated to be physically significant in the presence of matter coupling or within the non-perturbative regime.

When $\g$ is scaled to small while the geometry are held fixed,  we show that the matrix $K$ scales as $1/\g$, so the $\eps^2$-correction is of order $\co(\g)$, whereas the leading order Regge action is purely geometrical and independent of $\g$. 
In other words, the $\eps^2$ correction to the Regge action appears at higher order in $\g$. 
This aligns with what has been observed in previous analytic and numerical studies \cite{claudio,claudio1,Han:2023cen,propagator2,propagator3,lowE2}.

Combined with the large-$J$ stationary phase expansion, where $J$ is the scale of large spins, this organizes the spinfoam amplitude into a \emph{two-parameter expansion} in $1/J$ and $\g$, whose leading order is exactly the Regge action. The expansion in $\g$ relates to the curvature expansion in terms of $\eps_h$. 
In the long-wavelength regime of geometric 
fluctuations, Regge geometric quantities can be effectively approximated by their continuum limits. The quadratic Regge action is known to approach the action for linearized gravity in this limit \cite{Rocek:1981ama,Williams:1991cd,Williams:1986hx,Hohn:2014nwv}. This shows that, at leading order in large $J$ and small $\g$, the effective dynamics derived from the spinfoam amplitude coincides with the dynamics of a massless spin-2 graviton.

The results of this paper should be viewed as part of the program initiated in \cite{Han:2026hyq} for the covariant LQG. It was shown there that the ultraviolet (small-spin) regime of the theory is controlled by a candidate fixed point, at which the complete amplitude reduces at leading order to a triangulation-independent topological theory. 
The present work investigates the infrared regime of the theory, characterized by large spins (corresponding to areas much greater than the Planck scale) and long-wavelength fluctuations.
Our result shows that the theory at this regime behaves as a free graviton theory, with the $\g$-suppressed curvature corrections as a subleading correction. 
The ultraviolet fixed point identified in \cite{Han:2026hyq} and the infrared graviton dynamics established in this work are the two pivotal regimes of covariant LQG.

We also would like to mention a few limitations of the present analysis: we focus on the perturbative expansion around a single real critical point (corresponding to flat spacetime) and neglect non-perturbative contributions from other critical points as well as from the boundary and infinity of the integration domain, and we ignore $\co(1/J)$ corrections in the effective action.

The results in this paper are derived under the working hypotheses that certain matrices (e.g. the Hessian) appearing in the stationary phase analysis are invertible. We do not have a general proof of these nondegeneracy conditions and adopt them as working hypotheses. 
Nonetheless, all numerical examples examined confirm these nondegeneracy conditions, as detailed in the companion paper \cite{companionNumer}.

The paper is organized as follows. Section \ref{Purely integral formula of spinfoam amplitude} derives the pure integral formula of the spinfoam amplitude and the new action. Section \ref{Large boundary spins and stationary phase analysis} discusses the large spin scaling and the stationary phase integration, as well as the gauge symmetries of the new action $S$. Section \ref{Real critical point and flat Regge geometry} analyzes the real critical points of $S$ and their correspondence with flat Regge geometries, and sets up the perturbative expansion around the flat background. Section \ref{Regge and non-Regge areas} introduces the change of variables from areas to lengths and non-Regge areas, and defines the effective action $\ci_{\rm eff}(\ell)$. Section \ref{Scheme for computing the quadratic effective action} outlines the scheme for computing the quadratic effective action and explains the two-parameter expansion. Section \ref{A toy integral and Regge action} studies the toy integral with the leading-order action $S_0$ and shows that it reproduces the Regge action, while Section \ref{Perturbative geometrical identities} derives the perturbative geometric identities relating spinfoam Hessian data to Regge-geometric quantities. Section \ref{Higher curvature corrections} contains the main result: the derivation of the quadratic effective action with the $\eps^2$-correction, and the small-$\g$ scaling analysis.

\section{Pure integral formula of spinfoam amplitude}\label{Purely integral formula of spinfoam amplitude}

Let $\ck$ be a 2-complex dual to a simplicial complex. The 2-complex $\ck$ contains faces $f$, edges $e$, and vertices $v$. The faces are classified as internal faces $h$ and boundary faces $b$, with total counts $N_h$ and $N_b$ respectively. For each face $f$, we assign a half-integer spin $j_f$ and denote $k_f=2j_f \in \mathbb{N}_0$. The spinfoam amplitude associated to $\ck$ is defined by
\be
\sa=\sum_{\{k_h\in\mathbb{N}_0\}}\prod_h d_{\Lambda}(k_h)\int \rmd\mu\lt(g,z\rt) \prod_f E_f(g,z)^{k_f},\label{spinfoamAmpliutdeA}
\ee
where $g=\{g_{ve}\}_{(v,e)}$ is a collection of $\Slc$ group elements and $z=\{z_{vf}\}_{(v,f)}$ is a collection of $\mathbb{CP}^1$ spinors. In order that the integral is finite, group variables $g$ are subject to the gauge fixing that an edge $e_0$ is chosen at every vertex $v$ such that $g_{ve_0}=\mathbb{I}$ \cite{finite}. For a boundary face $b$, the function $E_b$ also depends on the boundary normalized spinor $\xi_{eb}$ in addition to $g$ and $z$, but this dependence is left implicit in the formula. We define $Z_{vef}=g_{ve}^\dagger z_{vf}$. The expressions for $E_f$ are given by
\be
&&E_f=\prod_{e\in\partial f} \ce_{f,e}\ ,\qquad f=h,b\ ,\\
\text{internal edge $e$:}\quad &&\ce_{f,e}
=\frac{\left\langle Z_{vef},Z_{v'ef}\right\rangle}{\left\langle Z_{vef},Z_{vef}\right\rangle^{\frac{1-i\gamma}{2}}\left\langle Z_{v'ef},Z_{v'ef}\right\rangle^{\frac{1+i\gamma}{2}}}\ ,\qquad f=h,b\ ,\\
\text{boundary edge $e$:}\quad && \ce_{b,e}
=\frac{\left\langle\xi_{eb},Z_{v'eb}\right\rangle}{\left\langle Z_{v'eb},Z_{v'eb}\right\rangle ^{\frac{1+i\gamma}{2}}}\ ,
\qquad \text{or}\qquad 
{\cal E}_{b,e}
=\frac{\left\langle Z_{veb},\xi_{eb}\right\rangle}{\left\langle Z_{veb},Z_{veb}\right\rangle^{\frac{1-i\gamma}{2}}}\ .
\ee
The boundary $\partial f$ of each face is oriented from the source vertex $v$ to the target vertex $v'$. The sesquilinear form $\lag z_1,z_2\rag=z_1^{\dagger} z_2$ is the Hermitian inner product on $\C^2$. The positive measure $\rmd\mu(g,z)$ is given by
\be
\rmd\mu(g,z)=\prod_{(v,e)}\rmd g_{ve}\prod_v\delta(g_{ve_0})\prod_{(v,f)}\frac{\rmd^2 z_{vf}}{\left\langle Z_{vef},Z_{vef}\right\rangle\left\langle Z_{ve'f},Z_{ve'f}\right\rangle},\label{measuredmugz}
\ee
where $e$ and $e'$ are the two edges of $\partial f$ at $v$, $\rmd g_{ve}$ is the Haar measure on $\Slc$ and $\rmd^2 z=\frac{i}{2\pi}(z^0\rmd z^1-z^1\rmd z^0)\wedge (z^0\rmd z^1-z^1\rmd z^0)^*$ is the $\Slc$ invariant homogeneous 2-form on $\C^2\setminus\{0\}$, normalized such that $\rmd^2 z/\langle z,z\rangle^2$ is the $\mathrm{SU}(2)$ invariant measure on $\mathbb{CP}^1$ with unit volume.

The face amplitude $d_{\Lambda}(k_h)$ is defined as
\be
d_{\Lambda}(k_h) = \chi\left({\Lambda}^{-1}{k_h}\right)\,\mathrm{Pol}(k_h)
\ee
where $\chi(x)$ is a smooth function with compact support, satisfying $\chi(x) = 1$ for $x \in [0,1]$ and vanishing outside a compact interval. The function $\mathrm{Pol}(k_h)$ denotes a polynomial in $k_h$. 
The function $\chi({\Lambda}^{-1}{k_h})$, with $\Lambda \in \mathbb{R}_+$, serves as a cut-off regulator for the spin-sum. When $k_h$ is already bounded by the triangle inequality, the parameter $\L$ is chosen to be larger than the bound, so that $\chi$ does not restrict the sum. However, in presence of bubbles which allow unrestricted sums over spins, $\chi$ imposes a cut-off and ensures finiteness.

One may ask how the amplitude depends on the choice of this cut-off function. There are two possible scenarios depending on a choice of boundary data and the 2-complex. In the first, all semiclassical contributions (we will specify them later on) come from points of the integral where $\chi=1$ which means that the bulk spins are comparable in size to the boundary spins. In this scenario, the influence of the cut-off on the amplitude is semiclassically suppressed. The second option is that some of the semiclassical stationary points are located in the region where the cut-off applies. In this case cut-off influence is visible also in the semiclassical regime. In the spin foam literature, such configurations are indication of so-called bubble divergences \cite{Christodoulou2013,Riello:2013bzw,Bonzom:2010ar,Frisoni:2021uwx}.

We now list some facts about the objects occurring in the amplitude:
\begin{enumerate}
    \item A useful property is the bound $|E_f(g,z)|\leq 1$ \cite{hanPI,HZ}. In the standard treatment of spinfoam models, where the integrand is written as $e^{S_{\rm SF}}$ with $S_{\rm SF}=\sum_fk_f\log E_f$, this property ensures that the real part of the action is non-positive.
    \item The measure $\rmd\mu\lt(g,z\rt)$ is positive and finite. Namely,
    \be
    V\equiv\int \rmd\mu\lt(g,z\rt),\qquad V<\infty
    \ee
     The finiteness of $\int \rmd\mu\lt(g,z\rt)$ can be established by explicitly integrating over the $z_{vf}$ variables and following the same arguments as those used in proving the finiteness of vertex amplitudes \cite{finite,Kaminski:2010qb,Han:2021bln}. Additional details are provided in Appendix \ref{sec:CP1 integral}.
\end{enumerate}
The sum in $\sa$ is over a finite range, consequently amplitude $\sa$ is finite and absolutely convergent and it satisfies a bound:
\be
\lt|\sa\rt|\leq \sum_{\{k_h\in\mathbb{N}_0\}}\prod_h \lt|d_{\Lambda}(k_h)\rt|\int \rmd\mu\lt(g,z\rt) <\infty\ .
\ee

The idea of the pure integral formula is to insert, for every internal face $h$, Euler's gamma integral 
$$
\int_0^\infty\rmd\eta_h\,e^{-\eta_h}\frac{\eta_h^{k_h}}{k_h!}=1
$$ 
into \eqref{spinfoamAmpliutdeA}, and to carry out the sum over $k_h$ at fixed $\eta_h$. For a function $F:\mathbb{N}_0\to\C$ of at most polynomial growth, we define its Poisson average (equivalently, $e^{c}\PT[F](c)$ is the Borel transform of the generating function $\sum_{k=0}^\infty F(k)x^k$)
\begin{equation}
\PT[F](c)=e^{-c}\sum_{k=0}^{\infty}F(k)\,\frac{c^{k}}{k!},\qquad c\in\C .\label{Poissontransform}
\end{equation}
The series converges absolutely for every $c\in\C$, so $\PT[F]$ is an entire function. The name refers to the Poisson distribution: for $c\geq0$, $\PT[F](c)$ is the expectation value of $F(k)$ in the Poisson distribution with mean $c$ (cf.\ \eqref{Q:Poisson} in Appendix \ref{app:Q}). We denote by $\mathcal{D}_{\Lambda}$ the Poisson average of the restriction of $d_{\Lambda}$ to $\mathbb{N}_0$:
\begin{equation}
\mathcal{D}_{\Lambda}(c)=\PT[d_{\Lambda}](c)=e^{-c}\sum_{k=0}^{\infty}\frac{c^{k}}{k!}\,d_{\Lambda}(k),\qquad c\in\C .\label{exactZL:D}
\end{equation}


Let $E\in\C$ with $|E|\leq1$. For any function $F:\mathbb{N}_0\to\C$ with $\sum_{k\geq0}|F(k)|<\infty$, and for $\eta\geq0$, we have
\begin{equation}
\PT[F](\eta E)\,e^{-\eta(1-E)}
=e^{-\eta}\sum_{k\geq0}F(k)\,E^k\,\frac{\eta^k}{k!},\label{PTexpansion}
\end{equation}
whose modulus has the bound 
\be
\lt|\PT[F](\eta E)\,e^{-\eta(1-E)}\rt|\leq \Phi_F(\eta)\equiv e^{-\eta}\sum_{k\geq0}|F(k)|\frac {\eta^k}{k!}\ ,\label{PoissonBound}
\ee 
because $|E|\leq1$. By $\int_0^\infty e^{-\eta}\eta^k/k!\,\rmd\eta=1$, we have $\int_0^\infty\Phi_F(\eta)\,\rmd\eta=\sum_k|F(k)|<\infty$ (the sum and integral can be interchanged due to non-negativity of summand). Hence, we have $\sum_k\int_0^\infty\rmd\eta \, e^{-\eta}|F(k)E^k|\eta^k/k!<\infty$, 
so the integral below converges absolutely, and the sum and integral can be interchanged in any order
\be
	\int_0^\infty\rmd\eta\;\PT[F](\eta E)\,e^{-\eta(1-E)}=\sum_{k\geq0}F(k)\,E^k .\label{Poissonidentity}
\ee

Taking $F = d_{\Lambda}$ and $E = E_h$ for each internal face and using \eqref{Poissonidentity}, we obtain the pure integral representation for the spinfoam amplitude.

\begin{theorem}[pure integral formula]\label{exactZL:theorem}
The spinfoam amplitude \eqref{spinfoamAmpliutdeA} is given by
\begin{eqnarray}
\mathscr{A}&=&\int_{\mathbb{R}_{\geq0}^{N_h}}\prod_h\mathrm{d}\eta_h\int\mathrm{d}\mu(g,z)\,\prod_h\mathcal{D}_{\Lambda}\left(\eta_h E_h(g,z)\right)e^{S},\label{pathintegral}\\
S&=&-\sum_h\eta_h\left[1-E_h(g,z)\right]+\sum_b k_b\log E_b(g,z,\xi),\label{theaction}
\end{eqnarray}
where $e^{\sum_b k_b\log E_b}$ stands for $\prod_b E_b^{k_b}$, which is single valued because $k_b\in\mathbb{N}_0$. The integral in \eqref{pathintegral} converges absolutely, so the integrations can be carried out in any order.
\end{theorem}

\begin{proof}
Since $d_{\Lambda}$ has compact support, $\sum_k|d_{\Lambda}(k)|<\infty$. By \eqref{PTexpansion} with $F=d_{\Lambda}$ and $E=E_h$, the integrand of \eqref{pathintegral} is
\begin{equation}
\prod_h\lt[\mathcal{D}_{\Lambda}(\eta_hE_h)\,e^{-\eta_h(1-E_h)}\rt]\prod_bE_b^{k_b}=e^{-\sum_h\eta_h}\sum_{\{k_h\in\mathbb{N}_0\}}\prod_h\lt[d_{\Lambda}(k_h)\,\frac{\eta_h^{k_h}E_h^{k_h}}{k_h!}\rt]\prod_bE_b^{k_b},\label{integrandSeries}
\end{equation}
a finite sum of products of continuous functions of $\eta$ and measurable functions of $(g,z)$. Since $|E_h|\leq1$ and $|E_b|\leq1$, \eqref{PoissonBound} with $F=d_{\Lambda}$ bounds its modulus by $\prod_h\Phi_{d_{\Lambda}}(\eta_h)$, and
\begin{equation}
\int_{\mathbb{R}^{N_h}_{\geq0}}\prod_h\rmd\eta_h\int\rmd\mu(g,z)\,\prod_h\Phi_{d_{\Lambda}}(\eta_h)=V\Big(\sum_{k\geq0}|d_{\Lambda}(k)|\Big)^{N_h}<\infty .
\end{equation}
Hence the integral \eqref{pathintegral} converges absolutely, so the integrations can be carried out in any order. At fixed $(g,z)$, the integral of $\prod_h[\mathcal{D}_{\Lambda}(\eta_hE_h)\,e^{-\eta_h(1-E_h)}]$ over $\eta$ factorizes over the internal faces, and it equals $\prod_h\sum_{k_h\geq0}d_{\Lambda}(k_h)\,E_h^{k_h}$ by \eqref{Poissonidentity} with $F=d_{\Lambda}$ and $E=E_h(g,z)$. We multiply by $\prod_bE_b^{k_b}$, integrate over $(g,z)$, and exchange $\int\rmd\mu(g,z)$ with the finite sum over $\{k_h\}$. The result is the original expression of $\sa$ in \eqref{spinfoamAmpliutdeA}.

\end{proof}

For $F\equiv1$, which violates $\sum_k|F(k)|<\infty$, the relation \eqref{Poissonidentity} still holds for $|E|<1$, by the Schwinger proper-time representation (see e.g.\ \cite{Krajewski:2011zzu} for a prior use of this technique in spinfoam and group field theory)
\begin{equation}
\sum_{k=0}^{\infty}E^k=\frac{1}{1-E}=\int_0^\infty\rmd\eta\;e^{-\eta(1-E)},\qquad|E|<1 .\label{properTime}
\end{equation}
Eq.\eqref{Poissonidentity} is the counterpart of \eqref{properTime} for summable weights $F(k)$: the exponent $-\eta(1-E)$ is the same, and the weight enters only through the prefactor $\PT[F](\eta E)$. 


The integration variable $\eta_h$ is a continuous counterpart of the spin $k_h$. For every $k$, the weight $w_k(\eta)=e^{-\eta}\eta^k/k!$ in \eqref{PTexpansion} is a probability density on $[0,\infty)$ peaked at $\eta=k$, with mean $k+1$ and standard deviation $\sqrt{k+1}$:
\be
&&\int_0^\infty \rmd\eta\, w_k(\eta) = 1,\qquad \frac{\rmd}{\rmd\eta} \log w_k(\eta)=\frac{k}{\eta}-1,\qquad \int_0^\infty \rmd\eta\, \eta w_k(\eta)=k+1,\\ 
&&\qquad \qquad \quad \lt[\int_0^\infty \rmd\eta\, \eta^2 w_k(\eta)-\lt(\int_0^\infty \rmd\eta\, \eta w_k(\eta)\rt)^2\rt]^{1/2}=\sqrt{k+1},
\ee 
so $\eta$ and $k$ agree up to a relative deviation of order $k^{-1/2}$ for large spins. This is consistent with the identification of $\eta_h$ with $k_h$ at the critical points (Section \ref{Real critical point and flat Regge geometry}) and with the interpretation of $\eta_h$ as areas (Section \ref{Regge and non-Regge areas}) in the large spin regime.

A key advantage of the integral formulation in \eqref{pathintegral} is that it rewrites the spinfoam amplitude as a pure integral, entirely free of discrete summations. This allows for a direct application of stationary phase methods to analyze the large-$j$ asymptotics of the amplitude, given that the action is linear in $\eta_h$ and $k_b$ and that the prefactor $\mathcal{D}_{\Lambda}$ does not modify the phase (see the next section). 
This approach differs from earlier treatments such as in \cite{Han:2021kll,Han:2023cen}, where the Poisson summation formula (which is unrelated to the Poisson average \eqref{Poissontransform}) is applied to the sum over spins of the amplitude \eqref{spinfoamAmpliutdeA}.
In the form appropriate to a sum over $k\in\mathbb{N}_0$, the Poisson summation formula reads
$\sum_{k=0}^{\infty}f(k)=\frac12f(0)+\sum_{m\in\Z}\int_0^\infty\rmd k\,f(k)\,e^{2\pi i km}$. For face amplitudes that vanish at $k_h=0$, the end-point terms are absent and one obtains, schematically,
\be
\sa=\sum_{\{m_h\in\Z\}}\int_{\R^{N_h}_{\geq0}} \prod_h\rmd k_h\prod_h d_{\Lambda}(k_h)\int \rmd\mu(g,z)\, e^{S_{\rm SF}+2\pi i \sum_h k_h m_h},\quad S_{\rm SF}=\sum_h k_h\log E_h+\sum_b k_b\log E_b,\label{poissonSumSa}
\ee
where for non-integer $k_h$ a branch of $\log E_h$ has to be chosen. This expression retains a discrete sum $\sum_{\{m_h\in\Z\}}$ and is not a pure integral, 
so the stationary phase analysis can be implemented only for each individual term in the sum, while carrying out the sum has been difficult. Additionally, as $k_h$ is treated as a continuous variable, the action $S_{\rm SF}$ is multi-valued; the single-valuedness (of the integrand) is recovered only upon summing over $\{m_h \in \mathbb{Z}\}$. In contrast, $e^{S}$ in \eqref{pathintegral} is manifestly single-valued, and the bulk part of $S$ is a single-valued function of $E_h$.

This comparison illustrates that the pure integral formula \eqref{pathintegral} effectively incorporates the discrete sum $\sum_{\{m_h\in\Z\}}$, and the resulting difference between $S$ and $S_{\rm SF}$ arises from this incorporation. 
At each $m$, the actions $S$ and $S_{\rm SF}+2\pi i \sum_h k_h m_h$ are related by an $m$-dependent change of variables $k_h=\eta_h c_m$ where $c_m=\frac{E_h-1}{\log E_h +2\pi i m}$, which gives $k_h(\log E_h+2\pi i m)=-\eta_h(1-E_h)$. 
Therefore, the critical points of $S$, real or complex, are the union over $\{m_h\in\Z\}$ of those of $S_{\rm SF}+2\pi i \sum_h k_h m_h$: each critical point of $S$ belongs to exactly one $m$, the one with $\log E_h=-2\pi i m_h$ for all $h$ (where $c_m=1$), and there the critical-point data of the two actions are related by the change of variables\footnote{
Fixing $m$, the change of variable at fixed $m$ is well-defined in the neighborhood of critical points with $\log E_h+2\pi i m_h=0$, i.e. $c_m=1$, whereas at a critical point of $S$ with $\log E_h+2\pi i m_h\neq0$ for some $h$, the change of variables degenerates $c_m=0$.}.
At every critical point, $\partial S/\partial\eta_h=E_h-1=0$ or $\partial S_{\rm SF}/\partial k_h=\log E_h=-2\pi i m$ implies $k_h=\eta_h$. The difference between two actions affects only off-shell quantities.

\section{Large boundary spins and asymptotics}\label{Large boundary spins and stationary phase analysis}

We consider the regime in which both the boundary spins $k_b$ and the cut-off parameter $\Lambda$ are taken to be large: We define a scaling parameter $J\gg1$ and make the replacement 
\be
k_b\to Jk_b,\qquad \L\to J\L\label{kscaling}
\ee
The following analysis investigates the behavior of the ampltiude under the double-scaling defined by \eqref{kscaling}, with $J\gg 1$.
In this regime, the amplitude admits an asymptotic expansion in powers of $1/J$, which will be derived using the stationary phase analysis.

To study the integral \eqref{pathintegral} in the above regime, we change the integration variables
\be
\eta_h\to J\eta_h.
\ee
The changes of integration variables introduce a Jacobian factor $J^{N_h}$. Since $S$ is linear in $k_b,\eta_h$, the integral formula of $\sa$ becomes
\be
\mathscr{A}&=&J^{N_h}\int_{\mathbb{R}_{\geq0}^{N_h}}\prod_h\mathrm{d}\eta_h\int\mathrm{d}\mu(g,z)\,\prod_h\mathcal{D}_{J\Lambda}\left(J\eta_h E_h(g,z)\right)e^{JS} .\label{integraletamu}\\
S&=&-\sum_h\eta_h\left[1-E_h(g,z)\right]+\sum_b k_b\log E_b(g,z,\xi).\label{theaction2}
\ee

Since $J\gg1$ multiplies the action $S$ in \eqref{integraletamu}, it is natural to analyze this integral with the stationary phase method. However, the prefactor $\prod_h\mathcal{D}_{J\Lambda}(J\eta_hE_h)$ also depends on $J$, and it has to be shown that it does not modify the phase. This is done in Appendices \ref{app:polynomial approximation} and \ref{app:Q}. We summarize the results in the following two paragraphs.

If the cut-off function $\chi$ were removed from the series \eqref{exactZL:D} that defines $\mathcal{D}_{J\Lambda}$, the series would be summed exactly: $e^{-c}\sum_{k=0}^{\infty}\frac{c^{k}}{k!}\,\mathrm{Pol}(k)=P(c)$, where $P$ is a polynomial with the same degree and the same leading coefficient as $\mathrm{Pol}$. Intuitively, in the region where all $\eta_h$ are smaller than $\Lambda$, the effect of the cut-off is exponentially small. Indeed, it is shown in Appendix \ref{app:polynomial approximation} that for every fixed $\bar\eta\in(0,\Lambda)$ there is a constant $\nu>0$ such that uniformly in $(g,z)$ and in $\eta_h\in[0,\bar\eta]$,
\begin{equation}
\prod_h\mathcal{D}_{J\Lambda}\left(J\eta_hE_h\right)e^{JS}=\prod_hP\left(J\eta_hE_h\right)e^{JS}+O\left(e^{-\nu J}\right).\label{prefactor:main}
\end{equation}
The integrand of \eqref{integraletamu} is therefore, up to an exponentially small error, $e^{JS}$ times an amplitude that is a polynomial in $J$ whose coefficients are functions of $(\eta_h,g,z)$ independent of $J$.

If we relax the restriction $\eta_h<\Lambda$, the cut-off function $\chi$ cannot be neglected. Nevertheless, $\mathcal{D}_{J\Lambda}$ does not modify the phase there either.
To show this, let $\psi(\eta,g,z)$ be any smooth function with compact support on the domain of integration, with no restriction on the location of its support relative to $\Lambda$, and define
\begin{equation}
\mathscr{A}_\psi=J^{N_h}\int_{\mathbb{R}_{\geq0}^{N_h}}\prod_h\mathrm{d}\eta_h\int\mathrm{d}\mu(g,z)\,\psi(\eta,g,z)\prod_h\mathcal{D}_{J\Lambda}\left(J\eta_hE_h(g,z)\right)e^{JS}.\label{prefactor:Apsi}
\end{equation}
By expanding $\chi(k/(J\L))$ in $\mathcal{D}_{J\Lambda}(c)$ at $k=|c|$, we define for an integer $M\geq1$
\begin{equation}
Q^{(M)}_{J\Lambda}(c)=\sum_{r=0}^{M-1}\frac{\chi^{(r)}\big(|c|/(J\Lambda)\big)}{r!\,(J\Lambda)^r}\,\mathcal{P}_r(|c|,c),\qquad \mathcal{P}_r(a,c)=e^{-c}\sum_{k=0}^{\infty}\frac{c^{k}}{k!}\,(k-a)^r\,\mathrm{Pol}(k),\label{prefactor:QM}
\end{equation}
where $\chi^{(r)}$ is the $r$-th derivative of $\chi$. Like $\mathrm{Pol}$ and $P$, the functions $\mathcal{P}_r$ and $Q^{(M)}_{J\Lambda}$ do not depend on the face $h$. Let $x_*\geq1$ be such that $\chi(x)=0$ for $|x|>x_*$. By Lemma \ref{Q:lemQ}, each $\mathcal{P}_r(a,c)$ is a polynomial in $(a,c)$ with $\mathcal{P}_0(a,c)=P(c)$, and $Q^{(M)}_{J\Lambda}(c)=P(c)$ for $|c|\leq J\Lambda$, i.e. below the cut-off, while $Q^{(M)}_{J\Lambda}(c)=0$ for $|c|>x_*J\Lambda$. It is shown in Appendix \ref{app:Q} that, with $n$ the degree of $\mathrm{Pol}$,
\begin{equation}
\mathscr{A}_\psi=J^{N_h}\int_{\mathbb{R}_{\geq0}^{N_h}}\prod_h\mathrm{d}\eta_h\int\mathrm{d}\mu(g,z)\,\psi(\eta,g,z)\prod_hQ^{(M)}_{J\Lambda}\left(J\eta_hE_h(g,z)\right)e^{JS}+O\left(J^{(n+1)N_h-M/2}\right).\label{prefactor:approx}
\end{equation}
For any given $N$, the choice $M\geq2(N+(n+1)N_h)$ makes the error $O(J^{-N})$. Moreover, by Corollary \ref{Q:corSmooth}, 
\be
\psi\prod_hQ^{(M)}_{J\Lambda}(J\eta_hE_h)=\sum_{p=-(M-1)N_h}^{n N_h}J^{p}a_p(\eta,g,z),
\ee 
where the $a_p$ are smooth functions with compact support, independent of $J$, which vanish unless $\eta_h|E_h|\leq x_*\Lambda$ for all $h$. Therefore, up to $O(J^{-N})$, $\mathscr{A}_\psi$ is a finite sum of terms $J^{N_h+p}\int\prod_h\mathrm{d}\eta_h\int\mathrm{d}\mu\,a_p\,e^{JS}$, i.e. of integrals with the $J$-independent phase $S$ and the $J$-independent smooth amplitudes $a_p$. The stationary phase analysis of $\mathscr{A}_\psi$, below or above the cut-off, is thus based on the variation of $S$ alone, while $\mathcal{D}_{J\Lambda}$ does not contribute to the action.

One important property of $S$ is that for $\eta_h\geq\eta_0>0$
\be
\re\lt(S \rt)=-\sum_h\eta_h\lt[1-\re\lt(E_h(g,z)\rt)\rt]+\sum_b k_b\log \lt|E_b(g,z,\xi)\rt|\leq 0\ ,
\ee
since $\lt|E_h(g,z)\rt|\leq 1$ and $\lt|E_b(g,z,\xi)\rt|\leq 1$ (since $|\ce_{f,e}|\leq 1$ for $f=b,h$). The dominant contribution to the integral $\sa$ arises from the regime where $\re(S)$ is close to zero. This condition implies that $\re\lt(E_h(g,z)\rt)$ is close to $1$, which means $E_h(g,z)$ itself must also be close to $1$. 
For $\sa_\psi$, we require that the support of $\psi$ includes critical points where $E_h(g,z)=1$ and $\left|E_b(g,z,\xi)\right|=1$ and does not include a neighborhood of $\eta_h=0$.

The large-$J$ asymptotic analysis in the rest of this paper will mainly focus on $\sa_{\psi}$. Before we come to the analysis, below are a few remarks on the integral $\sa$:
\begin{enumerate}
    \item Unlike $\sa_\psi$, the integration domain for $\sa$ is unbounded, which raises questions about whether the stationary phase method yields a valid asymptotic expansion for this integral, because contributions from the region at infinity cannot generally be excluded. This subtlety currently remains unresolved. Since our focus is on perturbative aspects of the amplitude, as captured by $\sa_\psi$, we will not address this issue here.
    
    \item The integral has a boundary due to the condition $\eta_h \geq 0$. This boundary is organized into a stratification of submanifolds given by
    \be
       M_{I}=\{\eta_h=0,\ h\in I;\ \eta_h>0,\ h\notin I\}
    \ee
    where $I$ ranges over nonempty subsets of the set of faces. If $\sa$ admits an asymptotic expansion, it can receive contributions from stationary points of the action restricted to any $M_I$. The stationary phase analysis should be suitably adapted in the presence of such boundaries.

    \item The $M_\emptyset$ is the interior of $\mathbb{R}^{N_h}_{\geq0}$. By the insertion of compactly supported function $\psi$, we will only consider in this paper contributions from stationary points in $M_\emptyset$.
\end{enumerate}

We compare the action \eqref{theaction2} with the spinfoam action $S_{\rm SF}$ standard in the semiclassical analysis of spinfoam amplitudes (see e.g. \cite{hanPI,Han:2023cen,Han:2024lti,Han:2021kll}):
\begin{eqnarray}
S_{\rm SF}=\sum_h k_h\log E_h(g,z)+\sum_b k_b\log E_b(g,z,\xi),\qquad k_h,k_b\in\mathbb{N}_0.\label{traditionalspinfoamaction}
\end{eqnarray}
In this expression, the presence of the logarithm implies that the single-valuedness of $e^{S_{\rm SF}}$ relies on the discreteness of the spins $k_h, k_b \in \mathbb{N}_0$. In contrast, for the bulk part of the new action $S$ (the contribution from the internal faces $h$), the action depends linearly on $E_h$ rather than through its logarithm. Consequently, the single-valuedness of the bulk contribution no longer relies on the discreteness of the variables.

The new action $S$ inherits the continuous gauge symmetries of $S_{\rm SF}$ (see for example, \cite{Han:2023cen,Han:2024lti,semiclassical}):

\begin{enumerate}
    \item Vertex $\SLDC$ gauge symmetry: For each vertex $v$, the action is invariant under $g_{ve} \mapsto X_v g_{ve}$ and $z_{vf} \mapsto (X_v^\dagger)^{-1} z_{vf}$ with $X_v \in \SLDC$. This transformation leaves $Z_{vef} = g_{ve}^\dagger z_{vf}$ invariant. 
	To fix this gauge freedom, we set one of the $g_{ve}$ adjacent to each vertex $v$ equal to the identity.
    
    \item Edge $\Su$ gauge symmetry: For each internal edge $e$, the action is invariant under $g_{ve} \mapsto g_{ve} H_e$ with $H_e \in \Su$. This transformation acts on $Z_{vef}$ by a global $\Su$ rotation, leaving the inner products in $\mathcal{E}_{f,e}$ invariant. By the Iwasawa decomposition of $\Slc$ matrix, we gauge fix one of $g_{ve}$ and $g_{v'e}$ to upper triangular matrix with positive diagonals for every internal edge $e$. 
    \item Spinor scaling symmetry: For each vertex-face pair $(v,f)$, the action is invariant under $z_{vf} \mapsto \lambda_{vf} z_{vf}$ with $\lambda_{vf} \in \mathbb{C} \setminus \{0\}$. This scaling cancels out in ${E}_{f}$. 
    This gauge redundancy is fixed by setting one of the components of $z_{vf}$ to 1.

\end{enumerate}

We observe that while $S_{\rm SF}$ features a discrete symmetry $g_{ve} \to -g_{ve}$ for integer $k_h,k_b$ compatible to non-empty intertwiner space \cite{semiclassical}. This transformation is not generally a gauge symmetry of $S$. The reason is that such a sign flip can change the sign of $E_h$. Nevertheless, the transformation becomes a gauge symmetry of $S$ if the set of sign flips among the various $g_{ve}$'s is chosen such that the sign changes cancel within each $E_h$, leaving all $E_h$ invariant.

Before turning to the next section, we add a few remarks:
\begin{itemize}

\item By the expansion $e^{we^{-i\zeta}}=\sum_{k=0}^\infty\frac{w^k}{k!}e^{-ik\zeta}$, one can show the following iterated integral
\begin{equation}
\int_\R\frac{\rmd\zeta}{2\pi}\int_\R \rmd\lambda\,d_{\Lambda}(\lambda)\,e^{i\zeta\lambda-\eta\left(1-e^{-i\zeta}E\right)}=\mathcal{D}_{\Lambda}\left(\eta\, E\right)e^{-\eta\left[1-E\right]},\label{exactZL:zetalambda}
\end{equation}
where the integral over $\l$ has to be carried out first. With the auxiliary variable $\l$ and $\zeta$, the pure integral formula of $\sa$ can be expressed as 
\be
\sa&=&J^{2N_h}\int_{\R_{\geq0}^{N_h}} \prod_h\rmd\eta_h\int \rmd\mu\lt(g,z\rt)\int\prod_h\frac{\rmd\zeta_h}{2\pi}\int \prod_h\rmd \l_h\prod_hd_{J\Lambda}(J\l_h) \, e^{J S_1},\label{integralLambdaZeta37}\\
S_{1}&=& i\sum_h\zeta_h\l_h-\sum_h\eta_h\lt[1-e^{-i\zeta_h}E_h(g,z)\rt]+\sum_b k_b\log E_b(g,z,\xi).
\ee
Although this integral depends on the order of integration, it becomes independent of that order when an arbitrary compactly supported function is inserted into the integrand.
In $d_{J\Lambda}(J\l_h) = \chi(\Lambda^{-1} \l_h)\mathrm{Pol}(J\l_h)$, the $J$ dependent part $\mathrm{Pol}(J\l_h)$ is a polynomial, so it clearly does not influence the stationary phase analysis with the action $S_1$. Moreover, $\partial_g S_1=\partial_\eta S_1=\partial_z S_1=0$ gives the same critical point equations as $S$, with additional $\zeta_h=0$ and $ \l_h=\eta_h E_h(g,z)$ from the varying $\zeta_h$ and $\l_h$. 
The stationary phase integral \eqref{integralLambdaZeta37} and the on-shell equivalence of $S_1$ and $S$ provide a justification for the stationary phase analysis using $S$, bypassing the need for the polynomial approximation of $\cd_{J\L}$.

\item The sum over internal spins in \eqref{spinfoamAmpliutdeA} includes $k_h=0$. The formalism in \cite{Han:2026hyq} defines the amplitude on a fixed complex with the sum over $k_h\in\mathbb{Z}_+$ rather than $k_h\in\mathbb{N}_0$. Excluding $k_h=0$ replaces $d_{\Lambda}(k)$ by $d_{\Lambda}(k)(1-\delta_{k,0})$, which also has finite support, so the proof of Theorem \ref{exactZL:theorem} applies with $\mathcal{D}_{\Lambda}(c)$ replaced by $\mathcal{D}_{\Lambda}(c)-d_{\Lambda}(0)\,e^{-c}$. This adds $-d_{\Lambda}(0)\,e^{-\eta_h}$ to the factor $\mathcal{D}_{\Lambda}(\eta_hE_h)\,e^{-\eta_h(1-E_h)}$ of each internal face $h$ in the integrand of \eqref{pathintegral}. After the large spin scaling, the added term is $-\mathrm{Pol}(0)\,e^{-J\eta_h}$, since $d_{J\Lambda}(0)=\mathrm{Pol}(0)$, while the face factors are bounded by a polynomial in $J$, uniformly in $\eta_h$ and $(g,z)$ (see Theorem \ref{Q:thmApprox}). Hence, for the integral $\sa_\psi$ in \eqref{prefactor:Apsi} with $\psi$ supported in $\{\eta_h\geq\eta_0\ \text{for all}\ h\}$, $\eta_0>0$, the change is bounded by a polynomial in $J$ times $e^{-J\eta_0}$, and it does not affect the stationary phase analysis around critical points with $\eta_h>0$. If $\mathrm{Pol}(0)=0$, the two conventions coincide. In this work, for simplicity, we include the $k_h=0$ term in the sum.

\end{itemize}

\section{Real critical point and flat Regge geometry}\label{Real critical point and flat Regge geometry}

Since $|E_f| \leq 1$ for $f=h,b$, the real parts of both $S$ and $S_{\rm SF}$ are non-positive:
\begin{eqnarray}
	\re(S)\leq 0,\qquad \re(S_{\rm SF})\leq 0
\end{eqnarray}
so the large-$J$ spinfoam amplitude is suitable for stationary phase analysis. The dominant contribution of the integral comes from the configurations near $\re(S)=0$, i.e.\ $E_h=1$ for the internal faces (with $\eta_h>0$) and $|E_b|=1$ for the boundary faces.
Furthermore, taking the derivative of $S$ with respect to each integration variable $\eta_h,\,g_{ve},\,z_{vf}$ yields the critical-point equations:
\be
E_h\approx 1,\qquad \sum_h\eta_h \partial_{g,z} E_h+\sum_b k_b \frac{ \partial_{g,z} E_b}{E_b}\approx 0.\label{criticalEqnNew}
\ee
We use the symbol $\approx$ to denote equality at critical points. We refer to solutions of the critical-point equations that lie on the integration cycle of \eqref{integraletamu} as \emph{real critical points}.
We use this terminology to distinguish these real critical points from the complex critical points, which will be discussed in later sections.

In this work, our analysis is restricted to perturbations around a background real critical point. We consider only the perturbative expansion from the path integral, while neglecting all non-perturbative effects. Specifically, we disregard non-perturbative contributions coming from other real critical points that have finite distances from the background, as well as possible boundary contributions from the region $\eta_h = 0$ (see \cite{stationaryphase} for a discussion of the boundary contributions).

The critical-point equations remain invariant under the gauge symmetries discussed above, so critical points are naturally identified as gauge equivalence classes. The discrete transformation $g_{ve}\to-g_{ve}$ does not, in general, preserve a critical point: flipping the sign of $g_{ve}$ may change the sign of $E_h$, and thus may violate $E_h\approx 1$. Nonetheless, this transformation does preserve the critical point in cases where it acts as a gauge symmetry and leaves all $E_h$ invariant.

The critical-point equations \eqref{criticalEqnNew} are equivalent to the familiar critical point equations obtained from the standard spinfoam action $S_{\rm SF}$, upon identifying $\eta_h = k_h$:
\be
\log(E_h)\approx 2\pi i\Z,\qquad \sum_h k_h \frac{ \partial_{g,z} E_h}{E_h}+\sum_b k_b \frac{ \partial_{g,z} E_b}{E_b}\approx0.\label{criticalEqnOld}
\ee
Any real critical point of \eqref{criticalEqnOld} subject to a suitable nondegeneracy condition corresponds to a Regge geometry that satisfies the following constraint \cite{flatness,frankflat,LowE1}: Let us introduce $\eps^{(s)}_h$, the Lorentzian (boost) angle associated to the hinge $h$ and weighted by the orientations of 4-simplices: $\eps^{(s)}_h = \sum_{v\in \partial h} s_v\, \Theta_h(v)$.
Here, $s_v = \pm 1$ labels 4-dimensional orientation for each 4-simplex dual to vertex $v$, and $\Theta_h(v)$ is the dihedral angle at $h$ in the simplex. When all $s_v = 1$, $\eps^{(s)}_h$ coincides with the deficit angle $\eps_h$ in Regge calculus. 
At a nondegenerate, time-oriented solution of the second equation in \eqref{criticalEqnOld} with $|E_f|=1$, one has $E_h=\pm e^{i\g\eps^{(s)}_h/2}$, where the sign depends on the lift of the solution to $\Slc$ \footnote{
	The sign is $+$ for the lifts used in Appendix \ref{Deficit angle and dihedral angle}, where one can also find the discussion about time-oriented solutions.}.
Since a real critical point has in addition $E_h\approx1$, the Regge geometries arising from such real critical points must obey $\g\, \eps^{(s)}_h \in 2\pi \mathbb{Z}$ for every internal triangle $h$ \cite{frankflat}.
In the case where the critical point corresponds to uniform 4-simplex orientation, this is a flatness condition modulo $2\pi\Z$:
\be
\frac{\g \eps_h}{ 2\pi}\in \Z\ .
\ee

The equivalence between \eqref{criticalEqnNew} and \eqref{criticalEqnOld} establishes a correspondence between their real critical points, and thus relates the real critical points of \eqref{criticalEqnNew} to flat Regge geometry. Among these, we single out a real critical point, with critical-point data denoted by $(\mathring{\eta},\mathring{g},\mathring{z})$, as the one corresponding to the time-oriented flat Regge geometry which has $\eps_h=0$ and a uniform 4-simplex orientation. 
We denote the boundary data compatible to this critical point by $(\mathring{k}_b,\mathring{\xi}_{eb})$. Both the critical-point data and the boundary data satisfy the equations \eqref{criticalEqnNew}. The boundary data are Regge-like, in the sense that they specify a definite 3-dimensional Regge geometry for the boundary triangulation. 

Here we consider a finite triangulation with boundary. The existence of real critical points---including those satisfying the flatness condition---depends on the choice of compatible boundary data, e.g. $(k_b,\xi_{eb})=(\mathring k_b,\mathring \xi_{eb})$. For other generic boundary data, real critical points may not exist. This reflects a general property of integrals with complex actions: for $n$ real integration variables, the critical point equations are $n$ complex equations, equivalent to $2n$ real equations, so the system is over-constrained on the real integration variables. 

For boundary data that does not admit any real critical point but is sufficiently close to $(\mathring k_b,\mathring\xi_{eb})$, we analytically continue the $n$ critical point equations and solve them for the $n$ complexified variables. The system is generically not over-constrained. The complex solution is called the \emph{complex critical point}\footnote{
If its Hessian at the real critical point is nondegenerate on the gauge slice,
the holomorphic implicit function theorem provides a unique solution close to $(\mathring\eta,\mathring g,\mathring z)$, depending analytically on the boundary data.
This strong nondegeneracy assumption may fail, e.g.\ for triangulations with an internal vertex: moving such a vertex in the flat geometry gives a continuous family of real critical points with the same boundary data. The analysis of the following sections uses only a weaker nondegeneracy assumption: the nondegeneracy of Hessian of \eqref{tgzintegral} at fixed lengths $\ell$. Moving the internal vertex, which change $\ell$, are not directions of the integrals in \eqref{tgzintegral}. The weaker nondegeneracy can be confirmed by numerical experiments \cite{companionNumer}
}.
Recent advancements in the numerical analysis of spinfoam amplitudes have revealed that curved geometries emerge from complex critical points of spinfoam amplitude \cite{Han:2023cen,Han:2021kll,Han:2024ydv}.

In the following, we take the real critical point and boundary data $(\mathring{\eta},\mathring{g},\mathring{z},\mathring{k},\mathring{\xi})$ of flat geometry as the background configuration, and analyze perturbations :
\be
({\eta}_h,{g}_{ve},{z}_{vf},{k}_b,{\xi}_{eb})=(\mathring{\eta}_h,\mathring{g}_{ve},\mathring{z}_{vf},\mathring{k}_b,\mathring{\xi}_{eb})+ (\delta{\eta}_h,\delta{g}_{ve},\delta{z}_{vf},\delta{k}_b,\delta{\xi}_{eb}).\label{perturbSetup}
\ee 
By \eqref{kscaling} and in the large-$J$ limit, we can treat $k_b$ as a continuous variable to a good approximation, justifying the validity of considering small perturbations $\delta k_b$. 
The lengths of the background flat Regge geometry are denoted by $\mathring{\ell}$.

\section{Regge and non-Regge areas}\label{Regge and non-Regge areas}

The variables $\eta_h$ are naturally interpreted as areas, as they play the same role as the spins in earlier analyses of spinfoam asymptotics. This interpretation is confirmed both by the analysis of the real critical points discussed above and by the behavior of perturbations considered later. The variables $\{\eta_h\}_h$ range over the integration domain $\R^{N_h}_{\geq0}$ of \eqref{integraletamu}.

\begin{figure}[h]
	\centering
	\includegraphics[width=0.3\textwidth]{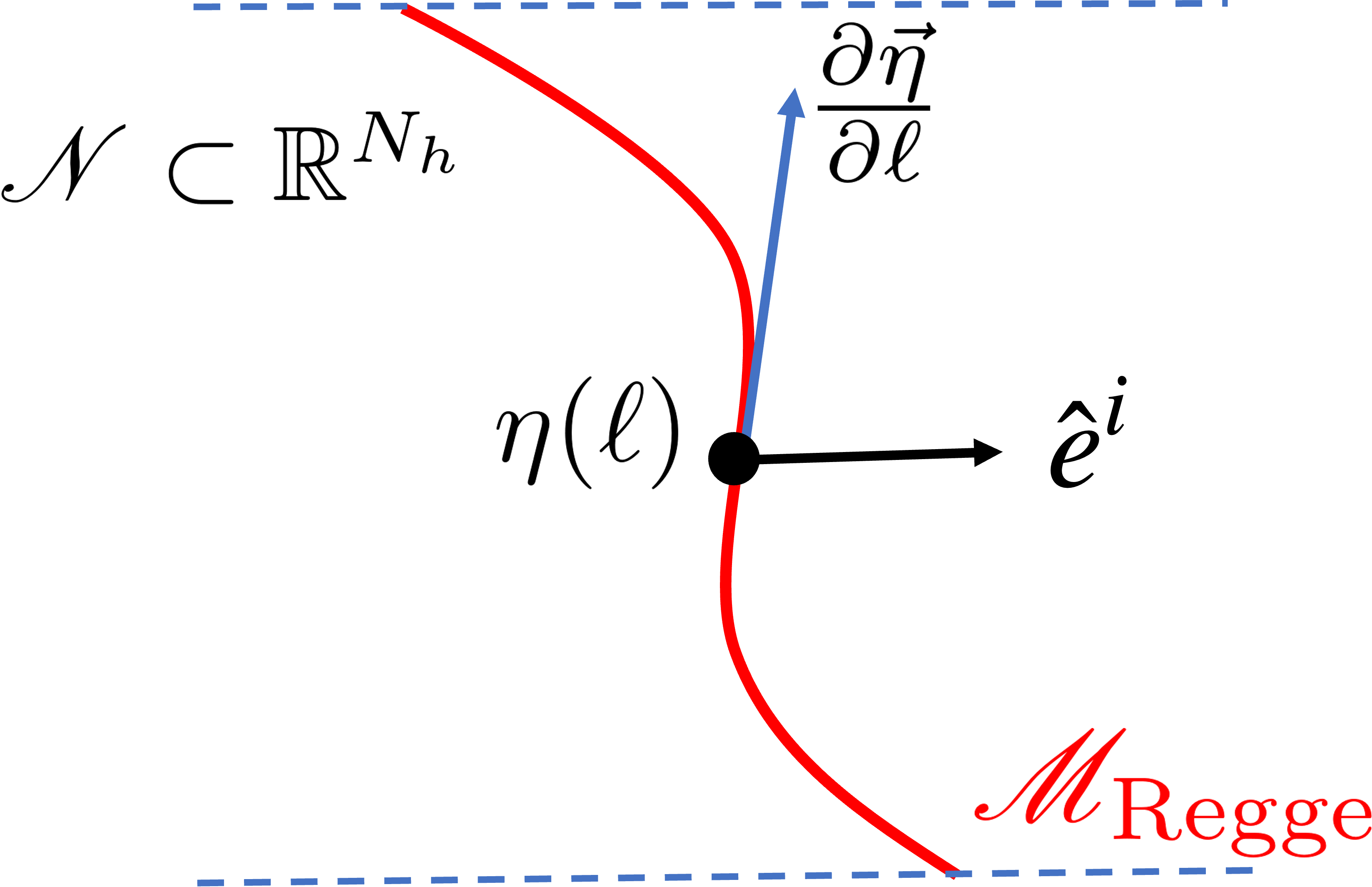}
	\caption{The Regge submanifold $\sm_{\rm Regge}$ and the transverse vectors $\hat e^i$, $i=1,\cdots,N_h-N_\ell$.}
	\label{fig:regge_submanifold}
\end{figure}

Recall that $\ck$ is a 2-complex dual to a simplicial complex $\ck^*$. Let $N_\ell$ be the number of internal segments $s$ of $\ck^*$, and let $\cu_\ell\subset\R^{N_\ell}_{>0}$ be the open set of length assignments $\ell_s$ to the internal segments which,
together with the fixed lengths of the boundary segments, define a nondegenerate Regge geometry with spacelike triangles.
Heron's formula defines a smooth map from $\cu_\ell$ to the space of areas $\R^{N_h}_{\geq}$, sending each collection of lengths $\ell$ to a collection of (spacelike) triangle areas $\eta_h$:
\be
\fa_h(\ell)=\frac{\g}{2}\eta_h(\ell)=\frac{1}{4} \sqrt{2\left(\ell_{i j}^2 \ell_{j k}^2+\ell_{i k}^2 \ell_{j k}^2+\ell_{i j}^2 \ell_{i k}^2\right)-\ell_{i j}^4-\ell_{i k}^4-\ell_{j k}^4} ,
\ee
where $\ell_{ij},\ell_{jk},\ell_{ik}$ are the lengths of the three segments bounding the triangle $h^*$ dual to $h$ (some of them may be boundary segments, whose lengths are fixed by the boundary data),
and lengths are measured in units in which $\fa_h=\g\eta_h/2$.
We assume that the Jacobian matrix $\partial\eta_h/\partial\ell_s$ has maximal rank $N_\ell$ at the background $\mathring\ell$ \footnote{
	One always has $N_h\geq\frac{4}{3}N_\ell$, so that $N_h>N_\ell$ as soon as $N_\ell\geq1$. Indeed, let us count the set of pairs:
	$$P=\{(s,t)|\ s\ \text{an internal segment},\ t\ \text{a triangle},\ s\subset t\}.$$ 
	A triangle containing an internal segment is internal (the boundary is a subcomplex).
	Every internal segment $s$ is contained in at least four triangles: three of them lie in a 4-simplex $\sigma\supset s$, and a tetrahedron of $\sigma$ containing $s$ is not a boundary tetrahedron, hence is shared with a second 4-simplex, whose fifth vertex spans with $s$ a fourth triangle. Therefore $|P|\geq 4N_\ell$.
	Every triangle contains at most three internal segments, so $|P|\leq 3N_h$. Combining two bounds gives $4N_\ell\leq |P|\leq 3N_h$.
	The rank condition is an additional nondegeneracy assumption.
}.
Then the map restricted to a neighborhood of $\mathring\ell$ is an embedding. Its image, denoted by $\sm_{\rm Regge}$, is an $N_\ell$-dimensional submanifold of the space of areas, called the \emph{Regge submanifold} (see FIG.\ref{fig:regge_submanifold}).


The flat geometry corresponding to the real critical point $(\mathring{\eta},\mathring{g},\mathring{z})$ determines a point $\mathring{\eta}=\eta(\mathring\ell)\in \sm_{\rm Regge}$, where $\mathring{\ell}$ represents the set of lengths for the flat geometry determined by the real critical point $(\mathring{\eta},\mathring{g},\mathring{z})$ and boundary data $(\mathring{k},\mathring{\xi})$. 
Since $\ell$ are functions defined on the Regge submanifold $\sm_{\rm Regge}$, let us select the compact neighborhood $\sn$ of the point $\eta(\mathring\ell)$ in $\eta$-space, and arbitrarily extend the functions $\ell$ off $\sm_{\rm Regge}$ to all of $\sn$. With such an extension chosen, we introduce the delta function $\delta_{\ell'}(\eta)$ supported on the set where $\ell=\ell'$, together with a measure $\rmd\rho(\ell)$ normalized such that $\int \rmd\rho(\ell)\,\delta_\ell(\eta)=1$ for all $\eta\in\sn$. 
We then insert this delta function into \eqref{integraletamu} so as to define the effective action $\ci_{\rm eff}$ by
\be
e^{J\ci_{\rm eff}(\ell)}=J^{N_h}\int \prod_h\rmd\eta_h\delta_\ell(\eta)\int \rmd\mu\lt(g,z\rt)\,\psi(\eta,g,z)\prod_h\mathcal{D}_{J\Lambda}\left(J\eta_hE_h\right)   e^{J S}.\label{expEffectiveAction}
\ee
The function $\psi$ has compact support on $\sn\times U$, with $U$ a compact neighborhood of $(\mathring g,\mathring z)$.
Then the integral $\sa_\psi$ is given by
\be
\mathscr{A}_{\psi}=\int \rmd\rho(\ell)\, e^{J \ci_{\rm eff} (\ell)}.
\ee
In this analysis, we restrict the integration to the compact neighborhood $\sn\times U$, as our primary interest lies in examining perturbations in the vicinity of the critical point $(\mathring{\eta},\mathring{g},\mathring{z})$.

The effective theory from the spinfoam amplitude is given by the effective action $\ci_{\rm eff} (\ell)$ as a function of lengths. 
In what follows, we aim to expand $\ci_{\rm eff}(\ell)$ around the flat background geometry, keeping terms up to second order in the perturbations $\delta \ell = \ell - \mathring{\ell}$.
Through this expansion, we will see that $\ci_{\rm eff}(\ell)$ provides corrections to the classical Regge action, incorporating higher-curvature contributions.

The construction of the effective action $\ci_{\rm eff}$ as given in \eqref{expEffectiveAction} depends on how we extend the definition of $\ell$ away from the Regge submanifold. One way to specify this extension is to select $(N_h-N_\ell)$ constant vectors $\{\hat{e}^i\}_{i=1}^{N_h-N_\ell}$ whose span is complementary to the tangent space of $\sm_{\rm Regge}$ at $\eta(\mathring\ell)$. With this choice, any point in the neighborhood $\sn$ can be expressed as
\be
\vec{\eta}(\ell,t)=\vec{\eta}(\ell)+\sum_{i=1}^{N_h-N_\ell} t_i \hat{e}^i,
\ee
and thus $(\ell,t)$ forms the local coordinates of $\sn$. This means $\ell$'s are extended as constants along the plane spanned by $\hat{e}^i$. The variables $t_i$ parametrize the perturbations corresponding to the non-Regge areas (the areas that are not given by lengths). The integral over $\eta_h$ can be re-written by the change of variable $\vec\eta\to (\ell,t)$
\be
\int_\sn\rmd^{N_h}\eta \cdots= \int_\sn \rmd^{N_\ell} \ell\,\rmd^{N_h-N_\ell}t\,\cj(\ell)\cdots,\qquad \cj(\ell)=\lt|\det\lt(\frac{\partial \vec{\eta}(\ell)}{\partial \ell},\hat{e}^i\rt)\rt|.\label{Vc:changevar}
\ee
The jacobian $\cj$ only depends on $\ell$ since $\hat{e}^i$ are constant vectors. 
The method of changing variables from areas to lengths and non-Regge areas was proposed in \cite{Han:2017xwo,Han:2018fmu} then applied to the numerical analysis of spinfoam amplitudes \cite{Han:2023cen,Han:2024ydv} and the area-Regge calculus \cite{zichangnote,Dittrich:2022yoo,Dittrich:2021kzs}.

We obtain the following expression of $\sa_\psi$
\be
\mathscr{A}_{\psi}&=&\int \rmd^{N_\ell} \ell\, \cj(\ell)\, e^{J \ci_{\rm eff} (\ell)},\label{saNepsintegral}\\
e^{J \ci_{\rm eff}(\ell)}&=&J^{N_h}\int \rmd^{N_h-N_\ell}t \int \rmd\mu\lt(g,z\rt) \,\psi\lt(\eta(\ell,t),g,z\rt)\prod_h\mathcal{D}_{J\Lambda}\left(J\eta_h(\ell,t)E_h\right) e^{J S}.\label{tgzintegral}
\ee
The function $\psi$ has compact support on $\sn\times U$, which encloses the background critical point.

In \eqref{tgzintegral}, the dependence of the effective action $\ci_{\rm eff}(\ell)$ on the way that $\ell$ is extended off the Regge submanifold reduces to the choice of the transverse plane spanned by the vectors $\hat{e}^i$.
However, the quadratic effective action at the leading order in $1/J$ is independent of the choice of basis within a given transverse plane, as will be seen explicitly in Section \ref{Higher curvature corrections}. Moreover, as shown in Appendix \ref{Change of transverse plane and field redefinition}, an infinitesimal change of the transverse plane amounts to a ``field redefinition'' $\ell\to\tilde\ell+\Delta\tilde\ell$ in $\ci_{\rm eff}$. Finally, as is manifest from the outset, the path integral $\mathscr{A}_{\psi}$ is independent of the choice.

\section{Scheme for computing the quadratic effective action}\label{Scheme for computing the quadratic effective action}

With the change of variable $\vec{\eta}\to (\ell,t)$, we classify the perturbations of the degrees of freedom as follows:
\be
\lt(\delta \ell_s,\delta t_i,\delta g_{ve},\delta z_{vf}\rt)=\lt(\ell_s, t_i,g_{ve},z_{vf}\rt)-(\mathring\ell_s, \mathring{t}_i,\mathring g_{ve},\mathring z_{vf}),\qquad \mathring{t}_i=0.
\ee
The background configuration is the real critical point $(\mathring\eta =\eta (\mathring \ell),\mathring g,\mathring z)$ corresponding to the flat geometry. 
For the perturbation of boundary data, we restrict to Regge-like perturbations, so that $({k}_b,{\xi}_{eb})=(\mathring{k}_b,\mathring{\xi}_{eb})+(\delta{k}_b,\delta{\xi}_{eb})$ are always parametrized by perturbations of the boundary segment lengths.

Although the perturbations are small, their range is assumed to be large compared with the scale of the $1/J$ corrections. We denote by $\Delta\ell$ the range of length perturbations, i.e. $\|\delta\ell\|\leq \Delta\ell$. The one-loop contributions to $\ci_{\rm eff}$ are of the order $J^{-1}\Delta\ell$ and $J^{-1}\Delta\ell^2$, while the Regge term is of order $\Delta\ell^2$ and the $\eps^2$-correction derived in Section~\ref{Higher curvature corrections} is at most of order $\g\,\Delta\ell^2$ for small $\g$ (Section~\ref{Small gamma scaling}). We therefore require $\Delta\ell\gg(J\g)^{-1}$ and $J\g\gg 1$, and ignore the $\co(1/J)$ corrections to the effective action in the following discussion\footnote{
	We ignore the constant term in the action. If $\ell$-dependent part of the one-loop term is of $O(\g^m)$, $m\geq 0$, we need $\g^m\Delta \ell^a\ll J\Delta\ell ^2$ and $\g^m\Delta \ell^a\ll \g J\Delta\ell ^2 $ with $a=1,2$, i.e. $\Delta\ell \gg \g^m J^{-1}$, $J\gg \g^m$ and $\Delta \ell \gg \g^m(\g J)^{-1}$, $\g J\gg \g^m$, which are implied by $\Delta \ell \gg (\g J)^{-1} $, $\g J\gg 1$ for $\g\leq 1$.
}.



Let $U$ be a small neighborhood around the background configuration $(\mathring{g},\mathring{z})$. This background configuration satisfies
\be
\log(E_h)\big|_{\mathring{g},\mathring{z}}=i\g\mathring\eps_h/2=0.
\ee
We can make the following expansion of $S$ within $U$:
\be
S= \sum_h\eta_h\log E_h(g,z)+\frac{1}{2}\sum_h\eta_h\lt[\log E_h(g,z)\rt]^2+\sum_b k_b\log E_b(g,z,\xi)+\co\lt(\lt(\log E_h\rt)^3\rt).\label{expansionSprimeinE}
\ee
A key feature of $S$, in contrast to the traditional spinfoam action $S_{\rm SF}$ in~\eqref{traditionalspinfoamaction}, is the presence of a series of higher-order terms in $\log E_h$. This arises because $S$ is expressed in terms of $E_h$ rather than directly in terms of $\log E_h$. Given the relation between $E_h$ and the deficit angle $\eps_h$ (see also \eqref{angleCorrespond} below), 
the difference between $E_h$ and $\log E_h$ reflects the difference between holonomy and curvature, so we refer to the higher-order terms in $\log E_h$ as holonomy corrections. 
As we will show in Section \ref{Higher curvature corrections}, these terms provide a part of $\eps^2$-correction to the Regge action in $\ci_{\rm eff}(\ell)$.

We now proceed to compute the effective action $\ci_{\rm eff}(\ell)$ up to quadratic order in the perturbations $\delta\ell$. This requires expanding the action $S$ perturbatively and retaining terms up to second order in the fluctuations.
We use the following collective notation for bulk integration variables and boundary variables in the integral \eqref{tgzintegral} 
\be
\cx_A=(t_i, X_I),\qquad X_{I}\equiv\left( g_{\alpha}, z_{j}\right)\qquad \cy_M = (k_b(\ell), \xi_m(\ell)).
\ee 
where $\alpha$, ${j}$ and $m$ are short-hand notations for $(v,e)$, $(v,f)$ and $(e,b)$ respectively.
The perturbations are denoted by $\delta \cx_A$, $\delta X_I$ and $\delta\cy_M$. Here the boundary perturbations are Regge-like, i.e. $\delta\cy_M=\lt( \delta k_b(\ell),\delta\xi_{m}(\ell)\rt)$ are determined by $\delta\ell$. From now on $\delta\ell$ includes the boundary lengths. 
Here $X_I$ are real local coordinates on the gauge-fixed integration domain (e.g.\ the real and imaginary parts of the free entries of the $g_{ve}$ and of the affine coordinates of the $z_{vf}$), and $\partial/\partial X_I$ denotes derivatives with respect to these real coordinates. Complex critical points refer to the holomorphic extension of the real-analytic action $S$.
We have already completely fixed the continuous gauge symmetries, so the perturbations do not contain any gauge redundancies.

We expand $S$ around the flat background geometry and ignore terms higher than quadratic order:
\be
S -\mathring{S} \simeq \cb_A \delta \cx_A + \frac{1}{2}{\cal H}_{AB}\delta{\cal X}_{A}\delta{\cal X}_{B} + \ci^{(2)}_{\rm bdy}.\label{quadraticS'-S00000}
\ee
In the linear term $\cb_A$ is proportional to length variations $\delta\ell$, which are external parameters of the integral \eqref{tgzintegral}. 
The term $\mathcal{I}^{(2)}_{\text{bdy}}$ collects all contributions only involving the fluctuations $\delta\cy_M$ of the boundary variables.
Here, we focus on the bulk contributions. Additional discussion and explicit expressions for the boundary terms can be found in the subsequent sections and in Appendix \ref{Perturbative computation: boundary action}.

Assuming $\ch$ is nondegenerate, we solve the linearized critical-point equations $\delta S/\delta \cx_A=0$ and substitute the solution back into $S$. The solution is a complex critical point of the integral \eqref{tgzintegral}. At the level of the quadratic perturbation of action, this is equivalent to performing the Gaussian integral over $\delta\cx_A$. We obtain the following quadratic effective action $\ci_{\rm eff}(\ell)$:
\be
\lt[\ci_{\rm eff}(\ell)  -\mathring{S}\rt]^{(2)} = -\frac{1}{2} \cb(\ell)^\top  \ch^{-1}  \cb(\ell) + \ci^{(2)}_{\rm bdy}\ .\label{quadraticEffectiveAction0000}
\ee

Let us examine in more detail the quantities $\cb_A$ and $\ch_{AB}$ that appear in the effective action~\eqref{quadraticEffectiveAction0000}.
For $\cb_A$, we note that $\cb_t = 0$ since $\log(E_h) = 0$ at the background configuration. For the components associated with $X$, we have $\cb_X = B + C$, where $C$ depends only on the boundary fluctuations $\delta\cy_M$, and
\be
B_I = \sum_h v_h V_{I,h},\qquad V_{I,h} = \frac{\partial \log E_h}{\partial X_I}=\frac{\partial E_h}{\partial X_I},\qquad v_h=\sum_s\frac{\partial \eta_h}{\partial \ell_s} \delta\ell_s,\label{BVv49} 
\ee
where all quantities are evaluated at the background critical point.
The Hessian matrix $\ch_{AB}$ of the integrals in \eqref{tgzintegral} is given by:
\be
\mathcal{H}_{tt} = 0, \qquad \mathcal{H}_{tX} =  \hat{e}^\top V^\top, \qquad \mathcal{H}_{XX} =\mathfrak{h} + V A V^\top,\qquad A_{h_1, h_2} = \mathring{\eta}_{h_1} \delta_{h_1, h_2},\label{HHHH50}
\ee
where the entries of the matrix $\mathcal{H}_{tX}$ are $(\mathcal{H}_{tX})_{iI}= \sum_h V_{I,h}\hat{e}^i_h$. The matrix $\fh_{IJ}$ is given by
\be
\mathfrak{h}_{IJ} = \sum_h \mathring{\eta}_h \frac{\partial^2 \log E_h}{\partial X_I \partial X_J} + \sum_b \mathring{k}_b \frac{\partial^2 \log E_b}{\partial X_I \partial X_J},
\ee
which is the Hessian matrix of $S_0$ in $X_I$, where $S_0$ represents the leading-order term in the expansion \eqref{expansionSprimeinE} of the full action $S$:
\be
S_0= \sum_h\eta_h\log E_h(g,z)+\sum_b k_b\log E_b(g,z,\xi).
\ee
The expression $\mathcal{H}_{XX} =\mathfrak{h} + V A V^\top$ in \eqref{HHHH50} arises from the expansion in \eqref{expansionSprimeinE}. 

Two properties of these background quantities will be used below. Since $|E_f|\leq1$ achieves its maximum $|E_f|=1$ at the background, $\partial|E_f|/\partial X_I=0$ there, so that $V_{I,f}=\partial\log E_f/\partial X_I$ is purely imaginary. Since $\re S_0=\sum_h\mathring\eta_h\log|E_h|+\sum_b\mathring k_b\log|E_b|\leq0$ achieves its maximum at the background, $\re\,\fh$ is negative semi-definite.

In the following sections, we will clarify how the effective action $\ci_{\rm eff}$ connects to Regge geometric variables, specifically areas and deficit angles. Understanding this connection is key for interpreting the effective action as it governs the dynamics of discrete geometries. To this end, it suffices to examine how the matrices $V_{I,f}$ and $\fh_{IJ}$ are related to geometric quantities, while other ingredients of $\cb_A$ and $\ch_{AB}$ in \eqref{BVv49} and \eqref{HHHH50} already admit geometric interpretations.
In Sections \ref{A toy integral and Regge action} and \ref{Perturbative geometrical identities}, we show that the relation between $V_{I,f}, \fh_{IJ}$ and geometry can be obtained by the stationary phase analysis of the leading-order action $S_0$.

Before coming to the explicit derivation of the effective action, it is useful to mention its connection to a two-parameter expansion. In particular, the asymptotic form of $e^{J\ci_{\rm eff}}$ appearing in \eqref{tgzintegral} can be described schematically as
\be
e^{J\ci_{\rm eff}(\ell)}=J^{\varsigma}{\cn}(\ell)\, e^{J\lt[\mathring{S}-\frac{1}{2}\cb(\ell)^\top  \ch^{-1} \cb(\ell) + \ci^{(2)}_{\rm bdy}+\co(\delta \ell^3)\rt]}\lt[1+\co(1/J)\rt]+\calr_1,\label{expansionJR}
\ee
where $\varsigma\in\frac12\Z$ and $\calr_1$ is a remainder of $\co(J^{-m})$ for every $m>0$ \cite{stationaryphase}.
As we will demonstrate explicitly in Section \ref{Higher curvature corrections}, the term $\cb(\ell)^\top  \ch^{-1} \cb(\ell)$ in the exponent can be expanded in powers of the Barbero-Immirzi parameter $\g$. In the regime of small $\g$, the leading term reproduces the Regge action, while the subleading term yields higher-curvature corrections. 
Therefore, the expansion \eqref{expansionJR} should be understood as a two-parameter expansion: in $1/J$ and in $\g$. The expansion in $\g$ is associated with the inclusion of higher-curvature corrections to the gravitational effective action.

As we see in Section \ref{Higher curvature corrections}, the exponent equals, up to second order in $\delta\ell$, $J\big[\mathring{S}+iS^{(2)}_{\rm Regge}-\frac{\g^2}{8}\sum_{h_1h_2}K_{h_1,h_2}\eps_{h_1}\eps_{h_2}\big]$. Since $\mathring{S}=\frac{i\g}{2}\sum_b\mathring{k}_b\mathring{\Theta}_b$ is purely imaginary and $S^{(2)}_{\rm Regge}$ is real, its real part is $-\frac{J\g^2}{8}\,\re\big(\eps^\top K\eps\big)$, which is non-positive by the remark following Eq.~\eqref{Kinverse} and has the range of order $J\g\,\Delta\ell^2$ by Section~\ref{Small gamma scaling}. For the leading exponential term not to be suppressed below the remainder $\calr_1$, $J\g^2\,\re(\eps^\top K\eps)$ must be at most of order $\log J$: $J\g\,\Delta\ell^2\lesssim \log J$. On the other hand, the ``1-loop correction'' arising from $\log\cn$ can contribute at order $\co(\delta\ell)$ (a constant term in the action does not affect the equations of motion, so we ignore the constant part of $\log\cn$). Assuming, as power counting suggests, that the $\ell$-dependent part of $\log\cn$ is of order one in $\g$, the $\eps^2$-correction dominates it when $J\g\,\Delta\ell^2\gg\Delta\ell$, i.e.\ $\Delta\ell\gg(J\g)^{-1}$ (recall that $J\g\gg1$). Both conditions hold in the window $(J\g)^{-1}\ll\Delta\ell\lesssim(J\g)^{-1/2}(\log J)^{1/2}$, which is non-empty provided $(J\g)^{1/2}(\log J)^{1/2}\gg1$. Since $J$ and $\g$ can scale independently, the expansion of the effective action proceeds in terms of these two parameters.

\section{A toy integral and Regge action}\label{A toy integral and Regge action}


Let us consider the following ``toy integral'' for the stationary phase analysis of the leading-order action $S_0$:
\be
Z(\ell):=\int \rmd\mu\lt(g,z\rt) \,\psi(g,z)\, e^{J S_0},\qquad S_0= \sum_h\eta_h(\ell)\log E_h(g,z)+\sum_b k_b\log E_b(g,z,\xi),\label{gzintegralS0}
\ee
where $\psi$ has the support on a sufficiently small ball around the background $(\mathring g,\mathring z)$ on which $|1-E_h|<\delta<1$ for all $h$. Here $\log E_h$ is the principal branch. Specifying the branch is needed because $J\eta_h(\ell)$ are in general not integers. 

In the integral $Z(\ell)$, the areas $\eta_h$ are treated as external variables and are constrained to lie on the Regge submanifold.
Real critical points of this integral have Regge-geometric interpretations. In fact, the integral above is precisely the one underlying the analysis of the spinfoam large-$j$ asymptotics at fixed spins (see, e.g.,~\cite{hanPI}), since here $S_0$ reduces to the standard action $S_{\rm SF}$ upon identifying $k_h = \eta_h(\ell)$.

For clarity, the toy integral $Z(\ell)$ differs from the actual integral \eqref{tgzintegral} used to compute the effective action in following ways:
\begin{itemize}
\item We disregard perturbations in $t$ by fixing $t_i = \mathring{t}_i = 0$, and focus exclusively on the effects of perturbations in the remaining degrees of freedom;
\item We ignore terms of order $[\log E_h(g,z)]^2$ and higher in $S$;
\item The non-exponential prefactor $J^{N_h}\psi\prod_h\mathcal{D}_{J\Lambda}$ of \eqref{tgzintegral} is replaced by $\psi$. But such a factor does not contribute to the action, so it affects neither the critical points nor the value of the exponent.
\end{itemize}

Although the toy integral $Z(\ell)$ is different from the actual integral, it turns out to be useful for deriving the relations between matrices $V_{I,h}, \fh_{IJ}$ and Regge-geometrical quantities. 

In the rest of this section, we present the correspondence between the critical point data of the integral $Z(\ell)$ and Regge geometric quantities. 
As explained in Section~\ref{Perturbative geometrical identities}, this correspondence results in that $V_{I,h}$ and $\fh_{IJ}$---matrices closely related to $S_0$---are connected to Regge geometric data.


The critical point equations from $\partial_{g,z} S_0\approx 0$ are given by 
\be
\sum_h\eta_h(\ell) \frac{\partial_{g,z} E_h}{E_h}+\sum_b k_b \frac{\partial_{g,z} E_b}{E_b}\approx 0\label{crticalEqgandz}
\ee
We will only consider the linearized equations (see \eqref{linearCritEqn}) since we only expand the action to quadratic order in perturbations. 

In addition to \eqref{crticalEqgandz}, the critical points we consider satisfy
\be
|E_h|=|E_b|=1.\label{moduleUnit}
\ee
At linear order this is automatic: $|E_f|\leq1$ attains its maximum value $1$ at the background, so $\delta|E_f|=0$ for every first-order perturbation of $(g_{ve},z_{vf},\xi_{eb})$. By Appendix~\ref{Geometrical correspondence of spinfoam critical point}, for every $\ell$ in a neighbourhood of $\mathring\ell$, there are data $\Phi(\ell)=(\eta_h(\ell),g_{ve}(\ell),z_{vf}(\ell),k_b(\ell),\xi_{eb}(\ell))$ solving \eqref{crticalEqgandz} and \eqref{moduleUnit}. On the other hand, if the Hessian $\fh$ of $S_0$ at the background is nondegenerate (after gauge-fixing), the holomorphic implicit function theorem, applied to \eqref{crticalEqgandz} continued holomorphically in the real coordinates $X_I$ of $(g,z)$, implies that, for external data $(\eta(\ell),k(\ell),\xi(\ell))$ close to the background, \eqref{crticalEqgandz} has a unique, a priori complex, solution near $(\mathring g,\mathring z)$, depending continuously on $\ell$. The gauge-fixed real point $(g(\ell),z(\ell))$ of $\Phi(\ell)$ solves \eqref{crticalEqgandz} with the same external data, depends continuously on $\ell$ and equals $(\mathring g,\mathring z)$ at $\ell=\mathring\ell$, so it lies in this neighbourhood for $\ell$ close to $\mathring\ell$. This unique solution is therefore the real point $(g(\ell),z(\ell))$ of $\Phi(\ell)$; in particular, it is real and satisfies \eqref{moduleUnit}.


The system of equations \eqref{crticalEqgandz} and \eqref{moduleUnit} matches the one studied in \cite{hanPI,HZ} for the asymptotic analysis of the spinfoam amplitude with large, fixed Regge-like spins. As a result, there is a one-to-one correspondence between perturbative Regge geometries on the flat background and the critical points of $Z(\ell)$, i.e.\ the spinfoam data $(\eta_h, g_{ve}, z_{vf}, k_b, \xi_{eb})$ that solve \eqref{crticalEqgandz} and \eqref{moduleUnit} (modulo gauge freedom, with boundary frames and phases fixed).
A self-contained review of the correspondence is provided in Appendix \ref{Geometrical correspondence of spinfoam critical point}. Perturbative Regge geometries are characterized by the lengths $\ell=\mathring\ell+\delta\ell$ of all segments both in the bulk and on the boundary. 
Consequently, the segment lengths $\ell$ provide a complete parametrization of the corresponding spinfoam configurations, allowing us to write $(\eta_h(\ell), g_{ve}(\ell), z_{vf}(\ell), k_b(\ell), \xi_{eb}(\ell))$.
This correspondence provides a geometric interpretation for the spinfoam data that satisfy the equations: specifically, $\eta_h(\ell)$ and $k_b(\ell)$ correspond to the areas of triangles, given by
\be
\fa_h\approx\lt(\g \eta_h(\ell)/2\rt)J\ell_P^2,\qquad \fa_b\approx\lt(\g k_b(\ell)/2\rt)J\ell_P^2,\qquad \ell_P^2=8\pi G\hbar. \label{areaCorrespond}
\ee 
We evaluate $E_h$ and $E_b$ at the spinfoam data determined by $(\eta_h(\ell), g_{ve}(\ell), z_{vf}(\ell), k_b(\ell), \xi_{eb}(\ell))$. These quantities relate to the deficit angle $\eps_h(\ell)$ and the boundary dihedral angle $\Theta_b(\ell)$ \footnote{
	Let $\Theta_f(v)$ denote the dihedral angle at triangle $f$ within the 4-simplex dual to vertex $v$. The deficit angle associated with an internal face $h$ is then $\eps_h(\ell) = \sum_{v \in \partial h} \Theta_h(v)$, while for a boundary face $b$, the total dihedral angle is $\Theta_b(\ell) = \sum_{v \in \partial b} \Theta_b(v)$. In the geometric interpretation of the EPRL spinfoam critical points, all triangles are spacelike, so all these dihedral angles $\Theta_f(v)$ correspond to Lorentzian boost angles.
}
via
\be
E_h\approx e^{i\g \eps_h(\ell)/2},\qquad E_b\approx e^{i\g \Theta_b(\ell)/2}.\label{angleCorrespond}
\ee
See Appendix \ref{Deficit angle and dihedral angle} for explanations of these relations. In the second relation the phases of the boundary spinors $\xi_{eb}(\ell)$ are chosen, smoothly in $\ell$, such that the boundary phase $\phi_b$ of Appendix~\ref{Deficit angle and dihedral angle} vanishes for all $\ell$, not only at the background; this is possible because a change of phase of $\xi_{eb}$ multiplies $E_b$ by a constant phase. 


At leading order, the stationary phase approximation of $Z(\ell)$ is obtained by evaluating the exponent at the critical point. Substituting the expressions above into $S_0$ yields the Regge action $S_{\rm Regge}$,
\be
S_0\approx iS_{\rm Regge},\qquad S_{\rm Regge}=\frac{1}{2}\sum_h\g\eta_h(\ell)\eps_h(\ell)+\frac{1}{2}\sum_b\g k_b(\ell)\Theta_b(\ell),\label{S074}
\ee
in the phase convention just described.
To analyze perturbations around the flat background, we consider the difference between $S_0$ evaluated on the perturbed critical point and the background action $\mathring{S}_0 = i \sum_b \gamma \mathring{k}_b \mathring{\Theta}_b / 2 $. Here, $\delta\ell = \ell - \mathring{\ell}$ represents small deviations in both bulk and boundary segment lengths. We expand $S_0 - \mathring{S}_0$ to quadratic order in $\delta\ell$:
\be
S_0-\mathring{S}_0\approx iS_{\rm Regge}^{(2)}+\co\lt(\delta\ell^3\rt),\label{S-S=Sreggeequation47}
\ee
where $S_{\rm Regge}^{(2)}$ is the expansion of the Regge action to quadratic order, including the boundary term
\footnote{
	The Schl\"afli identity $\sum_h \eta_h(\ell)\delta \eps_h(\ell)+\sum_b k_b(\ell)\delta\Theta_b(\ell)=0$ is used in the derivation.
}:
\be
S_{\rm Regge}^{(2)} 
 = \frac{\gamma}{4}\sum_{r,s}\lt[\sum_h
   \frac{\partial\eta_h}{\partial\ell_s}\,
   \frac{\partial\varepsilon_h}{\partial\ell_r}\rt]\,
   \delta\ell_s\delta\ell_r+S_{\rm bdy}.
\ee
The boundary term $S_{\rm bdy}$ reads
\be
   S_{\rm bdy}=\frac{\gamma}{2}\sum_{b}\Delta k_{b}(\ell)\mathring{\Theta}_b+\frac{\gamma}{4}\sum_{r,s}\lt[\sum_b
   \frac{\partial k_b}{\partial\ell_s}\,
   \frac{\partial\Theta_b}{\partial\ell_r}\rt]\,
   \delta\ell_s\delta\ell_r,\label{ReggeBoundaryTerm}
\ee
where the operator $\Delta$ denotes the variations up to second order in $\delta \ell$:
\be
\Delta k_b(\ell) = \delta k_b(\ell) + \frac{1}{2} \delta^{(2)} k_b(\ell),\qquad 
\delta k_b(\ell)=\sum_s\frac{\partial  k_b}{\partial\ell_s}\delta\ell_s,\qquad \delta^{(2)}  k_b(\ell)=\sum_{s,r}\frac{\partial^2  k_b}{\partial\ell_s\partial\ell_r}\delta\ell_s\delta\ell_r\ .
\ee

\section{Perturbative geometrical identities}\label{Perturbative geometrical identities}

In this section, we recompute the action at the perturbed critical point using a perturbative expansion of $S_0$ followed by Gaussian integral. Although the result reproduces \eqref{S-S=Sreggeequation47}, the key result from this derivation is the connections between matrices $V_{I,f},\fh_{IJ}$ and Regge geometric data, as well as their connection to the Regge action. As the discussion below \eqref{quadraticEffectiveAction0000}, these results lay the groundwork for relating the effective action $\ci_{\rm eff}$ to Regge-geometrical quantities.
Expressing the Regge action in terms of $V_{I,f},\fh_{IJ}$ will be useful for recovering the Regge action as the leading order in $\ci_{\rm eff}$.

For convenience, we use the following collective notation for bulk and boundary variables in the spinfoam amplitude:
\be
X_{I}\equiv\left( g_{\alpha}, z_{j}\right),\qquad \cy_M\equiv \lt( k_b(\ell),\xi_{m}(\ell)\rt)
\ee
where $\alpha$, ${j}$ and $m$ stand for $(v,e)$, $(v,f)$ and $(e,b)$ respectively. The perturbations of boundary data $\delta\cy_M=\lt( \delta k_b(\ell),\delta\xi_{m}(\ell)\rt)$ are Regge-like.
We expand $S_0$ and ignore the terms higher than quadratic order in perturbations:
\be
S_0-\mathring{S}_0=\bigl(B_I + C_I\bigr)\,\delta X_I + \frac{1}{2} \fh_{IJ}\,\delta X_I\delta X_J + \mathcal{I}^{(2)}_{\rm bdy},\label{quadraPerb}
\ee
We use the following short-hand notations
\be
B_I &\equiv& \sum_h  V_{I,h}\,v_h,\qquad V_{I,f}=\frac{\partial\log E_{f}}{\partial X_{I}},
\qquad v_h=\sum_s\frac{\partial\eta_h}{\partial\ell_s}\delta\ell_s,
\qquad
C_I \equiv \sum_b \sum_M \frac{\partial^2(k_b\log E_b)}{\partial X_I\partial\mathcal{Y}_M}\,\delta\mathcal{Y}_M.\\
\fh_{IJ}&=&\sum_{h}\mathring\eta_{h}\frac{\partial^{2}\log E_{h}}{\partial X_{I}\partial X_{J}}+\sum_{b}\mathring k_{b}\frac{\partial^{2}\log E_{b}}{\partial X_{I}\partial X_{J}},\qquad \ci^{(2)}_{{\rm bdy}} = \sum_b \lt[ \frac{\partial(k_b\log E_{b})}{\partial\cy_{M}}\Delta\cy_{M}+\frac{1}{2}\frac{\partial^{2}(k_b\log E_{b})}{\partial\cy_{M}\partial\cy_{N}}\delta\cy_{M}\delta\cy_{N}\rt].
\ee 
where all coefficients are evaluated at the flat background critical point.
The term $\ci_{\rm bdy}^{(2)}$ depends only on the boundary variations.
Due to ${\partial \log E_{f}}/{\partial z_{vf}}=0$ at the background \footnote{
	At the critical point, the variation with respect to $z_{vf}$ leads to ${\partial \log E_{f}}/{\partial z_{vf}}\approx 0$, since for each $f = h, b$, only one $E_f$ contains dependence on the specific $z_{vf}$ variable.
}, nonzero entries of $V_{I,h}$ are at $I=\alpha$:
\begin{eqnarray}
	V_{\a,f}=\frac{\partial\log E_{f}}{\partial g_\a},\qquad V_{j,f}=0\quad\text{for all } j\ .
\end{eqnarray}

The equation of motion of $\delta X_I$ is
\be
\fh_{IJ}\delta X_{J}+\lt(B_I+C_I\rt)\approx 0.\label{linearCritEqn}
\ee
This is nothing but the linearization of critical point equations \eqref{crticalEqgandz}. We denote the solution by $\delta X_{I}(\ell)=-(\fh^{-1})_{IJ}\lt(B_J+C_J\rt)$.

At the level of quadratic perturbation, the integral \eqref{gzintegralS0} reduces to computing the following Gaussian integral: Assuming the Hessian matrix $\fh_{IJ}$ is invertible
\be
e^{J\lt[\mathring{S}_0+ \mathcal{I}^{(2)}_{\rm bdy}\rt]}\int \mathrm d(\delta X)\,
e^{J\lt[(B_I+C_I)\delta X_I + \tfrac12 \fh_{IJ}\delta X_I\delta X_J\rt]}
=\frac{1}{\sqrt{\det (-J\fh/(2\pi))}} e^{JS_{\rm 0,eff}},
\ee
where $S_{\rm 0,eff}$ is given by
\be
S_{\rm 0,eff}&=&\mathring{S}_0-\tfrac12 (B+C)^\top\,\fh^{-1}\,(B+C)+ \mathcal{I}^{(2)}_{\rm bdy}\label{bbbdyterms00000}
\ee
This result represents the stationary phase approximation of the integral \eqref{gzintegralS0}, evaluated at the critical point determined by the solution of \eqref{linearCritEqn}.


As discussed in the last section, this critical point describes the perturbative Regge geometry. By linearizing the first relation in \eqref{angleCorrespond}, we obtain the following correspondence between spinfoam critical point data and deficit angles. 
\be
\sum_I V_{I,h}\,\delta X_I 
= \frac{i\gamma}{2}\delta\eps_h
\label{eq:geometric}
\ee
We have $\eps_h=\delta\eps_h+\co(\delta\ell^2)$ since the background is flat. 
Here the variation is along the space of perturbative Regge geometries, which is equivalent to varying along the set of critical points of $S_0$, due to their one-to-one correspondence to the perturbative Regge geometries. So $\delta X_I$ in \eqref{eq:geometric} is evaluated at the critical point of a perturbative Regge geometry. As a consequence, $\delta X_I=\delta X_I(\ell)$ satisfies \eqref{linearCritEqn} and gives the following useful identity
\be
\frac{i\gamma}{2}\delta\eps_h= -\sum_{I,J} V_{I,h}(\fh^{-1})_{IJ}(B_J + C_J).\label{wv+tildev}
\ee
Since the right-hand side is linear in $\delta\ell$, this relation is nothing but expressing deficit angles in terms of length perturbations at the flat background: $\delta\eps_h\equiv \sum_s(\partial \eps_h/\partial \ell_s) \delta\ell_s$. 
However, this identity is significant because it establishes a connection between Regge-geometric quantities $\partial \eps_h/\partial \ell_s$ and the spinfoam quantities $V_{I,h},\fh_{IJ}$ evaluated at the background.
As an identity, equation \eqref{wv+tildev} can be carried over to the analysis of the full theory in the next section.

Inserting the relation \eqref{wv+tildev} to $S_{\rm 0, eff}$ and use $B=Vv$ gives
\be
S_{\rm 0,eff} - \mathring{S}_0 &=&\frac{i\g}{4} \sum_{h,r,s}\frac{\partial \eta_h}{\partial\ell_s} \frac{\partial\varepsilon_h}{\partial\ell_r}\delta\ell_s\delta\ell_r+i S'_{\rm bdy}
,\qquad i S'_{\rm bdy}=-\frac{1}{2}\left(B+C\right)^{\top}{\cal \mathfrak{h}}^{-1}C+{\cal I}^{(2)}_{{\rm bdy}}.\label{S-Sequation65}
\ee
The first term on the right reproduces the bulk contribution of $iS_{\rm Regge}^{(2)}$. Moreover, $S_{\rm 0,eff}$ is nothing but $S_0$ with the external data $\eta(\ell)$, $k(\ell)$, $\xi(\ell)$ and evaluated at the critical point $X(\ell)$, expanded up to $\co(\delta\ell^2)$ \footnote{
	For $\ell$ near $\mathring{\ell}$, let $X(\ell)$ be the critical point of $S_0(X;\ell)$ with external data $\eta(\ell), k(\ell), \xi(\ell)$. Define $f(\ell) \equiv S_0(X(\ell); \ell)$. By Section~\ref{A toy integral and Regge action}, we have $f(\ell) = i S_{\rm Regge}(\ell)$. Expanding to quadratic order in $\delta\ell$, $\big[ f(\ell) - \mathring{S}_0 \big]^{(2)} = i S_{\rm Regge}^{(2)}$. Expand $X(\ell) = \mathring{X} + \delta X^{(1)} + \delta X^{(2)} + \co(\delta\ell^3)$, where $\delta X^{(n)} \sim \co(\delta\ell^n)$, and expand $S_0(X;\ell)$ to quadratic order near $\mathring{X}$: $f(\ell) - \mathring{S}_0 = \mathcal{I}^{(2)}_{\rm bdy} + (B+C)^\top \delta X^{(1)} + \tfrac{1}{2} (\delta X^{(1)})^\top \fh\, \delta X^{(1)} + \co(\delta\ell^3)$, where $\delta X^{(1)} = -\fh^{-1}(B+C)$ by linearizing the critical-point equation. Therefore, $\big[f(\ell) - \mathring{S}_0 \big]^{(2)} = -\tfrac{1}{2}(B+C)^\top \fh^{-1}(B+C) + \mathcal{I}^{(2)}_{\rm bdy}=S_{\rm 0, eff}-\mathring{S}_0$. Comparing to $\big[ f(\ell) - \mathring{S}_0 \big]^{(2)} = i S_{\rm Regge}^{(2)}$, this gives the relations \eqref{relationtoRegge94} and \eqref{identityReggeaction}.
}. 
By \eqref{S074} and \eqref{S-S=Sreggeequation47}, we have
\be
S_{\rm 0,eff} - \mathring{S}_0=iS_{\rm Regge}^{(2)},\qquad S'_{\rm bdy} = S_{\rm bdy}.\label{relationtoRegge94}
\ee
Appendix \ref{Perturbative computation: boundary action} also provides an independent check for these two relations.

Combining \eqref{bbbdyterms00000} and \eqref{relationtoRegge94}, we obtain the following identity about the quadratic Regge action 
\be
iS_{\rm Regge}^{(2)}=-\tfrac12 (B+C)^\top\,\fh^{-1}\,(B+C)+ \mathcal{I}^{(2)}_{\rm bdy}.\label{identityReggeaction}
\ee
Similar to \eqref{wv+tildev}, this identity connects Regge action to spinfoam quantities evaluated at the background and can be applied to the analysis of the full theory.

In this section and Section \ref{A toy integral and Regge action}, we observed that the critical point remains real, but it describes a curved geometry. At first glance, this appears to conflict with our earlier statement that real critical points correspond to flat geometries. This apparent contradiction arises because the toy integral \eqref{gzintegralS0} is only a partial integral of the full theory.
Once the full integral is taken into account, the critical points identified above no longer persist as true critical points. Nevertheless, the preceding analysis remains valuable for deriving useful identities \eqref{wv+tildev} and \eqref{identityReggeaction} that can be carried over to the following analysis of the full theory.

\section{Higher curvature corrections}\label{Higher curvature corrections}

\subsection{Effective action $\ci_{\rm eff}(\ell)$}\label{Effective action ci eff}

In this section, we analyze the integral \eqref{tgzintegral}, which accounts for integration over the $t$ directions associated with non-Regge areas, and we now incorporate the full action $S$ rather than its previous approximation $S_0$. The external variables for this integral are the segment lengths $\ell$ and the boundary data $(k_b,\xi_{eb})$. We will introduce perturbations in all these variables, still assuming the boundary data remain Regge-like. Such perturbations correspond to curved geometries. Our objective is to evaluate the integral in \eqref{tgzintegral} perturbatively. Under this approach, the critical points are shifted away from the original, real, flat-geometry critical point. These new, perturbed critical points are generally complex.

We use the following collective notation for bulk and boundary variables 
\be
\cx_A=(t_i, X_I)\qquad \cy_M = (k_b(\ell), \xi_m(\ell)).
\ee 
We expand $S$ around the flat background geometry and ignore terms higher than quadratic order:
\be
S -\mathring{S} \simeq \cb_A \delta \cx_A + \frac{1}{2}{\cal H}_{AB}\delta{\cal X}_{A}\delta{\cal X}_{B} + \ci^{(2)}_{\rm bdy}.\label{quadraticS'-S0}
\ee
where the boundary-only term is 
\be
\mathcal{I}^{(2)}_{\text{bdy}} = \sum_b \left[ \frac{\partial(k_b \log E_b)}{\partial \mathcal{Y}_M} \Delta\mathcal{Y}_M + \frac{1}{2} \frac{\partial^2 (k_b \log E_b)}{\partial \mathcal{Y}_M \partial \mathcal{Y}_N} \delta\mathcal{Y}_M \delta\mathcal{Y}_N \right].
\ee
The linear coefficient $\cb_A$ has components $\cb_t = 0$ and $\cb_X = B + C$, where:
\be
B_I = \sum_h v_h V_{I,h},\qquad V_{I,h} = \frac{\partial \log E_h}{\partial X_I}=\frac{\partial E_h}{\partial X_I},\qquad v_h=\sum_s\frac{\partial \eta_h}{\partial \ell_s} \delta\ell_s \qquad C_I = \sum_{b, M} \frac{\partial^2 (k_b \log E_b)}{\partial X_I \partial \mathcal{Y}_M} \delta\mathcal{Y}_M
\ee
The Hessian matrix $\ch_{AB}$ is given by:
\be
\mathcal{H}_{tt} = 0, \qquad \mathcal{H}_{tX} \equiv D=  \hat{e}^\top V^\top, \qquad \mathcal{H}_{XX} \equiv H=\mathfrak{h} + V A V^\top,\qquad A_{h_1, h_2} = \mathring{\eta}_{h_1} \delta_{h_1, h_2}
\ee
where the entries of the matrix $D$ are $D_{iI}= \sum_h V_{I,h}\hat{e}^i_h$. Recall that $\mathfrak{h}$ is the Hessian of $S_0$. The difference between $H$ and $\fh$ is due to the holonomy correction $\co([\log E_h]^2)$ in the action $S$ (recall \eqref{expansionSprimeinE}).

Assuming that the Hessian $\ch$ is nondegenerate, we perform the Gaussian integral over $\delta\cx_A$. The quadratic perturbation of the effective action $\ci_{\rm eff}(\ell)$ is given by:
\be
\lt[\ci_{\rm eff}(\ell)  -\mathring{S}\rt]^{(2)} = -\frac{1}{2} \cb_X^\top \lt(\ch^{-1}\rt)_{XX} \cb_X + \ci^{(2)}_{\rm bdy}\ .\label{quadraticEffectiveAction}
\ee

To evaluate the $XX$-block of the inverse Hessian matrix $\ch^{-1}$, we apply the block matrix inversion formula:
\be
\lt(\ch^{-1}\rt)_{XX} = H^{-1} + H^{-1} D^\top G^{-1} D H^{-1},\qquad G = -D H^{-1} D^\top = \hat{e}^\top W \hat{e},\qquad W = -V^\top H^{-1} V\ ,
\ee
where both $G$ and $H$ are assumed to be invertible. 
The distinction between $(\ch^{-1})_{XX}$ and $H^{-1}$ arises because the presence of $t_i$ perturbations (associated with non-Regge areas) enlarges the space of integration variables in the full Hessian. 


Furthermore, we apply the Woodbury matrix identity to express the inverse of $H = \fh + V A V^\top$:
\be
H^{-1}=\fh^{-1}-\fh^{-1}V\rho V^\top \fh^{-1},\qquad \rho = (A^{-1} - w)^{-1},\qquad w = -V^\top \mathfrak{h}^{-1} V,
\ee
under the assumption that $A^{-1}-w$ is invertible (given that $\fh$ is invertible, this is equivalent to the invertibility of $H$, since $\det H=\det\fh\,\det A\,\det(A^{-1}-w)$). We obtain the following expressions of $W$
\be
W=w+w\rho w=w\rho A^{-1}=A^{-1}\rho w=A^{-1}\rho A^{-1} - A^{-1}.\label{WArho}
\ee

By applying the expressions of $(\ch^{-1})_{XX}$ and $H^{-1}$, the quadratic effective action \eqref{quadraticEffectiveAction} splits into three parts:
\be
\lt[\ci_{\rm eff}(\ell)  -\mathring{S}\rt]^{(2)} = \underbrace{-\frac{1}{2} (B+C)^\top \mathfrak{h}^{-1} (B+C) + \mathcal{I}^{(2)}_{\text{bdy}}}_{\text{Part 1}} + \underbrace{\frac{1}{2} (B+C)^\top \mathfrak{h}^{-1} V \rho V^\top \mathfrak{h}^{-1} (B+C)}_{\text{Part 2}} \nonumber\\
 \underbrace{-\frac{1}{2} (B+C)^\top H^{-1} D^\top G^{-1} D H^{-1} (B+C)}_{\text{Part 3}}\ .
\ee

\begin{itemize}
\item {Part 1:} This part matches $S_0$ evaluated at its linearized critical point \eqref{linearCritEqn}. By the analysis in the last section, this term yields the Regge action:
\be
-\frac{1}{2} (B+C)^\top \mathfrak{h}^{-1} (B+C) + \mathcal{I}^{(2)}_{\text{bdy}} = iS^{(2)}_{\text{Regge}}\ .
\ee

\item {Part 2:} This contribution arises from the difference between $H$ and $\fh$, due to the presence of the holonomy correction $\co([\log E_h]^2)$ in the action $S$.
In order to compute this term, we use the geometric identity \eqref{wv+tildev} established in the last section:
\be
V^\top \mathfrak{h}^{-1} (B+C)  =-\frac{i\gamma}{2}\varepsilon\ ,
\ee
Substituting this result into Part 2 gives:
\be
\frac{1}{2} \left(-\frac{i\gamma}{2}\varepsilon\right)^\top \rho \left(-\frac{i\gamma}{2}\varepsilon\right) = -\frac{\gamma^2}{8} \sum_{h_1, h_2} \rho_{h_1, h_2} \varepsilon_{h_1} \varepsilon_{h_2}\ .
\ee

\item {Part 3:} This term arises from the presence of area perturbations $t_i$ that are not aligned with Regge-geometric variations.
It originates in much the same way as the correction term found in area-Regge calculus \cite{Dittrich:2021kzs,Dittrich:2022yoo}.
We compute the quantity $D H^{-1} (B+C)$. Since $D = \hat{e}^\top V^\top$, we first evaluate $V^\top H^{-1}$:
\be
V^\top H^{-1} = V^\top \mathfrak{h}^{-1} - V^\top \mathfrak{h}^{-1} V \rho V^\top \mathfrak{h}^{-1} = (\mathbb{I} + w \rho) V^\top \mathfrak{h}^{-1}\ .
\ee
Using the relation $\rho = (A^{-1} - w)^{-1}$, we find $\mathbb{I} + w \rho = (\rho^{-1} + w)\rho = A^{-1} \rho$. Thus
\be
V^\top H^{-1} (B+C) = A^{-1} \rho V^\top \mathfrak{h}^{-1} (B+C) = -A^{-1} \rho \left(\frac{i\gamma}{2}\varepsilon\right)\ ,
\ee
where we again use the identity \eqref{wv+tildev}. This relation yields $D H^{-1} (B+C) = -\hat{e}^\top A^{-1} \rho \left(\frac{i\gamma}{2}\varepsilon\right)$. Part 3 then becomes:
\be
-\frac{1}{2} \left[ -\hat{e}^\top A^{-1} \rho \left(\frac{i\gamma}{2}\varepsilon\right) \right]^\top G^{-1} \left[ -\hat{e}^\top A^{-1} \rho \left(\frac{i\gamma}{2}\varepsilon\right) \right] = \frac{\gamma^2}{8} \sum_{h_1, h_2} \left[ \rho A^{-1} {\kappa} A^{-1} \rho \right]_{h_1, h_2} \varepsilon_{h_1} \varepsilon_{h_2}\ ,
\ee
where we define the $N_h \times N_h$ matrix $\kappa$ by
\be
\kappa_{h_1,h_2}=\sum_{i,j}\lt(G^{-1}\rt)_{ij}\hat{e}_{h_{1}}^{i}\hat{e}_{h_{2}}^{j} = \sum_{i,j}\lt[\sum_{h_{3}h_{4}}\hat{e}_{h_{3}}^{i} W_{h_{3}h_{4}}\hat{e}_{h_{4}}^{j}\rt]^{-1}_{ij}\hat{e}_{h_{1}}^{i}\hat{e}_{h_{2}}^{j}\ .
\ee
The matrix $\kappa$ is the only part in the effective action depending on the transverse basis $\hat{e}^i$. It is invariant under the change of basis $\hat e^i$, $i=1,\cdots,N_h-N_\ell$ within the plane transverse to $\sm_{\rm Regge}$: $\hat{e}^i\to O^i_{\ j}\hat{e}^j$, $O\in \mathrm{GL}(N_h-N_\ell)$. But it depends on the choice of the transverse plane. 

\end{itemize}

Combining all three parts, we conclude that the quadratic effective action corrects the Regge action by a term quadratic to the deficit angle $\eps_h$:
\be
\lt[\ci_{\rm eff}(\ell)  -\mathring{S}\rt]^{(2)}= iS_{\rm Regge}^{(2)}-\frac{\g^2}{8}\sum_{h_{1}h_{2}}K_{h_1,h_2}\eps_{h_1}\eps_{h_2}\ ,\label{Ieff2order}
\ee
where the correction matrix $K$ is given by
\be
K = \rho - \rho A^{-1} {\kappa} A^{-1} \rho\ .\label{Kmatrix}
\ee

An important aspect of the above derivation is the application of the identities \eqref{wv+tildev} and \eqref{identityReggeaction}. At the perturbative level, these identities connect Regge geometry with the spinfoam quantities, allowing us to express the effective action $\ci_{\rm eff}$ evaluated at the complex critical point in terms of Regge-geometric variables.

Eq. \eqref{WArho} gives $G=\hat e^\top A^{-1}\rho A^{-1}\hat e-\hat e^\top A^{-1}\hat e$. Applying the Woodbury identity to $K=\rho-\rho A^{-1}\hat e\,G^{-1}\hat e^\top A^{-1}\rho$ then gives 
\be
K^{-1}=\Pi-w,\qquad \Pi=A^{-1}-A^{-1}\hat e\,\big(\hat e^\top A^{-1}\hat e\big)^{-1}\hat e^\top A^{-1}.\label{Kinverse}
\ee
Since $A$ is positive definite ($\mathring\eta_h>0$), the matrix $\Pi$ is real, symmetric and positive semi-definite. The kernel of $\Pi$ is the plane spanned by the $\hat e^i$, and it depends on the $\hat e^i$ only through this plane \footnote{
	Set $B:=A^{-1/2}\hat e$ and $Q:=B\,(B^\top B)^{-1}B^\top$. Here $Q$ is an orthogonal projector: $Q^\top=Q$ and $Q^2=Q$. Moreover, $\Pi= A^{-1/2}(\mathbb I-Q )A^{-1/2}$. Since $QB=B$ and $Qx=x$ implies $x=By$ for some $y$, the kernel of $\mathbb I-Q$ is spanned by the columns of $B$, thus the kernel of $\Pi$ is the transverse plane spanned by $\hat e^i$.
}. Moreover \footnote{
	First, $\det\mathcal H=\det H\cdot\det\!\big(-DH^{-1}D^\top\big)=\det H\,\det G$. Then use $\det(\mathbb I_n+XY)=\det(\mathbb I_k+YX)$, we obtain $\det H=\det\mathfrak h\,\det(\mathbb I+\mathfrak h^{-1}VAV^\top)=\det\mathfrak h\,\det(\mathbb I+AV^\top\mathfrak h^{-1}V)=\det\mathfrak h\,\det(\mathbb I-Aw)=\det\mathfrak h\,\det A\,\det(\rho)^{-1}$, and similarly, $\det(\Pi-w)=\det(\rho^{-1}-PN^{-1}P^\top)=\det(\rho)^{-1}\,\det(\mathbb I_m-N^{-1}P^\top \rho P)=\frac{\det(\rho)^{-1}}{\det (N)}\,\det(N-P^\top\rho P)$ where $P=A^{-1}\hat e$, $ N=\hat e^\top A^{-1}\hat e$, and $N-P^\top\rho P=-G$.
}, 
\be
\det\lt(\Pi-w\rt)=\frac{(-1)^{N_h-N_\ell}\det\ch}{\det\fh\,\det A\,\det(\hat e^\top A^{-1}\hat e)}.
\ee 
Thus, $K=(\Pi-w)^{-1}$ is well-defined as long as both $\ch$ and $\fh$ are invertible. 

In the following, we show that $\re\big(\eps^\top K\eps\big)\geq0$ for every real $\eps$, so the $\eps^2$-correction never increases $|e^{J\ci_{\rm eff}}|$.
Write $V=i\Upsilon$ with $\Upsilon$ real (recall Section~\ref{Scheme for computing the quadratic effective action}). Then $w=\Upsilon^\top\fh^{-1}\Upsilon$, and for real $y$, with $x=\Upsilon y$ and $q=\fh^{-1}x$, one has 
\be
y^\top(\re\,w)\,y=\re\big(x^\top\fh^{-1}x\big)=q^\dagger(\re\,\fh)\,q\leq0,\label{ywyqhqBound}
\ee 
because $\re\,\fh$ is negative semi-definite. Hence $\re\,K^{-1}=\Pi-\re\,w$ is positive semi-definite. For real $\eps$ and $u=K\eps$ one has $\eps^\top K\eps=u^\dagger(K^{-1})^*u$, so that
\be
\re\big(\eps^\top K\eps\big)=u^\dagger\big(\Pi-\re\,w\big)u\geq 0\ .\nonumber
\ee


\subsection{$\eps^2$-correction}

In the $\eps^2$-correction to the Regge action in \eqref{Ieff2order}, two terms $\rho$ and $-\rho A^{-1} {\kappa} A^{-1}\rho$ in $K$ originate from different sources: 
\begin{itemize}

\item The term $\rho$ exists only because of the holonomy correction, i.e.\ the $\co([\log E_h]^2)$ terms in the expansion of $S$: if these terms are dropped, then $H=\fh$ and Part 2 vanishes. 
The holonomy correction arises because the new action $S$ is single-valued on holonomies rather than on their logarithms. This feature is reminiscent of similar effects in the effective dynamics of loop quantum cosmology \cite{Agullo:2016tjh}.

\item The term $-\rho A^{-1}\kappa A^{-1}\rho$ exists because of the integration over the non-Regge area variables $t_i$. In the absence of the holonomy correction it reduces to $K_{\rm SF}:=-\hat e(\hat e^\top w\,\hat e)^{-1}\hat e^\top$ \footnote{
The invertibility of $\hat e^\top w\hat e$ relies on its small-$\gamma$ version via $\bar\Gamma_{\hat e}$, see Section~\ref{Small gamma scaling}.}, the pure non-Regge-area contribution, whose origin is the same as that of the $\eps^2$-correction in area-Regge calculus \cite{Dittrich:2021kzs,Dittrich:2022yoo}; in the presence of the holonomy correction it is dressed by $A$ through $\rho$ and $W$.
	
\end{itemize}

In the regime where the wavelength of geometric fluctuations $\delta\ell$ greatly exceeds the typical lattice scale endowed by the background geometry $\mathring\ell$, the lattice effects are entirely disregarded, the effective action $\ci_{\rm eff}$ should reduce to its continuum limit.
In this limit, the leading order Regge action approaches the Einstein-Hilbert action, while the $\eps^2$-correction is expected to correspond to higher-curvature terms quadratic in the Riemann tensor.
The results on area-Regge calculus in \cite{Dittrich:2021kzs,Dittrich:2022yoo} show that the $\eps^2$-correction in area-Regge calculus gives rise to a $\cw^2$ term in the effective action, where $\cw$ denotes the Weyl tensor. 
The fact that the correction involves the Weyl tensor, rather than the full Riemann tensor, is directly connected to its coupling to area fluctuations that are independent of length fluctuations \cite{Borissova:2022clg}.
Since the undressed part $K_{\rm SF}$ of the term $-\rho A^{-1}\kappa A^{-1}\rho$ is exactly this area-fluctuation contribution, we expect a similar outcome: $K_{\rm SF}$ should also generate a $\mathcal{W}^2$ correction in the continuum limit.

In contrast, the $\varepsilon^2$-correction coming from the $\rho$ term together with the dressing, i.e. $K-K_{\rm SF}$, is not a pure area-fluctuation effect. It is originated by the holonomy correction $\co([\log E_h]^2)$ in the action $S$.
Since this contribution is unrelated to area fluctuations, the reasoning that restricts the correction to a Weyl-tensor-squared $\mathcal{W}^2$ form does not apply, and other curvature-squared invariants can appear. 
Indeed, in Section \ref{Small gamma scaling}, we show that for small $\g$
\be
\g \eps^\top (K-K_{\rm SF}) \eps=\Fg^\top N_{\hat e}^{-1}\Fg +\co(\g),\qquad \Fg_s= \sum_h(\partial\mathring a_h/\partial\ell_s)\eps_h,\label{Gsquare103}
\ee
where $N_{\hat e}$ is a positive definite matrix. 
The component $\Fg_s$ gives the left-hand side of the linearized Regge equation and thus represents the discrete Einstein tensor \cite{Alsing:2011zz,McDonald:2008js} ($\mathring a_h\equiv\gamma \mathring \eta_h$ is the area of the triangle in the unit $J\ell_P^2/2$). 
This term vanishes on shell $\Fg_s=0$, and it is the discrete analogue of quadratic forms of the Ricci tensor $\calr_{\mu\nu}$ and Ricci scalar $\calr$.

In the effective field theory of pure gravity on the continuum, $\calr^2$ and $\calr_{\mu\nu}\calr^{\mu\nu}$, quadratic in the vacuum field equations, are removed by field redefinition $g_{\mu\nu}\to g_{\mu\nu}+a_1\calr_{\mu\nu}+a_2\calr\,g_{\mu\nu}$. So both quadratic curvature terms do not affect on-shell quantities \cite{tHooft:1974toh,Burgess:2003jk,Baldazzi:2021orb,Falls:2024noj,Knorr:2022dsx}\footnote{ 
	In four dimensions, $\cw^2$ is also redundant up to the Gauss--Bonnet density $\mathrm{GB}$, due to $\cw^2=\mathrm{GB}+2\calr_{\mu\nu}\calr^{\mu\nu}-\frac23\calr^2$, and GB is a total derivative.}. 
However, they become physical when coupled to matter or in non-perturbative scenarios, as in \cite{Starobinsky:1980te}.

The $\eps^2$ correction in spinfoam has the similar situation. 
Recall that the $m$-dependent change of variables $k_h=\eta_h c_m$, introduced at the end of Section \ref{Purely integral formula of spinfoam amplitude}, transforms the action $S$ in the pure integral formula into $S_{\rm SF}+2\pi i\sum_h k_h m_h$, where the second term arises from Poisson summation. The flat real critical point $\eps_h=0$ and small perturbations only contribute dominantly to the $\{m_h=0\}_h$ term in the Poisson sum \eqref{poissonSumSa}. Restricting to this term and repeating the computation of Section \eqref{Effective action ci eff} for $S_{\rm SF}$ in place of $S$, we find that the resulting $\eps^2$ correction is just $\eps^T K_{\rm SF}\eps$; the holonomy correction term $\eps^\top (K-K_{\rm SF})\eps$ is absent. This indicates that, at the perturbative level, this latter term can be removed by a field redefinition, just as the quadratic curvature terms can be removed in the continuum theory. At the perturbative level, the term $\eps^\top (K-K_{\rm SF})\eps$ affects only off-shell quantities. We expect it to become physically relevant in the non-perturbative regime, where the restriction of the Poisson sum to a single term is no longer justified.

As we show in Section \ref{Small gamma scaling}, the $\eps^2$-correction in the effective action is of higher order in the Barbero--Immirzi parameter $\g$ than the Regge action:
\be
-\frac{\g^2}{8}\sum_{h_{1}h_{2}}K_{h_1,h_2}\eps_{h_1}\eps_{h_2}\sim \co(\g),\qquad S^{(2)}_{\rm Regge}\sim \co(1)\ .
\ee
Combining the stationary phase approximation with large $J$, our analysis has resulted in an expansion of the spinfoam amplitude in two parameters: $1/J$ and $\gamma$.
When all higher-order terms in this expansion are neglected, the leading contribution to the effective action $\ci_{\rm eff}$ reduces exactly to the Regge action, as expressed in \eqref{Ieff2order}.
This result illustrates how the spinfoam formulation leads to the emergence of the classical gravitational action. 
Specifically, in the long-wavelength regime of geometric fluctuations $\delta\ell$, the leading-order (quadratic) Regge action approaches the action of linearized gravity, i.e.\ of a massless spin-2 field (see e.g. \cite{Rocek:1981ama,Williams:1991cd,Williams:1986hx,Hohn:2014nwv}).

In the above analysis, we have assumed that various matrices such as $\ch$, $H$, $\fh$, and $G$ are invertible. 
Although we do not have a general proof establishing their invertibility, 
a numerical study of spinfoam amplitudes on several triangulated manifolds has been carried out, and the results presented in the companion paper \cite{companionNumer} confirm the invertibility of all these matrices in every numerical example analyzed.

As discussed in the derivation of Part 3 of the effective action, the $\eps^2$-correction term depends on the choice of the plane defined by vectors $\hat e^i$, $i=1,\ldots, N_h-N_\ell$, which are transverse to the Regge submanifold $\sm_{\rm Regge}$. This dependence has been pointed out in the definition of the effective action $\ci_{\rm eff}$ in \eqref{tgzintegral}.
However, as discussed below \eqref{tgzintegral}, the spinfoam amplitude $\sa_\psi=\int \rmd^{N_\ell} \ell\, \cj(\ell)\, e^{J \ci_{\rm eff} (\ell)}$, obtained by integrating the exponential of the effective action over the space of Regge geometries, is manifestly independent of the choice of transverse planes. Changing the transverse planes simply corresponds to selecting different parametrizations for integrating out the non-geometric (non-Regge) degrees of freedom. This may yield different forms for the effective action itself, but the resulting $\sa_\psi$ remains unchanged.

\subsection{Small-$\gamma$ scaling}\label{Small gamma scaling}

In this subsection, we demonstrate that the effective action in \eqref{Ieff2order} admits a systematic expansion in the Barbero-Immirzi parameter $\gamma$, provided the underlying Regge geometry is held fixed. Specifically, the leading-order Regge action is of order $\mathcal{O}(\gamma^0)$, while the $\varepsilon^2$-correction enters at order $\mathcal{O}(\gamma)$.
Recovering the Regge action as the leading term in the small-$\gamma$ expansion aligns with earlier expectations in \cite{claudio,claudio1,propagator3,propagator2,propagator1,lowE2,Han:2023cen}.

The flat Regge geometry corresponding to the background critical point is held fixed while scaling $\g$ to be small. The areas satisfy $\mathring\fa_h \approx (\gamma \mathring\eta_h / 2) J \ell_P^2$ and $\mathring\fa_b \approx (\gamma \mathring k_b / 2) J \ell_P^2$, so fixing the areas as $\gamma $ decreasing implies that $\mathring\eta_h$ and $\mathring k_b$ scale inversely with $\gamma$. Thus, we can write:
\be
\mathring\eta_h = \frac{\bar{\eta}_h}{\gamma}, \qquad \mathring k_b = \frac{\bar{k}_b}{\gamma} ,\qquad \bar{\eta}_h, \bar{k}_b \sim \mathcal{O}(1)
\ee
The support of the function $\psi$ in $\sa_\psi$ is taken large enough as $\g$ is decreased.

The values of $\mathring g_{ve}$, $\mathring z_{vf}$ and $\mathring \xi_{eb}$ are independent of $\g$, since they only relate to normals and angles. 
Furthermore, the geometric fluctuation $\delta\ell$ remains independent of $\gamma$. Since the Regge action $S^{(2)}_{\rm Regge}$ is purely geometric, it exhibits no dependence on $\gamma$. For the $\eps^2$-term in \eqref{Ieff2order}, the deficit angle $\eps_h$ does not scale with $\g$, so our aim is to determine the scaling of the matrix $K$ in \eqref{Kmatrix} with respect to $\gamma$.

We trace the scaling of all ingredients in $K$ step-by-step.
First, the diagonal area matrix $A_{h_1, h_2} = \mathring{\eta}_{h_1} \delta_{h_1, h_2}$ scales as:
\be
A^{-1} = \gamma \bar{A}^{-1}\ ,\qquad \bar{A}_{h_1, h_2} = \bar{\eta}_{h_1} \delta_{h_1, h_2} \sim \mathcal{O}(1).
\ee 
Second, the Hessian of the leading-order action $S_0$ evaluated at the background is given by:
\be
\mathfrak{h}_{IJ} = \sum_h \mathring{\eta}_h \frac{\partial^2 \log E_h}{\partial X_I \partial X_J} + \sum_b \mathring{k}_b \frac{\partial^2 \log E_b}{\partial X_I \partial X_J}\ .
\ee
Both $\mathring{\eta}_h$ and $\mathring{k}_b$ scale as $1/\gamma$. Additionally, $\log E_f=F^{(0)}_f+\g F^{(1)}_f$ with $F^{(0)}_f$ and $F^{(1)}_f$ independent of $\g$. Applying these scalings to the definition of the Hessian above yields
\be
\mathfrak{h} = \frac{1}{\gamma} \bar{\mathfrak{h}}+\fh^{(1)},\qquad \bar\fh_{IJ}=\sum_h\bar\eta_h\frac{\partial^2F^{(0)}_h}{\partial X_I\partial X_J}+\sum_b\bar k_b\frac{\partial^2F^{(0)}_b}{\partial X_I\partial X_J},\qquad \fh^{(1)}_{IJ}=\sum_h\bar\eta_h\frac{\partial^2F^{(1)}_h}{\partial X_I\partial X_J}+\sum_b\bar k_b\frac{\partial^2F^{(1)}_b}{\partial X_I\partial X_J},
\ee
so that $\mathfrak{h}^{-1} = \gamma \bar{\mathfrak{h}}^{-1}+\co(\g^2)$, where we assume that $\bar{\mathfrak{h}}$ is invertible. Similarly, $V=\bar V+\g V^{(1)}$ with $\bar V_{I,h}=\partial F^{(0)}_h/\partial X_I$ and $(V^{(1)})_{I,h}=\partial F^{(1)}_h/\partial X_I$ independent of $\g$. Substituting the inverse Hessian into $w = - V^\top \mathfrak{h}^{-1} V$ yields
\be
w = \gamma \bar{w}+\co(\g^{2})\ ,\qquad \bar{w} = - \bar V^\top \bar{\mathfrak{h}}^{-1} \bar V \ ,
\ee
so that $\bar w$, and hence $\bar\rho$, $\bar W$, $\bar G$, $\bar\kappa$ and $\bar K$ below, are independent of $\g$. Thus, $w \sim \mathcal{O}(\gamma)$.


The matrix $\rho$ is the inverse of $(A^{-1} - w)$. Substituting the scalings from the previous steps:
\be
A^{-1} - w = \gamma \bar{A}^{-1} - \gamma \bar{w} +\co(\g^{2}) = \gamma (\bar{A}^{-1} - \bar{w}+\co(\g))\ .
\ee
Inverting this expression yields $\rho$:
\be
\rho = (A^{-1} - w)^{-1} = \frac{1}{\gamma} (\bar{A}^{-1} - \bar{w})^{-1} +\co(1)= \frac{1}{\gamma} \bar{\rho}+\co(1)\ ,
\ee
where we assume $(\bar{A}^{-1} - \bar{w})$ is invertible. Since $\bar{\rho} = (\bar{A}^{-1} - \bar{w})^{-1} \sim \mathcal{O}(1)$, we obtain $\rho \sim \mathcal{O}(1/\gamma)$.

The matrix $W$ is given by $W = w + w \rho w$. Inserting the above scalings gives:
\be
W = \gamma \bar{w} + \lt[\gamma \bar{w}+\co(\g^2)\rt] \left[ {\gamma}^{-1} \bar{\rho} +\co(1)\right] \lt[\gamma \bar{w}+\co(\g^2)\rt] +\co(\g^2)=  \gamma \bar{W}+\co(\g^2)\ ,
\ee
where $\bar{W} = \bar{w} + \bar{w} \bar{\rho} \bar{w} \sim \mathcal{O}(1)$. The matrix $G$ is the projection of $W$ onto the constant transverse basis vectors $\hat{e}$:
\be
G = \hat{e}^\top W \hat{e} = \gamma \hat{e}^\top \bar{W} \hat{e}+\co(\g^2) = \gamma \bar{G} +\co(\g^2)\quad \implies \quad G^{-1} = \frac{1}{\gamma} \bar{G}^{-1}+\co(1)\ ,
\ee
where $\bar{G} = \hat{e}^\top \bar{W} \hat{e} \sim \mathcal{O}(1)$. The kernel matrix $\kappa$ is defined as $\kappa = \hat{e} G^{-1} \hat{e}^\top$, which evaluates to:
\be
\kappa = \hat{e} \left[\frac{1}{\gamma} \bar{G}^{-1} +\co(1)\right] \hat{e}^\top = \frac{1}{\gamma} \bar{\kappa}+\co(1)\ ,
\ee
where $\bar{\kappa} = \hat{e} \bar{G}^{-1} \hat{e}^\top \sim \mathcal{O}(1)$. Thus, $\kappa \sim \mathcal{O}(1/\gamma)$. 
The invertibility of $\bar{G}$ is assumed. 

Finally, we substitute all above scalings into the definition of $K$. Both terms in $K$ scale in the same way as $1/\g$, therefore
\be
K = \rho - \rho A^{-1} \kappa A^{-1} \rho =  \frac{1}{\gamma} \bar{K}+\co(1)\ ,
\ee
where $\bar{K} = \bar{\rho} - \bar{\rho} \bar{A}^{-1} \bar{\kappa} \bar{A}^{-1} \bar{\rho}$ is an $\mathcal{O}(1)$ matrix independent of $\gamma$. 

Similar to \eqref{Kinverse}, $\bar K^{-1}=\bar\Pi-\bar w$ with $\bar\Pi=\g^{-1}\Pi=\bar A^{-1}-\bar A^{-1}\hat e(\hat e^\top\bar A^{-1}\hat e)^{-1}\hat e^\top\bar A^{-1}$. Hence the $\eps^2$-correction is at most of order $\g$, and $\bar K$ is invertible. 
Consider purely bulk perturbations $\delta\ell$, one has 
\be
C=0,\qquad B=\g^{-1}V\,T\,\delta\ell,\qquad T_{hs}=\partial\bar\eta_h/\partial\ell_s
\ee 
by $\eta_h=\bar{\eta}_h/\g$, so \eqref{wv+tildev} reads $\frac{i\g}{2}\delta\eps=\g^{-1}w\,T\,\delta\ell$. Since $\delta\eps$ does not depend on $\g$ and $w=\g\bar w+\co(\g^2)$, its order-$\g^0$ part gives the identity
\be
\bar w\,T=0.
\ee 
This identity and $\bar\Pi\,\hat e=0$ give a closed form of $\bar K$: Since the $\hat e^i$ span a plane complementary to the tangent space of $\sm_{\rm Regge}$, the $N_h\times N_h$ matrix $\mathbb M:=(T\ \ \hat e)$ is invertible. Define the short-hand notation:
\be
N_{\hat e}:=T^\top\bar\Pi\,T,\qquad \bar\Gamma_{\hat e}:=\hat e^\top\bar w\,\hat e,\qquad \Fg:=T^\top\eps,\qquad x_{\hat e}:=\hat e^\top\eps,
\ee
so that $\mathbb M^\top\eps=(\Fg;x_{\hat e})$. Here $\Fg_s=\sum_h(\partial\bar\eta_h/\partial\ell_s)\eps_h$ is the left-hand side of the linearized Regge equations. The quantity $x_{\hat e}$ collects the components of $\eps$ along the transverse plane. Since $\bar w$ is symmetric, $\bar wT=0$ also gives $T^\top\bar w=0$, hence $\mathbb M^\top\bar w\,\mathbb M=\mathrm{diag}(0,\bar\Gamma_{\hat e})$. Since the kernel of $\bar\Pi=\g^{-1}\Pi$ is the plane spanned by the $\hat e^i$, $\mathbb M^\top\bar\Pi\,\mathbb M=\mathrm{diag}(N_{\hat e},0)$. Therefore
\be
\mathbb M^\top\bar K^{-1}\mathbb M=\mathbb M^\top\big(\bar\Pi-\bar w\big)\mathbb M=\begin{pmatrix}N_{\hat e}&0\\0&-\bar\Gamma_{\hat e}\end{pmatrix}.
\ee
The matrix $N_{\hat e}$ is positive definite: for real $y\neq0$, $y^\top N_{\hat e}\,y=(Ty)^\top\bar\Pi\,(Ty)\geq0$, with equality only if $Ty$ lies in the plane spanned by the $\hat e^i$, which is impossible because this plane is complementary to the range of $T$ and $Ty\neq0$ ($T$ has rank $N_\ell$). Since $\bar K^{-1}$ is invertible, so is $\bar\Gamma_{\hat e}$. Inverting the block-diagonal matrix gives $\bar K=\mathbb M\,\mathrm{diag}\big(N_{\hat e}^{-1},-\bar\Gamma_{\hat e}^{-1}\big)\mathbb M^\top$, i.e.
\be
\bar K=T\,N_{\hat e}^{-1}T^\top-\hat e\,\bar\Gamma_{\hat e}^{-1}\hat e^\top,\qquad
\eps^\top\bar K\eps=\Fg^\top N_{\hat e}^{-1}\Fg-x_{\hat e}^\top\bar\Gamma_{\hat e}^{-1}x_{\hat e}.\label{Kbardecomp}
\ee
The first term is real and positive whenever $\Fg\neq0$. On shell $\Fg=0$, the second term of \eqref{Kbardecomp} can still survive.

The above derivations establish a power-counting argument for how the $\eps^2$-correction to the Regge action behaves in the small-$\g$ limit:
\be
-\frac{\gamma^2}{8}\sum_{h_{1}h_{2}}K_{h_1,h_2}\eps_{h_1}\eps_{h_2} \sim \mathcal{O}(\gamma)\ .\label{smallgammaK}
\ee
In contrast, the quadratic Regge action $S_{\rm Regge}^{(2)}$ scales as $\mathcal{O}(1)$. This demonstrates that the $\eps^2$-correction is subleading in the regime $\gamma \ll 1$. This verifies that the spinfoam (quadratic) effective action is dominated by the Regge action in the small $\g$ regime.

The above power-counting argument depends on the invertibility of several matrices, including $\bar{\fh}$, $\bar{\rho}$, and $\bar{G}$. In the companion paper~\cite{companionNumer}, the scaling behavior predicted in Eq.~\eqref{smallgammaK} is directly verified through numerical analysis, providing independent confirmation beyond the analytic power-counting argument.

Since the Regge action is independent of $\gamma$, it stands out as the leading contribution in the effective action \eqref{Ieff2order} when $J$ is large and $\gamma$ is small. This demonstrates that the classical gravitational action naturally emerges in the appropriate regime of spinfoam. In particular, for long-wavelength geometric fluctuations $\delta\ell$, the leading-order (quadratic) Regge action approaches, in the continuum limit, the action of linearized gravity.

\begin{acknowledgements}

This work builds on an unpublished note by Wojciech Kami\'nski and Aldo Riello, in which the pure integral formula \eqref{integralLambdaZeta37} was first derived. The author is grateful to Wojciech Kami\'nski for sharing his ideas and for many illuminating discussions during the course of this research.

The author thanks Renata Ferrero, Hongguang Liu, Dongxue Qu, and Sean Tobin for numerous valuable discussions. This work receives supports from the National Science Foundation through grants PHY-2207763 and PHY-2512890, from the Center for SpaceTime and the Quantum (CSTQ), and from a sponsorship of renewed research stay in Erlangen from the Humboldt Foundation. This work is also supported by the WithOut SpaceTime (WOST) project (https://withoutspacetime.org), Grant ID\# 63683 from the John Templeton Foundation (JTF). The opinions expressed in this work are those of the author(s) and do not necessarily reflect the views of the John Templeton Foundation. The author acknowledges the use of Claude to proofread the manuscript and assist in simplifying an earlier derivation of the pure integral formula. All outputs from Claude were reviewed, verified, and edited by the author, who take full responsibility for the content of this manuscript.

\end{acknowledgements}

\appendix

\section{Integral on $\mathbb{CP}^1$}\label{sec:CP1 integral}

To begin, consider the integral over a single variable $z_{vf}$ on $\mathbb{CP}^1$. The corresponding measure is positive and given by
\be
\frac{\rmd^2 z_{vf}}{\left\langle Z_{vef},Z_{vef}\right\rangle\left\langle Z_{ve'f},Z_{ve'f}\right\rangle},\qquad Z_{vef}=g_{ve}^\dagger z_{vf},\quad Z_{ve'f}=g_{ve'}^\dagger z_{vf}
\ee
where $\rmd^2 z=\frac{i}{2\pi}(z^0\rmd z^1-z^1\rmd z^0)\wedge (z^0\rmd z^1-z^1\rmd z^0)^*$. Let us ignore the indices of variables and write 
\be
z=z_{vf},\qquad g_1=g_{ve},\qquad g_2=g_{ve'}.
\ee
We have the following result:

\begin{lemma}\label{integralCP1}
\be
\int_{\mathbb{CP}^{1}}\frac{\rmd^{2}z}{\langle g_{1}^{\dagger}z,g_{1}^{\dagger}z\rangle\langle g_{2}^{\dagger}z,g_{2}^{\dagger}z\rangle}=\frac{ r_{12}}{\sinh\left(r_{12}\right)},\label{r12behavior}
\ee
where $r_{12}=r\lt(({g}_{1}^\dagger)^{-1},({g}_{2}^\dagger)^{-1}\rt)\geq 0$ is the distance on hyperboloid between two points $({g}_{1}^\dagger)^{-1}$ and $({g}_{2}^\dagger)^{-1}$.
\end{lemma}

\begin{proof}
We make a change of variable $z\to Z=g^\dagger_2z$. The measure $\rmd^2 z$ is invariant by $z\to g z$ for any $g\in \Slc$. By the Cartan decomposition, $g_1^\dagger (g_2^\dagger)^{-1}=R_{12}d(r_{12})R'_{12}$ where $R_{12},R'_{12}\in\Su$ and
\be
d(r_{12})=\begin{pmatrix} e^{r_{12}/2} & 0\\
0 & e^{-r_{12}/2}
\end{pmatrix}.
\ee
By another change of variable $Z\to Z'=R_{12}'Z$ and the SU(2) invariance of the inner product, the integral reduces to
\be
\int_{\mathbb{CP}^{1}}\frac{\rmd^{2}Z^{\prime}}{\langle d(r_{12}) Z^{\prime}, d(r_{12}) Z^{\prime}\rangle\langle Z^{\prime},Z^{\prime}\rangle}.
\ee
By using the parametrization $Z'=\left(e^{\frac{i}{2}\varphi}\cos\left(\frac{\beta}{2}\right),-e^{-\frac{i}{2}\varphi}\sin\left(\frac{\beta}{2}\right)\right)^\top$ with $\b\in[0,\pi]$ and $\varphi\in[0,2\pi)$, the measure is $\rmd^2 Z'=\frac{1}{4\pi} \rmd\varphi\rmd\beta\sin(\beta)$, then the above integral reduces to
\be
\frac{1}{4\pi}\int^{2\pi}_{0}\rmd\varphi\int^{\pi}_{0}\rmd\beta \frac{\sin(\beta)}{\cos (\beta ) \sinh (r_{12})+\cosh (r_{12})}=\frac{r_{12}}{\sinh\left(r_{12}\right)}.
\ee

\end{proof}

The function on the right-hand side of \eqref{r12behavior} is the kernel $K_0(r)=r/\sinh r$ that bounds the face weights in the proof of the finiteness of the vertex amplitude \cite{finite,Kaminski:2010qb,Han:2021bln}. By factorizing the integral into vertices,
\be
V&=&\int \rmd\mu(g,z)=\prod_v V_v\nonumber\\
V_v&=&\int \prod_{e \text{ at }v}\rmd g_{ve}\,\delta (g_{v e_0})\prod_{(e,e')\text{ at }v }\int \frac{\rmd^2 z_{vf}}{\left\langle Z_{vef},Z_{vef}\right\rangle\left\langle Z_{ve'f},Z_{ve'f}\right\rangle}\nonumber\\
&=&\int \prod_{e \text{ at }v}\rmd g_{ve}\,\delta (g_{v e_0})\prod_{(e,e')\text{ at }v }\frac{ r_{ee'}}{\sinh r_{ee'}}\ ,\qquad
r_{ee'}\equiv r\lt((g_{ve}^\dagger)^{-1},(g_{ve'}^\dagger)^{-1}\rt).
\ee
In the above expression, each unordered pair of edges $(e,e')$ at $v$ identifies a face $f$ such that $v$, $e$, and $e'$ are all contained in the boundary $\partial f$.
The integrand depends on $g_{ve}$ only through the point $x_e$ of the hyperboloid $\mathbb{H}^3=\Slc/\Su$. Hence, up to a constant fixed by the normalization of the Haar measure,
$V_v$ is the integral of $\prod_{e<e'} K_0(r_{ee'})$ over four copies of $\mathbb{H}^3$, the fifth point $x_{e_0}$ being fixed.
For a 2-complex $\ck$ that is dual to a simplicial complex, this integral is finite. Indeed, since every face bounded by $e,e'$ belongs to the two tetrahedra dual to $e$ and $e'$,
\be
\prod_{e<e'}K_0(r_{ee'})=\prod_{e}\Big[\prod_{e'\neq e}K_0(r_{ee'})^{5/2}\Big]^{1/5}\leq\frac15\sum_{e}\ \prod_{e'\neq e}K_0(r_{ee'})^{5/2}
\ee
by the inequality between the geometric and the arithmetic mean. Each term on the right has four factors of $K_0(r_{ee'})^{5/2}$, each associated with a face bounded by $e$. By the homogeneity of $\mathbb{H}^3$, its integral over the four free points factorizes into $C^4$ with
$C=\int_{\mathbb{H}^3}K_0(r)^{5/2}\,\rmd x=4\pi\int_0^\infty r^{5/2}\,(\sinh r)^{-1/2}\,\rmd r<\infty$. Therefore $V_v<\infty$ and $V<\infty$.

\section{Polynomial approximation of \texorpdfstring{$\mathcal{D}_{J\Lambda}$}{D\_JLambda}}\label{app:polynomial approximation}

In Appendices \ref{app:polynomial approximation} and \ref{app:Q}, $\chi\in C^\infty(\mathbb{R})$, $\chi=1$ on $[0,1]$, $\chi(x)=0$ for $|x|>x_*$ (with $x_*\geq1$), $M_\chi=\sup_x|1-\chi(x)|$, $\|f\|_\infty=\sup|f|$, and the polynomial $\mathrm{Pol}(k)=\sum_{j=0}^{n}\beta_jk^j$, $\beta_n\neq0$, is the same for all internal faces. Let $\Lambda>0$, $J\geq1$, $k_b\geq0$ with $Jk_b\in\mathbb{N}_0$, $d_{J\Lambda}(k)=\chi(k/(J\Lambda))\,\mathrm{Pol}(k)$ and $\mathcal{D}_{J\Lambda}=\PT[d_{J\Lambda}]$. By \eqref{PTexpansion} with $\eta\to J\eta$, the integrand of \eqref{integraletamu} is $\prod_hu_J(\eta_h,E_h)\prod_bE_b^{Jk_b}$, where, for $\eta\geq0$ and $|E|\leq1$,
\begin{equation}
u_J(\eta,E)=\mathcal{D}_{J\Lambda}(J\eta E)\,e^{-J\eta(1-E)}=e^{-J\eta}\sum_{k=0}^{\infty}d_{J\Lambda}(k)\,E^k\,\frac{(J\eta)^k}{k!}.\label{loc:1}
\end{equation}
In the following, we will use the telescoping inequality $|\prod_h\alpha_h-\prod_h\gamma_h|\leq\sum_h|\alpha_h-\gamma_h|\prod_{h'\neq h}\max\{|\alpha_{h'}|,|\gamma_{h'}|\}$ ($\alpha_h,\gamma_h\in\mathbb{C}$). With $k^{\underline m}=k(k-1)\cdots(k-m+1)$, $k^{\underline 0}=1$, we write $\mathrm{Pol}(k)=\sum_{m=0}^{n}b_mk^{\underline m}$; as $k^{\underline m}/k!$ is $1/(k-m)!$ for $k\geq m$ and $0$ for $k<m$, we have $\sum_{k=0}^{\infty}\frac{y^k}{k!}\,k^{\underline m}=\sum_{k\geq m}\frac{y^k}{(k-m)!}=y^m\,e^{y}$, $y\in\mathbb{C}$. We introduce the notation:
\begin{equation}
P(c)=\sum_{m=0}^{n}b_mc^m,\qquad A(y)=\sum_{m=0}^{n}|b_m|\,y^m,\qquad \phi(s)=s-1-\log s .\label{loc:8}
\end{equation}
$A$ is non-decreasing on $[0,\infty)$, $|P(c)|\leq A(|c|)$, $A(Jy)\leq J^nA(y)$, and $\phi$ is positive and decreasing on $(0,1)$.

\begin{lemma}\label{loc:lemBelow}
(a) $P=\PT[\mathrm{Pol}]$, i.e.\ $e^{-c}\sum_k\frac{c^k}{k!}\mathrm{Pol}(k)=P(c)$, and $P$ has degree $n$ and leading coefficient $\beta_n$.

(b) Let $L>0$, $K\geq0$ and $\tau:\mathbb{N}_0\to\mathbb{C}$ with $\tau(k)=1$ for the integers $0\leq k<L$ and $|1-\tau|\leq K$. For $0\leq y<L$ and $|E|\leq1$, $$\lt|\PT[\tau\,\mathrm{Pol}](yE)-P(yE)\rt|\,e^{-y(1-\mathrm{Re}E)}\leq KA(L)\,e^{-L\phi(y/L)},$$ where the right side is read as $0$ for $y=0$.

(c) Let $0<\bar\eta<\Lambda$ and $\nu_{\bar\eta}=\Lambda\phi(\bar\eta/\Lambda)>0$. For all $J\geq1$ with $Jk_b\in\mathbb{N}_0$, $\eta\in[0,\bar\eta]^{N_h}$ and $(g,z)\in X$, $$\lt|\prod_h\mathcal{D}_{J\Lambda}(J\eta_hE_h)e^{JS}-\prod_hP(J\eta_hE_h)e^{JS}\rt|\leq N_h[(1+M_\chi)A(\Lambda)]^{N_h}J^{nN_h}e^{-\nu_{\bar\eta} J}.$$ Hence \eqref{prefactor:main} holds for every $\nu\in(0,\nu_{\bar\eta})$.
\end{lemma}

\begin{proof}
(a) Multiply the identity $\sum_{k=0}^{\infty}\frac{y^k}{k!}\,k^{\underline m}=y^m\,e^{y}$ with $y=c$ by $b_me^{-c}$ and sum over $m$. Comparing the coefficients of $k^n$ gives $b_n=\beta_n$.

(b) By (a), $P(yE)e^{-y(1-E)}=e^{-y}\sum_k\mathrm{Pol}(k)E^ky^k/k!$, and likewise for $\PT[\tau\,\mathrm{Pol}]$ with $\tau\,\mathrm{Pol}$ in place of $\mathrm{Pol}$. Hence the left side is $|e^{-y}\sum_{k\geq L}[1-\tau(k)]\mathrm{Pol}(k)E^ky^k/k!|$, which is $0$ for $y=0$. For $0<y<L$ we use $|\mathrm{Pol}(k)|\leq\sum_m|b_m|k^{\underline m}$ ($k\in\mathbb{N}_0$), $y^k\leq(y/L)^LL^k$ ($k\geq L$); this bounds the left side by $Ke^{-y}(y/L)^L\sum_m|b_m|L^me^L=KA(L)e^{-L\phi(y/L)}$.

(c) The left side is $|\prod_hu_h-\prod_hv_h|\,|\prod_bE_b^{Jk_b}|$ with $u_h=u_J(\eta_h,E_h)$ and $v_h=P(J\eta_hE_h)e^{-J\eta_h(1-E_h)}$. By (b) with $\tau(k)=\chi(k/(J\Lambda))$, $L=J\Lambda$, $K=M_\chi$, $y=J\eta_h$, and since $\phi$ decreases, $|u_h-v_h|\leq M_\chi A(J\Lambda)e^{-\nu_{\bar\eta} J}\leq M_\chi A(\Lambda)J^ne^{-\nu_{\bar\eta} J}$. Also $|v_h|\leq A(J\eta_h)\leq A(\Lambda)J^n$, so $|u_h|,|v_h|\leq(1+M_\chi)A(\Lambda)J^n$. The result follows from the telescoping inequality.

\end{proof}

\section{Approximation of \texorpdfstring{$\mathscr{A}_\psi$}{A\_psi} for arbitrary \texorpdfstring{$\eta$}{eta}}\label{app:Q}

We keep the setting of Appendix \ref{app:polynomial approximation}, but $\eta_h$ is not restricted relative to $\Lambda$. Let $\psi$ be bounded and measurable on $\mathbb{R}^{N_h}_{\geq0}\times X$, with $\psi(\eta,g,z)=0$ unless $\eta\in[0,R]^{N_h}$ (some $R>0$), and $\mathscr{A}_\psi$ as in \eqref{prefactor:Apsi}. Let $B=\sum_j|\beta_j|$, so that $|\mathrm{Pol}(k)|\leq B(1+k)^n$ for $k\geq0$. The functions $\mathcal{P}_r(a,c)=\PT[(k-a)^r\mathrm{Pol}(k)](c)$ and $Q^{(M)}_{J\Lambda}$, for integers $M\geq1$, are defined in \eqref{prefactor:QM}. For $\varrho\geq0$ and $l\in\mathbb{N}_0$,
\begin{equation}
\langle f(k)\rangle_\varrho=e^{-\varrho}\sum_{k=0}^{\infty}\frac{\varrho^k}{k!}\,f(k),\qquad\langle|k-\varrho|^l\rangle_\varrho\leq2e\,l!\,(1+\varrho)^{l/2},\label{Q:Poisson}
\end{equation}
where $\langle f(k)\rangle_\varrho$ is the expectation value of $f$ in the Poisson distribution with mean $\varrho$. Indeed, $\langle e^{t(k-\varrho)}\rangle_\varrho=e^{\varrho(e^t-1-t)}\leq e^{\varrho t^2}$ for real $|t|\leq1$; with $s=(1+\varrho)^{-1/2}$, the inequalities $e^{s|x|}\leq e^{sx}+e^{-sx}$ and $x^l\leq l!\,e^x$ ($x\geq0$) give $s^l\langle|k-\varrho|^l\rangle_\varrho\leq l!\,\langle e^{s|k-\varrho|}\rangle_\varrho\leq2e^{\varrho s^2}l!\leq2e\,l!$, i.e. 
\be
\langle|k-\varrho|^{q}\rangle_{\varrho}\leq 2eq!\left(1+\varrho\right)^{q/2}
\ee

\begin{lemma}\label{Q:lemQ}
(a) $\mathcal{P}_r$ is a polynomial in $(a,c)$ of total degree at most $r+n$, and $\mathcal{P}_0(a,c)=P(c)$. 

(b) $Q^{(M)}_{J\Lambda}(c)=P(c)$ for $|c|\leq J\Lambda$, and $Q^{(M)}_{J\Lambda}(c)=0$ for $|c|>x_*J\Lambda$. 

(c) There are $J$-independent $q_p\in C^\infty(\mathbb{C})$ (smooth in $\mathrm{Re}\,w,\mathrm{Im}\,w$), vanishing for $|w|>x_*\Lambda$, with $$Q^{(M)}_{J\Lambda}(Jw)=\sum_{p=-(M-1)}^{n}J^p\,q_p(w)$$ for all $J>0$, $w\in\mathbb{C}$; so $Q^{(M)}_{J\Lambda}$ is smooth.
\end{lemma}

\begin{proof}
(a) $(k-a)^r\mathrm{Pol}(k)=\sum_{i=0}^{r}\binom{r}{i}(-a)^{r-i}k^i\,\mathrm{Pol}(k)$, and $\PT[k^i\mathrm{Pol}]$ is a polynomial of degree at most $i+n$ by \eqref{loc:8}, since $k^i\mathrm{Pol}(k)$ is a linear combination of the $k^{\underline m}$ with $m\leq i+n$. $\mathcal{P}_0=\PT[\mathrm{Pol}]=P$.

(b) For $r\geq1$, $\chi^{(r)}=0$ on $[0,1]$ (on $(0,1)$, and at $0,1$ by continuity), so for $|c|\leq J\Lambda$ only $\mathcal{P}_0(|c|,c)=P(c)$ remains. All $\chi^{(r)}$ vanish on $(x_*,\infty)$.

(c) Write $\mathcal{P}_r(a,c)=\sum_{i+j\leq r+n}\varpi^{(r)}_{ij}a^ic^j$, where $\varpi^{(0)}_{ij}=0$ for $i>0$ by (a). Then 
$$Q^{(M)}_{J\Lambda}(Jw)=\sum_{r<M}\sum_{i+j\leq r+n}\frac{\varpi^{(r)}_{ij}}{r!\Lambda^r}J^{i+j-r}f_{ri}(w)\,w^j,$$
where $f_{ri}(w)=\chi^{(r)}(|w|/\Lambda)|w|^i$ and $-(M-1)\leq i+j-r\leq n$. The $f_{ri}$ that occur ($r\geq1$ or $i=0$) are smooth on $\{w\neq0\}$ and on $\{|w|<\Lambda\}$, where $f_{00}=1$ and $f_{ri}=0$ for $r\geq1$, and they vanish for $|w|>x_*\Lambda$. Collecting the powers of $J$ defines the $q_p$.

\end{proof}

\begin{theorem}\label{Q:thmApprox}
For every integer $M\geq1$ there are constants $C_u,C_M$, depending only on $M,\Lambda,\chi,\mathrm{Pol}$, such that for all $J\geq1$, $\eta\geq0$ and $|E|\leq1$
\begin{equation}
|u_J(\eta,E)|\leq C_uJ^n,\qquad\big|u_J(\eta,E)-Q^{(M)}_{J\Lambda}(J\eta E)\,e^{-J\eta(1-E)}\big|\leq C_MJ^{-M}(1+J\eta)^{n+M/2}.\label{Q:pointwise}
\end{equation}
Let $\mathscr{A}^{(M)}_\psi$ be the first term on the right side of \eqref{prefactor:approx}. Then $\mathscr{A}_\psi$ and $\mathscr{A}^{(M)}_\psi$ converge absolutely, and for some $C$ independent of $J$, 
\be 
|\mathscr{A}_\psi-\mathscr{A}^{(M)}_\psi|\leq C\,J^{(n+1)N_h-M/2},
\ee  
for all $J\geq1$ with $Jk_b\in\mathbb{N}_0$.
\end{theorem}

\begin{proof}
Since $d_{J\Lambda}(k)=0$ for $k>x_*J\Lambda$ and the Poisson weights in \eqref{loc:1} sum to $1$, $|u_J|\leq\|\chi\|_\infty B(1+x_*J\Lambda)^n\leq C_uJ^n$ with $C_u=\|\chi\|_\infty B(1+x_*\Lambda)^n$. Let $c=J\eta E$, $\varrho=|c|\leq J\eta$, $t_0=\varrho/(J\Lambda)$. By Taylor's theorem, 
$$\chi\lt(\frac{k}{J\Lambda}\rt)=\sum_{r<M}\frac{\chi^{(r)}(t_0)}{r!}\lt(\frac{k-\varrho}{J\Lambda}\rt)^r+\mathcal{R}_k,\qquad |\mathcal{R}_k|\leq\frac{\|\chi^{(M)}\|_\infty}{M!}\lt(\frac{|k-\varrho|}{J\Lambda}\rt)^M.$$ 
We insert this into $u_J=e^{-J\eta}\sum_k\frac{c^k}{k!}\chi(\frac{k}{J\Lambda})\mathrm{Pol}(k)$. As $e^{-J\eta}e^{c}=e^{-J\eta(1-E)}$ and $\varrho=|c|$, the Taylor polynomial gives $\sum_{r<M}\frac{\chi^{(r)}(t_0)}{r!(J\Lambda)^r}\mathcal{P}_r(\varrho,c)\,e^{-J\eta(1-E)}=Q^{(M)}_{J\Lambda}(c)\,e^{-J\eta(1-E)}$. As $|c^k|=\varrho^k$ and $e^{\varrho-J\eta}\leq1$, the remainder is bounded by $\langle|\mathcal{R}_k\mathrm{Pol}(k)|\rangle_\varrho\leq\frac{\|\chi^{(M)}\|_\infty B}{M!(J\Lambda)^M}\langle|k-\varrho|^M(1+k)^n\rangle_\varrho$. By $(1+k)^n\leq2^{n-1}[(1+\varrho)^n+|k-\varrho|^n]$ and \eqref{Q:Poisson}, $\langle|k-\varrho|^M(1+k)^n\rangle_\varrho\leq2^{n+1}e\,(M+n)!\,(1+\varrho)^{n+M/2}$, and $\varrho\leq J\eta$ gives \eqref{Q:pointwise} with $C_M=2^{n+1}e\,(M+n)!\,B\,\|\chi^{(M)}\|_\infty/(M!\,\Lambda^M)$.

For $\eta\in[0,R]^{N_h}$ let $u_h=u_J(\eta_h,E_h)$ and $\tilde u_h=Q^{(M)}_{J\Lambda}(J\eta_hE_h)\,e^{-J\eta_h(1-E_h)}$. By \eqref{Q:pointwise} and $1+J\eta_h\leq(1+R)J$, $|u_h-\tilde u_h|\leq C_{M,R}J^{n-M/2}$ with $C_{M,R}=C_M(1+R)^{n+M/2}$, and $|u_h|,|\tilde u_h|\leq(C_u+C_{M,R})J^n$. By \eqref{loc:1}, the integrands of $\mathscr{A}_\psi$ and $\mathscr{A}^{(M)}_\psi$ are $\psi\prod_hu_h\prod_bE_b^{Jk_b}$ and $\psi\prod_h\tilde u_h\prod_bE_b^{Jk_b}$. They are measurable and bounded, and vanish unless $\eta\in[0,R]^{N_h}$. So both integrals converge absolutely, and the telescoping inequality gives the claim with $C=N_h\|\psi\|_\infty R^{N_h}V\,C_{M,R}(C_u+C_{M,R})^{N_h-1}$.
\end{proof}

\begin{corollary}\label{Q:corSmooth}
As for the spinfoam integral, let $X$ be a smooth manifold, $\mu$ a Borel measure, the $E_h$ smooth on $X$, and $\psi$ smooth on $\mathbb{R}^{N_h}_{\geq0}\times X$ (the restriction of a smooth function on $\mathbb{R}^{N_h}\times X$) with compact support. Then 
$$\psi\prod_hQ^{(M)}_{J\Lambda}(J\eta_hE_h)=\sum_{p=-(M-1)N_h}^{nN_h}J^pa_p, \qquad a_p=\psi\sum_{(p_h)}\prod_hq_{p_h}(\eta_hE_h)$$ 
summed over the tuples $(p_h)_h$ with $-(M-1)\leq p_h\leq n$ and $\sum_hp_h=p$. The function $a_p$ are smooth with compact support, independent of $J$, and vanish unless $(\eta,g,z)\in\mathrm{supp}\,\psi$ and $\eta_h|E_h|\leq x_*\Lambda$ for all $h$. Hence $\mathscr{A}^{(M)}_\psi=\sum_pJ^{N_h+p}\int\prod_h\mathrm{d}\eta_h\int\mathrm{d}\mu\,a_p\,e^{JS}$, with absolutely convergent integrals.
\end{corollary}

\begin{proof}
Multiply Lemma \ref{Q:lemQ}(c) with $w=\eta_hE_h(g,z)$ over $h$; each $q_{p_h}(\eta_hE_h)$ is smooth as a composition of smooth maps. Each $a_pe^{JS}$ is bounded, since $|e^{JS}|\leq1$, and vanishes outside the compact set $\mathrm{supp}\,\psi\subset[0,R]^{N_h}\times X$, whose measure is at most $R^{N_h}V<\infty$.
\end{proof}

\section{Change of transverse plane and field redefinition}\label{Change of transverse plane and field redefinition}

Let us consider the path integral over a small neighborhood $\mathscr{N}$ around the flat background geometry:
\begin{eqnarray}
\sa_\psi&=&\int_{\mathscr{N}}\mathrm{d}^{N_{h}}\eta\int\mathrm{d}\mu\left(X\right)\Phi\left(J,\eta,X;\mathcal{Y}\right)
\end{eqnarray}
where $\Phi=J^{N_h}\,\psi\prod_h\mathcal{D}_{J\Lambda}\left(J\eta_hE_h\right)e^{JS}$ and $X\equiv(g,z)$. We use the following notation for the tangent vectors of the Regge submanifold $\sm_{\rm Regge}$:
\be
b_{s}(\ell)&\equiv&\frac{\partial\vec{\eta}(\ell)}{\partial\ell_{s}},\qquad b_{hs}(\ell)\equiv\frac{\partial\eta_{h}(\ell)}{\partial\ell_{s}},\qquad s=1,\ldots,N_{\ell},
\ee
We introduce a local coordinate system in $\sn$ adapted to $\sm_{\rm Regge}$ by separating the integration variables into lengths $\ell$ and non-Regge area fluctuations $t$. The first choice of the plane transverse to $\sm_{\rm Regge}$ is spanned by constant basis vectors $\hat{e}$:
\begin{eqnarray}
\eta&=&\eta(\ell)+\hat{e}t,\qquad\qquad\mathcal{J}(\ell)=|\det(b(\ell),\hat{e})|\nonumber\\
\sa_\psi&=&\int\mathrm{d}^{N_{\ell}}\ell\;\mathcal{J}(\ell)\;e^{J\mathcal{I}_{{\rm eff}}(\ell;\,\mathcal{Y})},\qquad e^{J\mathcal{I}_{{\rm eff}}(\ell;\,\mathcal{Y})}\equiv\int\mathrm{d}^{N_{h}-N_{\ell}}t\:\mathrm{d}\mu\left(X\right)\Phi\left(J,\eta(\ell)+\hat{e}t,X;\mathcal{Y}\right).
\end{eqnarray}
Alternatively, we can choose a different transverse plane spanned by another set of basis vectors $\tilde{e}$, leading to a new set of local coordinates $(\tilde{\ell}, \tilde{t})$:
\begin{eqnarray}
\eta&=&\eta(\tilde{\ell})+\tilde{e}\tilde{t},\qquad\qquad\tilde{\mathcal{J}}(\tilde{\ell})=|\det(b(\tilde{\ell}),\tilde{e})|\nonumber\\
\sa_\psi&=&\int\mathrm{d}^{N_{\ell}}\tilde{\ell}\;\tilde{\mathcal{J}}(\tilde{\ell})\;e^{J\tilde{\mathcal{I}}_{{\rm eff}}(\tilde{\ell};\,\mathcal{Y})},\qquad e^{J\tilde{\mathcal{I}}_{{\rm eff}}(\tilde{\ell};\,\mathcal{Y})}\equiv\int\mathrm{d}^{N_{h}-N_{\ell}}\tilde{t}\,\mathrm{d}\mu\left(X\right)\Phi\left(J,\eta(\tilde{\ell})+\tilde{e}\tilde{t},X;\mathcal{Y}\right).
\end{eqnarray}
Let us examine how an infinitesimal modification of the transverse basis affects the effective actions. Specifically, we analyze and relate the effective actions ${\mathcal{I}}_{{\rm eff}}({\ell};\,\mathcal{Y})$ and $\tilde{\mathcal{I}}_{{\rm eff}}(\tilde{\ell};\,\mathcal{Y})$ that arise from the two distinct choices of transverse planes. 
We consider the following change of basis parametrized by a small matrix $\o$,
\be
\tilde{e}=\hat{e}M(\tilde\ell)+b(\tilde{\ell})\o(\tilde\ell)^{\top},\qquad\tilde{e}^{\ i}_{h}=\hat{e}^{\ j}_{h}M(\tilde\ell)^{\ i}_{j}+b_{hs}(\tilde{\ell})\left(\o(\tilde\ell)^{\top}\right)^{\ i}_{s}
\ee
In the following, we will neglect the correction of $\co(\o^2)$. 
Both $\tilde e^i$ and $\hat e^i$ are constant basis vectors. The matrices $M$ and $\o$ depend on the parameters $\tilde\ell$, since the tangent vectors $b_s$ depend on $\tilde\ell$. Equating the area vector $\eta$ expressed in both coordinate systems $\eta(\ell)+\hat{e}t=\eta(\tilde{\ell})+\tilde{e}\tilde{t}$ and replacing $\eta(\ell)$ with its first-order Taylor expansion $\eta(\tilde{\ell}) + b_s(\tilde{\ell})(\ell_s - \tilde{\ell}_s)$, we obtain a relation between the two sets of local coordinates. The omitted $\co((\ell-\tilde{\ell})^2)$ term is proportional to $\o^2$ (as shown below) and is therefore negligible for the first-order variation with respect to $\o$:
\begin{eqnarray}
\eta(\tilde{\ell})+b_{s}(\tilde{\ell})\left(\ell_{s}-\tilde{\ell}_{s}\right)+\hat{e}^{i}t_{i}	&=&\eta(\tilde{\ell})+\hat{e}^{i}\left(M\tilde{t}\right)_{i}+b_{s}(\tilde{\ell})\left(\o(\tilde \ell)^{\top}\tilde{t}\right)_{s}\ .
\end{eqnarray}
By matching the coefficients of the basis vectors, we find the coordinate transformation connecting $(\ell, t)$ and $(\tilde{\ell}, \tilde{t})$:
\be
t=M(\tilde\ell)\tilde{t},\qquad\ell_{s}-\tilde{\ell}_{s}=\left(\o(\tilde\ell)^{\top}\tilde{t}\right)_{s}\ .
\ee
We substitute this linear coordinate transformation back into the expression for the area vector to identify the small shift $\delta$:
\begin{eqnarray}
\eta(\tilde{\ell})+\tilde{e}\tilde{t}	&=&\eta(\tilde{\ell})+\hat{e}M(\tilde\ell)\tilde{t}+b(\tilde{\ell})\o(\tilde\ell)^{\top}\tilde{t}=\eta(\tilde{\ell})+\hat{e}t+b(\tilde{\ell})\o(\tilde\ell)^{\top}M(\tilde\ell)^{-1}t\nonumber\\
	&\equiv&\eta(\tilde{\ell},t)+b(\tilde{\ell})\delta(\tilde{\ell},t)\ ,
\end{eqnarray}
where
\begin{eqnarray}
	\eta(\tilde{\ell},t)=\eta(\tilde{\ell})+\hat{e}t\ ,\qquad\delta(\tilde{\ell},t)=\o(\tilde\ell)^{\top}M(\tilde\ell)^{-1}t\ .
\end{eqnarray}
We now substitute this shifted area vector into the expression for the effective action derived from the new transverse plane $\tilde{e}$, and perform a Taylor expansion with respect to the small shift $\delta$ at the linear order in $\o$:
\begin{eqnarray}
e^{J\tilde{\mathcal{I}}_{{\rm eff}}(\tilde{\ell};\mathcal{Y})}	
	&=&\left|\det\left(M\right)\right|^{-1}\int\mathrm{d}^{N_{h}-N_{\ell}}t\mathrm{d}\mu\left(X\right)\left[\Phi\left(J,\eta(\tilde{\ell},t),X;\mathcal{Y}\right)+\frac{\partial}{\partial\eta_{h}}\Phi\left(J,\eta(\tilde{\ell},t),X;\mathcal{Y}\right)b_{hs}(\tilde{\ell})\delta_{s}(\tilde\ell,t)+O\left(\delta^{2}\right)\right]\ .\nonumber
\end{eqnarray}
Since $S$ is linear in $\eta$, $\Phi(\eta+b\delta)=\Phi(\eta)\,e^{J\,\partial_{\eta_h}S\,b_{hs}\delta_s}$ up to the slowly varying face-amplitude prefactor, so the second-order term of the expansion is suppressed relative to the first-order term by a factor of order $J\,|\partial_\eta S|\,|b\delta|\sim |\o|$. We work strictly to first order in $\o$, i.e.\ we take the infinitesimal limit of $\o$ prior to the asymptotic limit $J\to\infty$.

By employing the calculus chain rule, the variation of the integrand with respect to the area variables can be rewritten directly as a derivative with respect to the length variables $\ell_s$:
\begin{eqnarray}
\frac{\partial}{\partial\eta_{h}}\Phi\left(J,\eta(\tilde{\ell},t),X;\mathcal{Y}\right)b_{hs}(\tilde\ell)\delta_{s}(\tilde\ell,t)	&=&\frac{\partial}{\partial\eta_{h}}\Phi\left(J,\eta(\tilde{\ell},t),X;\mathcal{Y}\right)\frac{\partial\eta_{h}}{\partial\tilde\ell_{s}}(\tilde{\ell})\delta_{s}(\tilde\ell,t)\nonumber\\
	&=&\frac{\partial}{\partial\tilde\ell_{s}}\Phi\left(J,\eta(\tilde{\ell},t),X;\mathcal{Y}\right)\delta_{s}(\tilde\ell,t)\ .
\end{eqnarray}
Given that the transformation matrices $\o$ and $M$ are independent of the internal integration variables $(t, X)$, we extract the derivative with respect to $\tilde{\ell}_s$ outside the path integral. Identifying the remaining integrals as the expectation value of the transverse fluctuation $t$ in the original basis, we obtain the relation between the two effective actions:
\begin{eqnarray}
e^{J\tilde{\mathcal{I}}_{{\rm eff}}(\tilde{\ell};\mathcal{Y})}	&=&\left|\det\left(M\right)\right|^{-1}\int\mathrm{d}^{N_{h}-N_{\ell}}t\mathrm{d}\mu\left(X\right)\left\{ \Phi\left(J,\eta(\tilde{\ell},t),X;\mathcal{Y}\right)+\frac{\partial}{\partial\tilde\ell_{s}}\left[\Phi\left(J,\eta(\tilde{\ell},t),X;\mathcal{Y}\right)\delta_{s}(\tilde{\ell},t)\right]\rt.\nonumber\\
&&\lt.-\Phi\left(J,\eta(\tilde{\ell},t),X;\mathcal{Y}\right)\frac{\partial}{\partial\tilde\ell_{s}}\delta_{s}(\tilde{\ell},t)+O\left(\delta^{2}\right)\right\} \nonumber\\
	&=&\left|\det\left(M\right)\right|^{-1}\left\{e^{J\mathcal{I}_{{\rm eff}}(\tilde{\ell};\mathcal{Y})}+\frac{\partial}{\partial\tilde\ell_{s}}\left[e^{J\mathcal{I}_{{\rm eff}}(\tilde{\ell};\mathcal{Y})}\left(\o(\tilde{\ell})^{\top}M(\tilde{\ell})^{-1}\left\langle t\right\rangle _{\tilde{\ell}}\right)_{s}\right]\rt.\nonumber\\
&&\lt.-e^{J\mathcal{I}_{{\rm eff}}(\tilde{\ell};\mathcal{Y})}\left(\frac{\partial}{\partial\tilde\ell_{s}}\left[\o(\tilde{\ell})^{\top}M(\tilde{\ell})^{-1}\right]\left\langle t\right\rangle _{\tilde{\ell}}\right)_{s}+O\left(\delta^{2}\right)\right\}\nonumber\\
	&=&\left|\det\left(M\right)\right|^{-1}e^{J\mathcal{I}_{{\rm eff}}(\tilde{\ell};\mathcal{Y})}\left[1+J\frac{\partial\mathcal{I}_{{\rm eff}}}{\partial\tilde\ell_{s}}(\tilde{\ell};\mathcal{Y})\Delta\tilde{\ell}_{s}+\left[\o(\tilde{\ell})^{\top}M(\tilde{\ell})^{-1}\right]_s^{\ i}\frac{\partial}{\partial\tilde\ell_{s}}\left\langle t_i\right\rangle _{\tilde{\ell}}+O\left(\delta^{2}\right)\right]. \label{AppDc:relation}
\end{eqnarray}
The expectation value of the non-Regge fluctuations $\langle t_{i}\rangle_{\tilde{\ell}}$ and the corresponding shift in the length variables $\Delta\tilde{\ell}$ are explicitly given by:
\be
\left\langle t_{i}\right\rangle _{\tilde{\ell}}=\frac{\int\mathrm{d}^{N_{h}-N_{\ell}}t\, \mathrm{d}\mu\left(X\right)\Phi\left(J,\eta(\tilde{\ell},t),X;\mathcal{Y}\right)t_{i}}{\int\mathrm{d}^{N_{h}-N_{\ell}}t\, \mathrm{d}\mu\left(X\right)\Phi\left(J,\eta(\tilde{\ell},t),X;\mathcal{Y}\right)},\qquad\Delta\tilde{\ell}=\delta(\tilde\ell,\lag t\rag_{\tilde\ell})=\o(\tilde\ell)^{\top}M(\tilde\ell)^{-1}\left\langle t\right\rangle _{\tilde{\ell}}\ .
\ee
Note that $\langle t_i\rangle_{\tilde\ell}$ is a ratio of integrals of the complex-valued integrand $\Phi$ and is therefore complex in general: up to $\co(1/J)$ corrections it is the $t$-component of the (complex) critical point of $S$ at fixed $\tilde\ell$.
Correspondingly the shift $\Delta\tilde\ell$ is complex in general, and $\mathcal{I}_{\rm eff}(\tilde\ell+\Delta\tilde\ell)$ below stands for $\mathcal{I}_{\rm eff}(\tilde\ell)+\partial_{s}\mathcal{I}_{\rm eff}(\tilde\ell)\,\Delta\tilde\ell_s$.

To relate $\tilde{\mathcal{I}}_{{\rm eff}}(\tilde{\ell};\mathcal{Y})$ and ${\mathcal{I}}_{{\rm eff}}(\tilde{\ell};\mathcal{Y})$, we disregard the $1/J$ correction terms, consistent with working in the semiclassical regime. Therefore, the contribution from $[\o(\tilde{\ell})^{\top}M(\tilde{\ell})^{-1}]_s^{\ i}{\partial}\left\langle t_i\right\rangle _{\tilde{\ell}}/{\partial\tilde\ell_{s}}$ and $|\det(M)|^{-1}$ can be neglected. Furthermore, we only keep corrections linear to the small matrix $\o$ and omit $\delta^2 \simeq (\Delta\tilde\ell)^2$ corrections because it is of order $\o^2$. Therefore, we obtain
\be
\tilde{\mathcal{I}}_{\rm eff}(\tilde{\ell}) \simeq  \mathcal{I}_{\rm eff}(\tilde{\ell} + \Delta\tilde{\ell}).
\ee 
This confirms that \be
\ell\to \tilde{\ell} + \Delta\tilde{\ell}\ .
\ee 
acts as a field redefinition of the length variable inside the effective path integral. Thus, the residual dependence of the effective action $\mathcal{I}_{\rm eff}$ on the choice of the transverse plane reduces to a field redefinition of the length variables.

\section{Perturbative computation: boundary action}\label{Perturbative computation: boundary action}

In this appendix, we want to show that the boundary term $S_{\rm bdy}'$ matches exactly the boundary term $S_{\rm bdy}$ of Regge action:
\be
   S_{\rm bdy}=\frac{\gamma}{2}\sum_{b}\Delta k_{b}(\ell)\mathring{\Theta}_b+\frac{\gamma}{4}\sum_{r,s}\lt[\sum_b
   \frac{\partial k_b}{\partial\ell_s}\,
   \frac{\partial\Theta_b}{\partial\ell_r}\rt]\,
   \delta\ell_s\delta\ell_r.\label{App:ReggeBoundaryTerm}
\ee
In our notation, $\Delta$ denotes the variations up to second order in $\delta \ell$:
\be
\Delta k_b(\ell) = \delta k_b(\ell) + \frac{1}{2} \delta^{(2)} k_b(\ell),\qquad 
\delta k_b(\ell)=\sum_s\frac{\partial  k_b}{\partial\ell_s}\delta\ell_s,\qquad \delta^{(2)}  k_b(\ell)=\sum_{s,r}\frac{\partial^2  k_b}{\partial\ell_s\partial\ell_r}\delta\ell_s\delta\ell_r\ .
\ee

We begin by establishing the second-order Taylor expansion of the original boundary action $\sum_b k_b \log E_b(g, z, \xi)$ restricting the variations solely to the boundary data $\mathcal{Y}_M = (k_b, \xi_m)$. The complete variation up to second order in the perturbation parameters $\delta\ell$ is:
\be 
\mathcal{I}^{(2)}_{\rm bdy} = \frac{i\gamma}{2} \sum_b \Delta k_b(\ell) \mathring{\Theta}_b + \sum_{b,m} \mathring{k}_b \frac{\partial \log E_b}{\partial \xi_m} \Delta \xi_m(\ell) + \sum_{b,m} \delta k_b(\ell) \frac{\partial \log E_b}{\partial \xi_m} \delta \xi_m(\ell) + \frac{1}{2} \sum_{b,m,n} \mathring{k}_b \frac{\partial^2 \log E_b}{\partial \xi_m \partial \xi_n} \delta \xi_m(\ell) \delta \xi_n(\ell). 
\ee

We use the relation
\be
-\frac{1}{2}\left(B+C\right)^{\top}\mathfrak{h}^{-1}C=\frac{1}{2}C^{\top}\delta X(\ell),
\ee
where $\delta X(\ell)$ is the solution to the linear equation of motion \eqref{linearCritEqn}. The full boundary contribution to the action \eqref{S-Sequation65} is given by:
$$ iS'_{\rm bdy} = \frac{1}{2} C^\top \delta X(\ell) + \mathcal{I}^{(2)}_{\rm bdy} $$
where $C_I$ has the following expression
\be
C_I = \sum_b \delta k_b(\ell) V_{I,b} + \sum_{b,m} \mathring{k}_b \frac{\partial^2 \log E_b}{\partial X_I \partial \xi_m} \delta \xi_m(\ell).
\ee

By linearizing the second relation in \eqref{angleCorrespond}, we obtain the following perturbative identity relating boundary dihedral angles to spinfoam critical point data: 
\be
\sum_I V_{I,b}\delta X_I(\ell)+\sum_m \frac{\partial \log E_b}{\partial \xi_m}\delta \xi_m(\ell)=\frac{i\g}{2}\delta\Theta_b\ ,\label{geometicIdDihedral}
\ee
where $\delta X_I(\ell)$ solves the linear equation of motion \eqref{linearCritEqn} of $S_0$ in the toy integral. 
This identity can be understood in a similar way to \eqref{eq:geometric}.

We evaluate $\frac{1}{2} C^\top \delta X(\ell)$ by applying the identity \eqref{geometicIdDihedral},
\be 
\frac{1}{2} \sum_I C_I \delta X_I(\ell) = \frac{i\gamma}{4} \sum_b \delta k_b(\ell) \delta \Theta_b - \frac{1}{2} \sum_{b,m} \delta k_b(\ell) \frac{\partial \log E_b}{\partial \xi_m} \delta \xi_m(\ell) + \frac{1}{2} \sum_{b,m} \mathring{k}_b \sum_I \frac{\partial^2 \log E_b}{\partial X_I \partial \xi_m} \delta X_I(\ell) \delta \xi_m(\ell) 
\ee

Combining this with $\mathcal{I}^{(2)}_{\rm bdy}$, we note that $i S_{\rm bdy} = \frac{i\gamma}{2} \sum_b \Delta k_b(\ell) \mathring{\Theta}_b + \frac{i\gamma}{4} \sum_b \delta k_b(\ell) \delta \Theta_b$. The difference evaluates exactly to:
\be 
i S_{\rm bdy}' - i S_{\rm bdy} = \sum_{b,m} \mathring{k}_b \frac{\partial \log E_b}{\partial \xi_m} \Delta \xi_m(\ell) + \frac{1}{2} \sum_{b,m} \left( \delta k_b(\ell) \frac{\partial \log E_b}{\partial \xi_m} + \mathring{k}_b \sum_I \frac{\partial^2 \log E_b}{\partial X_I \partial \xi_m} \delta X_I(\ell) + \mathring{k}_b \sum_n \frac{\partial^2 \log E_b}{\partial \xi_n \partial \xi_m} \delta \xi_n(\ell) \right) \delta \xi_m(\ell) \nonumber
\ee
The term inside the parentheses is identically the first-order differential $\delta \left( k_b(\ell) \frac{\partial \log E_b}{\partial \xi_m} (\ell)\right)$. Substituting the variation $\Delta \xi_m(\ell) = \delta \xi_m(\ell) + \frac{1}{2} \delta^{(2)} \xi_m(\ell)$, we obtain
\be 
i S_{\rm bdy}' - i S_{\rm bdy} = \sum_{b,m} \mathring{k}_b \frac{\partial \log E_b}{\partial \xi_m} \delta \xi_m(\ell) + \frac{1}{2} \sum_{b,m} \mathring{k}_b \frac{\partial \log E_b}{\partial \xi_m} \delta^{(2)} \xi_m(\ell) + \frac{1}{2} \sum_{b,m} \delta \left( k_b(\ell) \frac{\partial \log E_b}{\partial \xi_m} (\ell)\right) \delta \xi_m(\ell) .\label{actionDifference}
\ee
In our notation, $\frac{\partial \log E_b}{\partial \xi_m} (\ell)$ means evaluating the derivative at the critical point (of the toy integral) corresponding to $\ell$, while $\frac{\partial \log E_b}{\partial \xi_m} $ means the evaluation at the background $\mathring\ell$.


For all $\ell$, which can be a curved geometry, the critical point equations of $S_0$ hold: $\sum_h \eta_h(\ell) V_{I,h}(\ell) + \sum_b k_b(\ell) V_{I,b}(\ell) = 0$.
Combining the Schl\"afli identity $\sum_h \eta_h(\ell) \delta \varepsilon_h(\ell) + \sum_b k_b(\ell) \delta \Theta_b(\ell) \equiv 0$ with the geometric relations \eqref{angleCorrespond} (generalizing straightforwardly \eqref{eq:geometric} and \eqref{geometicIdDihedral} to critical points that are not flat background) yields the identity:
\be 
\sum_I \delta X_I(\ell) \left( \sum_h \eta_h(\ell) V_{I,h}(\ell) + \sum_b k_b(\ell) V_{I,b}(\ell) \right) + \sum_{b,m} k_b(\ell) \frac{\partial \log E_b}{\partial \xi_m}(\ell) \delta \xi_m(\ell) = 0 
\ee
Since the first grouping is identically zero for all $\ell$, we obtain a constraint on the boundary parameters:
\be 
\sum_{b,m} k_b(\ell) \frac{\partial \log E_b}{\partial \xi_m}(\ell) \delta \xi_m(\ell)  = 0 \quad \text{for all } \ell .
\ee


Evaluating this continuous identity at the background $\mathring{\ell}$ where $\delta \ell = 0$, we find $\sum_{b,m} \mathring{k}_b \frac{\partial \log E_b}{\partial \xi_m} \delta \xi_m(\ell) = 0$, eliminating the first linear term in our action difference formula \eqref{actionDifference}.
Furthermore, because the identity holds universally for all $\ell$, we can differentiate it with respect to $\ell$ and evaluate at the background:
\be
\sum_{b,m} \delta \left( k_b(\ell) \frac{\partial \log E_b}{\partial \xi_m} \right) \delta \xi_m(\ell) +  \sum_{b,m} \mathring{k}_b \frac{\partial \log E_b}{\partial \xi_m} \delta^{(2)} \xi_m(\ell) = 0 
\ee
Therefore, the difference of boundary actions perfectly cancels, demonstrating that $ S_{\rm bdy}' = S_{\rm bdy} $.

\section{Geometrical correspondence of spinfoam critical point}\label{Geometrical correspondence of spinfoam critical point}

\begin{theorem}
Let $\mathcal{K}$ be a 2-complex dual to a finite simplicial triangulation $\mathcal{K}^*$ with boundary (each 4-simplex has exactly 5 boundary tetrahedra in its boundary, each tetrahedron has exactly 4 triangular faces). Let $(\mathring{\eta}_h,\mathring{g}_{ve},\mathring{z}_{vf},\mathring{k}_b,\mathring{\xi}_{eb})$ denote the spinfoam critical-point data (modulo gauge symmetries) corresponding to a nondegenerate flat Regge geometry $\mathring{\ell}$ (all deficit angles $\mathring{\eps}_h=0$, uniform 4-simplex orientation). Then there exist open neighborhoods $U_{\mathcal{R}}$ of $\mathring{\ell}$ and $U_{\mathcal{S}}$ of the background spinfoam data $(\mathring{\eta}_h,\mathring{g}_{ve},\mathring{z}_{vf},\mathring{k}_b,\mathring{\xi}_{eb})$, and there exists a bijection\footnote{
	Here $\mathcal{S}$ is the set of gauge classes that have a representative in $U_{\mathcal{S}}$. The neighbourhood $U_{\mathcal{S}}$ can be taken arbitrarily small, and then $U_{\mathcal{R}}=\{\ell\in W:\ \Phi(\ell)\in\mathcal{S}\}$, where $W$ is the neighbourhood of $\mathring\ell$ on which $\Phi$ is defined. Several critical points correspond to $\mathring\ell$. We have fixed one of them as the background: it has the given boundary data $(\mathring k_b,\mathring\xi_{eb})$, $\mu(v)=-1$ for all $v$, and the lift to $\mathrm{SL}(2,\mathbb{C})$ for which $\mathring E_h=1$ for all $h$ (see Appendix~\ref{Deficit angle and dihedral angle}). The map $\Phi$ is the branch through this background.}
\be
\Phi: \mathcal{R} \to \mathcal{S}
\ee
between:

$\bullet$ $\mathcal{R}$: the space of perturbative Regge geometries on the flat background, i.e., edge length assignments $\ell \in U_{\mathcal{R}}$ such that each 4-simplex is a nondegenerate flat Minkowski simplex and tetrahedra are spacelike.

$\bullet$ $\mathcal{S}$: the space of spinfoam critical-point data $(\eta_h, g_{ve}, z_{vf}, k_b, \xi_{eb}) \in U_{\mathcal{S}}$, modulo the spinfoam gauge symmetries (vertex $\mathrm{SL}(2,\mathbb{C})$, edge $\mathrm{SU}(2)$, and spinor scaling)
and modulo the changes of frame and of phase convention of the boundary data (for each boundary edge $e$: $g_{ve}\mapsto g_{ve}H_e$, $\xi_{eb}\mapsto H_e^{-1}\xi_{eb}$ for all $b\ni e$, with $H_e\in\mathrm{SU}(2)$; and $\xi_{eb}\mapsto e^{i\beta_{eb}}\xi_{eb}$, $\beta_{eb}\in\R$),
satisfying the critical-point equations
\be
\sum_h \eta_h \frac{\partial_{g,z} E_h}{E_h} + \sum_b k_b \frac{\partial_{g,z} E_b}{E_b} = 0 \label{App:B2}
\ee
and the modulus conditions
\be
|E_h| = |E_b| = 1. \label{App:B3}
\ee

\end{theorem}

The changes of frame and of phase convention of the boundary data map solutions of \eqref{App:B2} and \eqref{App:B3} to solutions (the first leaves every $E_f$ invariant, the second multiplies $E_b$ by a constant phase) and do not change the reconstructed geometry.
Equivalently, one can fix them by choosing once and for all, smoothly in $\ell$, a frame for every boundary tetrahedron and the phases of the boundary spinors; this is what is meant by the notation $\xi_{eb}(\ell)$ in the main text.

The proof of this theorem is in the following.

The 2-complex $\mathcal{K}$ has vertices $v$ (dual to 4-simplices $\sigma_v$), edges $e$ (dual to tetrahedra $t_e$), and faces $f$ (dual to triangles $\triangle_f$). Faces are either internal $h$ or boundary $b$. For each vertex-face pair $(v,f)$ with $f$ incident to $v$, we assign a spinor $z_{vf} \in \mathbb{CP}^1$. For each vertex-edge pair $(v,e)$ with $e$ incident to $v$, we assign a group element $g_{ve} \in \mathrm{SL}(2,\mathbb{C})$. Define $Z_{vef} = g_{ve}^\dagger z_{vf}$. For each boundary edge-face pair $(e,b)$, we assign a normalized boundary spinor $\xi_{eb}$.

The face amplitude is $E_f = \prod_{e \in \partial f} \mathcal{E}_{f,e}$ where, for an internal edge $e$ oriented from $v$ to $v'$:
\be
\mathcal{E}_{f,e} = \frac{\langle Z_{vef}, Z_{v'ef}\rangle}{\langle Z_{vef}, Z_{vef}\rangle^{\frac{1-i\gamma}{2}} \langle Z_{v'ef}, Z_{v'ef}\rangle^{\frac{1+i\gamma}{2}}},\label{App:cefe}
\ee
and for a boundary edge $e$ with boundary spinor $\xi_{eb}$:
\be
\mathcal{E}_{b,e} = \frac{\langle\xi_{eb}, Z_{veb}\rangle}{\langle Z_{veb}, Z_{veb}\rangle^{\frac{1+i\gamma}{2}}} \quad\text{or}\quad \mathcal{E}_{b,e} = \frac{\langle Z_{veb}, \xi_{eb}\rangle}{\langle Z_{veb}, Z_{veb}\rangle^{\frac{1-i\gamma}{2}}},\label{App:cefeforb}
\ee
depending on the orientation. Here $\langle a,b\rangle = a^\dagger b$ and $\gamma \in \mathbb{R}$ is the Barbero-Immirzi parameter.

The action $S_0$ (restricted to the linear term in $\log E_h$) is given by
\be
S_0 = \sum_h \eta_h \log E_h + \sum_b k_b \log E_b,\qquad \eta_h, k_b \in \mathbb{R}_{\geq 0}.
\ee
The action $S_0$ is invariant under:
(1) $\mathrm{SL}(2,\mathbb{C})$ vertex gauge: $g_{ve} \mapsto x_v^{-1} g_{ve}$, $z_{vf} \mapsto x_v^{\dagger} z_{vf}$ ($x_v \in \mathrm{SL}(2,\mathbb{C})$), which leaves $Z_{vef}=g_{ve}^\dagger z_{vf}$ invariant.
(2) $\mathrm{SU}(2)$ edge gauge (for spacelike internal edges): $g_{ve} \mapsto g_{ve} H_e$, $g_{v'e} \mapsto g_{v'e} H_e$ ($H_e \in \mathrm{SU}(2)$).
(3) Spinor scaling: $z_{vf} \mapsto \lambda_{vf} z_{vf}$ ($\lambda \in \mathbb{C}^*$).
$\mathcal{S}$ is the quotient of the solution set by these gauge equivalences intersected with $U_{\mathcal{S}}$.

A Regge geometry on $\mathcal{K}^*$ assigns positive lengths $\ell_s$ to all edges $s$ such that: For each 4-simplex $\sigma_v$, the 10 edge lengths ensure a nondegenerate flat 4-simplex embedding in $\mathbb{R}^{1,3}$ with spacelike boundary tetrahedra.
A perturbative Regge geometry is $\ell = \mathring{\ell} + \delta\ell$ with $\|\delta\ell\|$ sufficiently small. $U_{\mathcal{R}}$ is a neighborhood of $\mathring{\ell}$ in $\mathbb{R}^{N_\ell}$.

For a 4-simplex $\sigma_v$ embedded in $\mathbb{R}^{1,3}$ and tetrahedron $t_e \subset \sigma_v$, we denote by $N_{ve} \in \mathbb{R}^{1,3}$ the timelike outward-pointing unit 4d normal of the tetrahedron. $N_{ve}$ may be future or past pointing. The unit vector $N_{\mathrm{ref}} = (\pm1,0,0,0)$ is the reference normal for the tetrahedron, and the group element $\Lambda_{ve} \in \mathrm{SO}^+(1,3)$ is a pure boost with $\Lambda_{ve} N_{\mathrm{ref}} = N_{ve}$.
$\vec{n}_{ef} \in S^2$ is the outward-pointing unit 3d normal of triangle $\triangle_f$ within $t_e$, computed in the local frame where $t_e$ lies in the $x$-$y$-$z$ hyperplane.

The map $\pi: \mathbb{CP}^1 \to S^2$ is $\pi(z) = \frac{\langle z | \vec{\sigma} | z \rangle}{\langle z | z \rangle}$ with Pauli matrices $\vec{\sigma} = (\sigma_1,\sigma_2,\sigma_3)$. The covering map $\Lambda: \mathrm{SL}(2,\mathbb{C}) \to \mathrm{SO}^+(1,3)$ is the 2-to-1 surjective homomorphism with kernel $\{\pm \mathbb I\}$, given by $\Lambda(g)^I_{\ J} = \frac{1}{2} \mathrm{Tr}({\sigma}_I g {\sigma}_J g^\dagger)$ where $\sigma_I = (\mathbb{I},\vec\sigma)$.

\subsection{Construction of $\Phi: \mathcal{R} \to \mathcal{S}$}

Equations \eqref{App:B2} and \eqref{App:B3} are equivalent to the following three conditions (see~\cite{hanPI} for derivations):

\begin{itemize}
    \item For every tetrahedron $t_e$ in every 4-simplex $\sigma_v$,
    \begin{equation}
        \sum_{\Delta_f \subset t_e} \eta_f \, \kappa_{ef}(v) \, \frac{\langle Z_{vef}, \vec{\sigma} Z_{vef}\rangle}{\langle Z_{vef}, Z_{vef}\rangle} = 0,\qquad \eta_b=k_b,\label{closureB7}
    \end{equation}
where $\vec{\sigma}$ are the Pauli matrices. The signs $\kappa_{ef}(v)=+1$ if the oriented boundary $\partial f$ leaves $v$ through $e$, and $\kappa_{ef}(v)=-1$ if $\partial f$ enters $v$ through $e$. It satisfies $\kappa_{ef}(v)=-\kappa_{e'f}(v)$ when $t_e,t_{e'}\subset\sigma_v$ share $\Delta_f$, and $\kappa_{ef}(v')=-\kappa_{ef}(v)$ when the internal tetrahedron $t_e$ is shared by $\sigma_v$ and $\sigma_{v'}$.
Eq.\eqref{closureB7} is equivalent to the closure of weighted 3D normals $\vec{n}_{ef}(v) = \kappa_{ef}(v)\pi\left(g_{ve}^\dagger z_{vf}\right)$ in the reference frame of $t_e$:
    \begin{equation}
        \sum_{\Delta_f \subset t_e} \eta_f \, \vec{n}_{ef}(v) = 0.
    \end{equation}

    \item For every face $f$ shared by $\sigma_v, \sigma_{v'}$ across $t_e$,
    \begin{equation}
        \frac{Z_{vef}}{\|Z_{vef}\|} = e^{i\alpha_{vv'}^f} \frac{Z_{v'ef}}{\|Z_{v'ef}\|},\label{ZZvvprimeB9}
    \end{equation}
    for some phase $\alpha_{vv'}^f \in \mathbb{R}$. For every boundary face $b$ of a boundary tetrahedron $t_e\subset \sig_v$, 
	\be
	\xi_{eb} = e^{i\a^b_{ev}}\frac{Z_{veb}}{\| Z_{veb}\|}.\label{App:xiZrelation}
	\ee
	for some phase $\a^b_{ev} \in \mathbb{R}$.

    \item For every 4-simplex $\sigma_v$ and a pair of tetrahedra $t_e,t_{e'}\subset \sigma_v$ sharing a triangle $\Delta_f$:
    \be
    \frac{g_{ve}Z_{vef}}{\langle Z_{vef},Z_{vef}\rangle}=\frac{g_{ve'}Z_{ve'f}}{\langle Z_{ve'f},Z_{ve'f}\rangle}\ .\label{App:ZZrelationB11}
	\ee

\end{itemize}

In the following, we construct $\Phi(\ell)$ for any $\ell \in \mathcal{R}$ and verify $\Phi(\ell) \in \mathcal{S}$:

\emph{Areas and spins.} For each triangle $\triangle_f$, let its three edge lengths be $\ell_{ij},\ell_{jk},\ell_{ik}$. The area is given by Heron's formula
\be
\mathfrak{a}_f(\ell) = \frac{1}{4} \sqrt{2\left(\ell_{i j}^2 \ell_{j k}^2+\ell_{i k}^2 \ell_{j k}^2+\ell_{i j}^2 \ell_{i k}^2\right)-\ell_{i j}^4-\ell_{i k}^4-\ell_{j k}^4}.
\ee
Define 
\be
\eta_h = \frac{2}{\gamma}\,\mathfrak{a}_h(\ell),\qquad k_b = \frac{2}{\gamma}\,\mathfrak{a}_b(\ell).
\ee 
Since $\ell \mapsto \fa_f(\ell)$ is smooth, for sufficiently small $\delta\ell$, we have $(\eta_h,k_b) \in U_{\mathcal{S}}$. 
For finite $J$, the boundary spin $Jk_b$ (see \eqref{kscaling}) is generally not an integer, but any deviation from integrality is small---of order $1/J$---and thus negligible at leading order in the $1/J$ expansion.

\emph{4D embedding.} Let us first focus on a single 4-simplex $\sigma_v$. 
Given the 10 edge lengths of a non-degenerate Lorentzian 4-simplex $\sigma_v$ that are small perturbations of those in the background geometry, one can embed $\sigma_v$ as a flat simplex in $\mathbb{R}^{1,3}$, with vertex positions denoted by $P_1^{(v)},\ldots,P_5^{(v)}$ \footnote{
	Given edge lengths $\{\ell_{ij}\}$ ($i,j=1,\ldots,5$), interpret $s_{ij}^2 = \ell_{ij}^2$ as squared spacetime intervals between vertices $P_i\in(\mathbb{R}^{1,3},\eta )$. By translation, set $P_1=0$, and write $X = (P_2\, P_3\, P_4\, P_5)$ as a $4\times4$ matrix of column vectors. For $a,b=2,3,4,5$, the Gram matrix $G_{ab} = \tfrac{1}{2}(s_{1a}^2 + s_{1b}^2 - s_{ab}^2)$ yields $X^T\eta X=G$. A (nondegenerate) Minkowski embedding exists iff $G$ has signature $(1,3)$. Uniqueness is up to Lorentz transformations: any two such $X,Y$ satisfy $Y=\Lambda X$ with $\Lambda=YX^{-1}\in\mathrm{O}(1,3)$;
	for embeddings in a sufficiently small neighborhood of a given one, $\Lambda$ lies in the identity component $\mathrm{SO}^+(1,3)$.
}.
The embedding is unique up to $\mathrm{ISO}(1,3) = \mathbb{R}^{1,3} \rtimes \mathrm{SO}^+(1,3)$. We fix the translation by fixing e.g. $P_1^{(v)} = 0$. 
Since this embedding is a perturbation around the background simplicial geometry, we disregard any discrete isometries.

\emph{4D normals and $\Slc$ elements.} For each tetrahedron $t_e \subset \sigma_v$ with vertices $\{P_i,P_j,P_k,P_l\}$ (consistent orderings of vertices are chosen for 4-simplices and tetrahedra \cite{HZ}), we obtain the outward-pointing 4d normal denoted by $N_{ve}^I$. Let $N_{\mathrm{ref}} = \pm(1,0,0,0)$. We construct the pure Lorentz boost $\Lambda_{ve} \in \mathrm{SO}^+(1,3)$ satisfying $\Lambda_{ve} N_{\mathrm{ref}} = N_{ve}$ by
\be
\Lambda_{ve} = e^{\Theta_{\mathrm{ref},e}\,\mathbf{B}_{ve}}, \qquad \mathbf{B}_{ve} = \frac{N_{\mathrm{ref}} \wedge N_{ve}}{|N_{\mathrm{ref}} \wedge N_{ve}|} ,\qquad \Theta_{\mathrm{ref},e}= \cosh ^{-1}\left(-N_{v e} \cdot N_{\rm ref}\right), 
\ee
Here $N_{\rm ref}$ has the same time orientation as $N_{ve}$, so that $-N_{ve}\cdot N_{\rm ref}\geq1$. We lift to $\Slc$ and set $g_{ve}^{\mathrm{boost}}  \in \mathrm{SL}(2,\mathbb{C})$ to satisfy $\Lambda(g_{ve}^{\mathrm{boost}}) = \Lambda_{ve}$. The sign ambiguity is fixed by comparing to the background configuration. The spinfoam critical-point data $g_{ve} \in \mathrm{SL}(2,\mathbb{C})$ is of the form $g_{ve}^{\mathrm{boost}} H_{ve}$ with $H_{ve} \in \mathrm{SU}(2)$ being the stabilizer of $N_{\mathrm{ref}}$. The group element $H_{ve}$ is arbitrary at this moment and will be fixed in a moment. 

\emph{Bivector.} In the local 3D frame of each tetrahedron $t_e\subset\sig_v$, the tetrahedron lies in the spatial hyperplane with 3D edge vectors $P_{e,ij}^{(4)} = \Lambda(g_{ve})^{-1} P_{ij}$, where $P_{ij}=P_i-P_j$ is the 4d edge vector. We define a bivector $X_{ef}\in sl(2,\bbc)$ for each face of $t_e$ by 
\be
\mathbb{X}_{e f}^{I J}=\frac{1}{2} \epsilon^{I J}{ }_{K L} P_{e, i j}^{(4)}{ }^K P_{e, i k}^{(4)}{ }^L ,\qquad
{X}_{e f}={\mathbb{X}_{e f}}/{\left|\mathbb{X}_{e f}\right|}=\pm \hat{t}\wedge \hat{n}_{ef},\label{App:bivector}
\ee
where $\hat{t}=(1,0,0,0)$, $\hat{n}_{ef}=(0,\vec{n}_{ef})$, and $\vec{n}_{ef}$ is the unit outward-pointing 3D normal of $\Delta_f\subset t_e$\footnote{
	Write $N_{\rm ref}=s_{ve}\,\hat t$ with $s_{ve}=\pm1$ (the time orientation of $N_{ve}$), and let $\nu_{ef}=\Lambda(g_{ve})\hat n_{ef}$ be the outward unit normal of $\Delta_f$ within $t_e$, so that $\Lambda(g_{ve})(s_{ve}\hat t\wedge\hat n_{ef})=N_{ve}\wedge\nu_{ef}$. We orient each triangle $\Delta_f$, i.e. we choose the order of $(i,j,k)$ in \eqref{App:bivector}, such that
    \be
      X_{ef}=-\kappa_{ef}(v)\,s_{ve}\,\hat t\wedge\hat n_{ef}.\label{App:bivectorSign}
    \ee
    The condition \eqref{App:bivectorSign} does not depend on the pair $(v,e)$ used to impose it. Inside $\sigma_v$, it is compatible with \eqref{paralleltraspBivec}. In 4D, \eqref{paralleltraspBivec} reads $\kappa_{ef}(v)\,N_{ve}\wedge\nu_{ef}=\kappa_{e'f}(v)\,N_{ve'}\wedge\nu_{e'f}$. This holds because $\kappa_{e'f}(v)=-\kappa_{ef}(v)$ and $N_{ve'}\wedge\nu_{e'f}=-N_{ve}\wedge\nu_{ef}$. To prove the latter, note that all four vectors lie in the timelike plane orthogonal to $\Delta_f$, and write $\nu_{e'f}=aN_{ve}+b\,\nu_{ef}$, $N_{ve'}=cN_{ve}+d\,\nu_{ef}$. Since $\nu_{e'f}$ and $N_{ve'}$ are outward, $a>0$ and $d<0$. From $N_{ve'}\cdot\nu_{e'f}=0$ and the normalizations, $N_{ve'}\wedge\nu_{e'f}=(cb-da)\,N_{ve}\wedge\nu_{ef}=(d/a)\,N_{ve}\wedge\nu_{ef}$ with $d^2=a^2$, i.e. $d/a=-1$. This holds for thin and thick wedges alike. 

	Across $t_e$, \eqref{App:bivectorSign} is compatible with the gluing: $X_{ef}$ and $\hat n_{ef}$ are the same from both sides, while $s_{v'e}=-s_{ve}$ and $\kappa_{ef}(v')=-\kappa_{ef}(v)$. With \eqref{App:bivectorSign} and \eqref{DefBivectorB18}, $\kappa_{ef}(v)\,\pi(\xi_{ef})=-s_{ve}\,\vec n_{ef}$. Hence \eqref{closureBivec} is $-s_{ve}/2$ times the closure $\sum_{\Delta_f\subset t_e}\fa_f\vec n_{ef}\cdot\vec\sigma=0$ of the tetrahedron. The vectors $\kappa_{ef}(v)\pi(Z_{vef})=\vec n_{ef}(v)$ in \eqref{closureB7} relates to the outward unit normals $\vec{n}_{ef}$ of $t_e$ by $\vec n_{ef}(v)=-s_{ve}\,\vec n_{ef}$. The overall sign in \eqref{App:bivectorSign} selects one of the two parity-related branches. With $+\kappa_{ef}(v)s_{ve}$ in place of $-\kappa_{ef}(v)s_{ve}$ one obtains the branch $g_{ve}\mapsto(g_{ve}^\dagger)^{-1}$ with the same boundary spinors, which has $\mu(v)=+1$. The minus sign gives $\mu(v)=-1$ for the dihedral-angle convention of \cite{semiclassical}; see the footnote to $\psi_f(v)=\mu(v)\Theta_f(v)/2$ in Appendix~\ref{Deficit angle and dihedral angle}. 
}.  We represent $X_{ef}$ as a 2-by-2 matrix. For any pair of $t_e,t_{e'}\subset\sig_v$ sharing the triangle $\Delta_f$, the bivectors $X_{ef},X_{e'f}$ satisfy the following parallel transport:
\be
g_{ve}X_{ef}g_{ve}^{-1}=g_{ve'}X_{e'f}g_{ve'}^{-1}\equiv X_f(v)\ , \label{paralleltraspBivec}
\ee
and they satisfy the closure condition:
\be
\sum_{\Delta_f\subset t_e}\kappa_{ef}(v) X_{ef}\fa_f(\ell)=0.\label{closureBivec}
\ee
The signs $\kappa_{ef}(v)=\pm1$ (same as in \eqref{closureB7}) define the orientations of the triangles in the 4-simplex, satisfying the relation $\kappa_{ef}(v)=-\kappa_{e'f}(v)$ for $t_e,t_{e'}$ sharing $\Delta_f$. 

\emph{Normalized spinor.} The bivector $X_{ef}$ is a traceless Hermitian 2-by-2 matrix, and it relates to a normalized 2-spinor $\xi_{ef}=(\xi_{ef}^1,\xi_{ef}^2)$ by 
\be
X_{ef}=\xi_{ef}\otimes\xi_{ef}^\dagger-\frac{1}{2}\mathbb{I}_{2\times 2},\qquad \langle \xi_{ef},\xi_{ef}\rangle=1.\label{DefBivectorB18}
\ee
The normalized spinor $\xi_{ef}$ is uniquely (up to a phase) specified as the eigenvector of $X_{ef}$ corresponding to the eigenvalue $1/2$. 
The eigenvector corresponding to the eigenvalue $-1/2$ is $J\xi_{ef}=(-\xi_{ef}^2,\xi_{ef}^1)^*$ (here the conjugate map $J$ should not be confused with the large spin parameter).
Geometrically, $\xi_{ef}$ encodes the unit 3D normal vector $\vec n_{ef}$ of $\Delta_f$ via the relation $\pm \vec n_{ef} = \langle \xi_{ef}| \vec{\sigma} | \xi_{ef} \rangle$.

\emph{Parallel transportation.} 
By \eqref{paralleltraspBivec}, both $g_{ve}\xi_{ef}$ and $g_{ve'}\xi_{e'f}$ are eigenvectors of $X_f(v)$ with eigenvalue $1/2$. Similarly, $g_{ve}J\xi_{ef}$ and $g_{ve'}J\xi_{e'f}$ are eigenvectors of $X_f(v)$ with eigenvalue $-1/2$. 
Each of the eigenspaces of $\pm1/2$ is 1-dimensional.
Therefore,
\be
g_{ve}\xi_{ef}\propto_\C g_{ve'}\xi_{e'f},\qquad g_{ve}J\xi_{ef}\propto_\C g_{ve'}J\xi_{e'f}.\label{parallelxiB19}
\ee

\emph{$\mathbb{CP}^1$ spinor.} For the 4-simplex $\sig_v$ and any face $f$, we define a $\mathbb{CP}^1$ spinor $z_{vf}$ by
\be
z_{vf} \propto_\C (g_{ve}^\dagger)^{-1}\xi_{ef}.\label{app:zvf}
\ee
The scale (overall normalization) of $z_{vf}$ is a gauge freedom. We fix this freedom by choosing the first component of $z_{vf}$ to be $1$. With this choice, the definition of $z_{vf}$ becomes independent of the choice of tetrahedron $e$ or $e'$ due to the second relation in \eqref{parallelxiB19} and the identity $-JgJ=(g^\dagger)^{-1}$ for any $g\in\Slc$.
The spinors $z_{vf},\xi_{ef}$ and the group element $g_{ve}$ satisfy the following relation:
\be
\xi_{ef}=e^{i\a^f_{ev}}\frac{Z_{vef}}{\|Z_{vef}\|},\label{relationXiandZ}
\ee
which coincides with \eqref{App:xiZrelation} for $f=b$. The first relation in \eqref{parallelxiB19} is equivalent to $g_{ve}\frac{Z_{vef}}{\|Z_{vef}\|^2}=C g_{ve'}\frac{Z_{ve'f}}{\|Z_{ve'f}\|^2}$, $C\in\C$. Acting $z_{vf}^\dagger$ on the left determines the ratio $C=1$. Therefore, we obtain the following relation of the spinor $z_{vf}$ and group elements $g_{ve},g_{ve'}$:
\be
\frac{g_{ve}Z_{vef}}{\|Z_{vef}\|^2}=\frac{g_{ve'}Z_{ve'f}}{\|Z_{ve'f}\|^2}\ ,
\ee
which coincides with \eqref{App:ZZrelationB11}. Furthermore, the closure condition \eqref{closureB7} is implied by the closure condition \eqref{closureBivec} of the bivectors. 

\emph{Gluing 4-simplices.} To glue two 4-simplices $\sigma_v$ and $\sigma_{v'}$ along a common tetrahedron $t_e$, we choose the SU(2) stabilizers $H_{ve}$ and $H_{v'e}$ so that for every face $\Delta_f \subset t_e$, the corresponding bivectors $X_{ef}$ computed from $\sigma_v$ and $\sigma_{v'}$ are identical\footnote{
	Although the tetrahedron $t_e$ may yield different sets of 3D face normals when computed from $\sigma_v$ and $\sigma_{v'}$, these sets differ only by a rotation. 
	This is due to the fact that, in the Regge geometry, the gluing of $\sigma_v$ and $\sigma_{v'}$ ensures that the shape of the shared tetrahedron matches on both sides. 
	We choose $H_{ve}$ or $H_{v'e}$ to account for this rotational difference and align the corresponding 3D normals.
}\footnote{
	Since the embedding of $\sigma_{v'}$ is unique only up to $\mathrm{ISO}(1,3)$, we may choose it such that $\sigma_v$ and $\sigma_{v'}$ lie in a common Minkowski space and are glued along $t_e$, with $\sigma_{v'}$ on the other side of the hyperplane of $t_e$ (this is possible at the background and persists by continuity). If $(\Lambda_0,a)$ is the corresponding Poincar\'e transformation and $\Lambda(x)=\Lambda_0$, this replaces $g_{v'e}$ by $x\,g_{v'e}$, a vertex gauge transformation, which leaves $P'{}^{(4)}_{e,ij}=\Lambda(g_{v'e})^{-1}P'_{ij}$ unchanged. In the common embedding the edge vectors $P_{ij}$ of $t_e$ are shared, and $N_{v'e}=-N_{ve}$. From $\Lambda(g_{ve})^{-1}(N_{ve},P_{ij})=(N_{\rm ref},P^{(4)}_{e,ij})$ and $\Lambda(g_{v'e})^{-1}(N_{v'e},P_{ij})=(N'_{\rm ref},P'{}^{(4)}_{e,ij})$ with $N'_{\rm ref}=-N_{\rm ref}$, it follows that $\Lambda(g_{ve}^{-1}g_{v'e})\in \mathrm{SO}^+(1,3)$ fixes $N_{\rm ref}$ and maps $P'{}^{(4)}_{e,ij}$ to $P^{(4)}_{e,ij}$. Hence its restriction $O_e$ to the hyperplane orthogonal to $N_{\rm ref}$ lies in $\mathrm{SO}(3)$, $O_eP'{}^{(4)}_{e,ij}=P^{(4)}_{e,ij}$, and the corresponding bivectors \eqref{App:bivector} are related by $O_e$. Replacing $H_{v'e}$ by $H_{v'e}u_e$, where $u_e\in\mathrm{SU}(2)$ satisfies $\Lambda(u_e)=O_e^{-1}$ (with the sign fixed by continuity from the background), completes the matching.
}, 
so that the resulting $\xi_{ef}$ does not depend on whether it is computed from $\sigma_v$ or $\sigma_{v'}$. This implies the equation \eqref{ZZvvprimeB9}. 

\emph{Gauge fixing.} For each 4-simplex $\sigma_v$ fix one tetrahedron $e_0(v)$ as follows: take a spanning tree of the dual graph rooted at a vertex $r$ adjacent to a boundary edge; let $e_0(r)$ be a boundary edge and, for every other vertex $v$, let $e_0(v)$ be the edge towards its parent. 
The gauge conditions are as follows. For each $e_0(v)$, we set $g_{ve_0(v)} = \mathbb{I}$ and require that, for the other end $v' \neq v$, $g_{v'e_0(v)}$ is upper-triangular with positive diagonal (Iwasawa decomposition $g = k u$ with $k$ upper-triangular and $u \in \mathrm{SU}(2)$). For each internal edge $e \neq e_0(v)$, we require that one of $g_{ve},g_{v'e}$ (where $v$ and $v'$ are the endpoints of $e$) is upper-triangular with positive diagonal.
Every configuration near the background is gauge equivalent to exactly one configuration near the background satisfying these conditions. Indeed, under $g_{ve}\mapsto x_v^{-1}g_{ve}H_e$, proceeding from the root: $x_r=g_{re_0(r)}$; for every other vertex, $g_{ve_0(v)}=\mathbb I$ gives $x_v=g_{ve_0(v)}H_{e_0(v)}$, and the Iwasawa condition at the parent vertex determines $H_{e_0(v)}$. For the remaining internal edges the Iwasawa conditions determine $H_e$. This also fixes the $\mathrm{SO}^+(1,3)$ isometry in the 4D embedding of each 4-simplex. In general, the resulting $g_{ve}$ may still have a residual sign ambiguity arising from the double cover $\mathrm{SL}(2,\mathbb{C})\to\mathrm{SO}^+(1,3)$; this ambiguity is uniquely resolved by referencing the background configuration, since each $g_{ve}$ is a small perturbation of the fixed background $\mathring g_{ve}$.

The above procedure defines the map $\Phi:\mathcal{R}\to\cs$, which uniquely determines $\Phi(\ell)$ as the set of spinfoam data $(\eta_h, k_b, g_{ve}, z_{vf}, \xi_{eb})$ satisfying the relations \eqref{closureB7}--\eqref{App:ZZrelationB11}. 
The procedure depends smoothly on variations of the edge lengths $\delta\ell$, so that $(\eta_h, k_b, g_{ve}, z_{vf}, \xi_{eb})$ vary smoothly as perturbations of the background configuration. The numerical program implementing $\Phi$ is provided in \cite{Han:2024lti}.

\subsection{Surjectivity and injectivity of $\Phi$}\label{Surjectivity and injectivity of Phi}

First, we prove the surjectivity: Let $(\eta_h, g_{ve}, z_{vf}, k_b, \xi_{eb})$ be a spinfoam critical-point data in $\mathcal{S}$, satisfying \eqref{closureB7}--\eqref{App:ZZrelationB11}. We reconstruct a Regge geometry $\ell \in \mathcal{R}$ and prove that $\Phi(\ell)=(\eta_h, g_{ve}, z_{vf}, k_b, \xi_{eb})$.

\emph{Reconstruction of tetrahedra.} Given the spinfoam data $(\eta_h, g_{ve}, z_{vf}, k_b, \xi_{eb})$ and for each $t_e \subset \sigma_v$, the face normals are given by $\vec{n}_{ef}(v) =\kappa_{ef}(v) \pi(Z_{vef})$ and $\vec{n}_{eb} = \kappa_{eb}(v)\pi(\xi_{eb})$. 
By Minkowski's theorem, the closure condition $\sum_{\Delta_f\subset t_e} (\g\eta_f/2) \vec{n}_{ef}(v)=0$ uniquely specifies (up to rigid motions in $\mathbb{E}(3)$) a tetrahedron whose faces have the prescribed outward area vectors $(\g\eta_f/2)  \vec{n}_{ef}(v)$. Consequently, the edge lengths $\ell$ of the tetrahedron are determined. The nondegeneracy condition holds for the spinfoam data because the background configuration is nondegenerate. 

For any edge in the simplicial complex that is shared by several tetrahedra and simplices, we must verify that the edge length determined by the reconstruction is independent of which tetrahedron or simplex is used in the calculation.

\emph{Shape-matching between simplices.} Condition \eqref{ZZvvprimeB9} implies $\pi(Z_{veh})=\pi(Z_{v'eh})$ for every face $h\subset t_e$. Since $\kappa_{eh}(v')=-\kappa_{eh}(v)$, the normals reconstructed from the two sides are related by $\vec n_{eh}(v')=-\vec n_{eh}(v)$ for all four faces of $t_e$, and the areas $\gamma\eta_h/2$ are the same. The overall sign-flip of the normals does not change the closure condition. By Minkowski's theorem, the tetrahedra $t_e$ reconstructed from $\sigma_v$ and from $\sigma_{v'}$ are isometric and have the same six edge lengths.

\emph{4D reconstruction of simplices.} For each $\sigma_v$, we use \eqref{DefBivectorB18} and the relation \eqref{relationXiandZ} to define the bivectors $X_{ef}$. The parallel transport equations \eqref{App:ZZrelationB11} imply the parallel transport \eqref{paralleltraspBivec} of the bivectors\footnote{We derive $g_{ve}X_{ef}g_{ve}^{-1}=\frac{g_{ve}Z_{vef}}{\|Z_{vef}\|^2}\otimes z_{vf}^\dagger-\frac{1}{2}\mathbb{I}$ and use \eqref{App:ZZrelationB11}.} \footnote{In addition, $X_{ef}$ is independent of $v$ due to shape-matching between simplices}. 
The closure condition \eqref{closureB7} gives $\sum_{\Delta_f\subset t_e}\kappa_{ef}(v)\,\frac{\gamma\eta_f}{2}\,X_{ef}=0$, i.e.\ \eqref{closureBivec} with $\fa_f(\ell)$ replaced by $\gamma\eta_f/2$.


Consequently, for each 4-simplex $\sigma_v$, we obtain ten bivectors $X_f(v)$, one associated to each triangle $\Delta_f$, which together satisfy the closure condition \eqref{closureBivec}. These bivectors are simple, i.e., $X_f(v)\wedge X_f(v)=0$, and they obey the orthogonality condition $N_{ve}\cdot (*X_f)(v)=0$ for all $\Delta_f\subset t_e$ (from the definition \eqref{DefBivectorB18}, the bivector satisfies $X_{ef} = \pm \hat{t} \wedge \hat{n}_{ef}$), where $N_{ve}\propto \L(g_{ve})\cdot\hat{t}$. Provided certain nondegeneracy conditions hold, these properties ensure the set of bivectors meets the hypotheses of Barrett's reconstruction theorem \cite{semiclassical,BC}. According to this theorem, such a collection of bivectors determines a unique geometric 4-simplex (up to translation and inversion), and in particular, uniquely specifies the edge lengths $\ell$ of the simplex. 
We apply the theorem to the bivectors $B_{ee'}(v)=\kappa_{ef}(v)\,\frac{\gamma\eta_f}{2}\,X_f(v)$, $\Delta_f=t_e\cap t_{e'}$, which satisfy $B_{e'e}(v)=-B_{ee'}(v)$. The resulting 4-simplex has triangle areas $\gamma\eta_f/2$, and in the frame defined by $g_{ve}$ its tetrahedron $t_e$ has outward area vectors $\pm\frac{\gamma\eta_f}{2}\vec n_{ef}(v)$, with one sign for all four faces. By the uniqueness in Minkowski's theorem, every boundary tetrahedron of this 4-simplex is isometric to the tetrahedron $t_e$ reconstructed above. Hence the edge lengths obtained from different tetrahedra of the same 4-simplex agree with each other and with those of the 4-simplex.

Equation \eqref{App:bivector} relates the bivectors to the simplex geometry.

The nondegeneracy conditions required by Barrett's reconstruction theorem are satisfied since the spinfoam data considered here are small perturbations of the nondegenerate background configuration, which already meets these criteria.

\emph{Gluing to a Regge geometry.} Due to the shape-matching between simplices, adjacent 4-simplices assign identical geometry to their shared tetrahedron $t_e$, and the Regge geometries of the 4-simplices are consistently glued to a Regge geometry $\ell=:\Psi(s)$ of the entire complex, where $s\in\cs$ denotes the given critical-point data\footnote{
	The map $\Psi$ is well defined on $\mathcal{S}$. The vertex gauge leaves $Z_{vef}$ invariant. The spinor scaling and the boundary phases leave $\pi(Z_{vef})$ and $\pi(\xi_{eb})$ invariant. The edge gauge and the change of boundary frame act as $Z_{vef}\mapsto H_e^{\dagger}Z_{vef}$ for all $\Delta_f\subset t_e$, which rotates the four area vectors of $t_e$ by an element of $\mathrm{SO}(3)$. The areas $\eta_h,k_b$ are invariant. $\Psi$ is also smooth, because the edge lengths of a tetrahedron with outward area vectors $A_1,\dots,A_4$ ($\sum_fA_f=0$, $A_1,A_2,A_3$ linearly independent) are $\ell_{ff'}=2\,|A_f\times A_{f'}|/(3V)$ with $V=\big(\tfrac{2}{9}\,|\det(A_1,A_2,A_3)|\big)^{1/2}$. Here $\ell_{ff'}$ is the edge shared by the faces $f$ and $f'$.
}. The reconstruction map $\Psi$ is continuous, and by the injectivity argument below $\Psi\circ\Phi=\mathrm{id}$ on the neighbourhood $W$ of $\mathring\ell$ on which $\Phi$ is defined. We choose $U_{\cs}$ so small that $\Psi(\cs)\subset W$, and set $U_{\calr}:=\Phi^{-1}(U_{\cs})$, an open neighbourhood of $\mathring\ell$. Then $\Phi$ maps $U_{\calr}$ into $\cs$, and $\ell=\Psi(s)\in U_{\calr}$ for any $s\in \cs$; hence $\ell\in\calr$.

\emph{Recovery of original spinfoam data.} Given the reconstructed lengths $\ell$, we denote $(\tilde\eta_h, \tilde g_{ve}, \tilde z_{vf}, \tilde k_b, \tilde\xi_{eb})\equiv \Phi(\ell)\in U_\cs$, and we want to verify the tilded spinfoam data are the same as the original critical-point data $\eta_h, g_{ve}, z_{vf}$. The identity $\tilde{\eta}_h=\eta_h$ holds because both $\tilde{\eta}_h, \eta_h$ relate to $\ell$ by Heron's formula. For the group variables, let $\tilde{g}_{ve}$ be generated from the reconstructed Regge geometry via $\Phi(\ell)$. Both $\tilde{N}_{ve} = \Lambda(\tilde{g}_{ve})N_{\text{ref}}$ and $N_{ve} = \Lambda(g_{ve})N_{\text{ref}}$ are 4D normals of the tetrahedron $t_e$
\footnote{
	$N_{ve}$ is the normal of the tetrahedron because $N_{ve}\cdot (*X_f)(v)=0$
}. 
By Barrett's reconstruction theorem, the geometric 4-simplex $\sigma_v$ determines the 4D normals up to a global isometry. Thus, the reconstructed normals $\tilde{N}_{ve}$ and original normals $N_{ve}$ are related by a Lorentz transformation $\Lambda(x_v^{-1}) \in \mathrm{SO}^+(1,3)$: $\tilde{N}_{ve} = \Lambda(x_v^{-1}) N_{ve}$, yielding $\tilde{g}_{ve} = x_v^{-1} g_{ve} H_{ve}$ with $x_v \in \mathrm{SL}(2,\mathbb{C})$ and $H_{ve} \in \mathrm{SU}(2)$
\footnote{
	Any 4-vector can be represented as a 2-by-2 matrix by $X=X^I\sig_I$ where $\sig_I=(\mathbb{I},\vec{\sig})$. The action of $g\in\Slc$ on the vector is given by $g X g^\dagger$. In this representation, $N_{\rm ref}=\pm\mathbb{I}$, and thus $\tilde{N}_{ve}=N_{ve}$ is equivalent to $\tilde g_{ve}\tilde g_{ve}^\dagger=g_{ve}g_{ve}^\dagger$, if we set $x_v=1$. If we denote by $H_{ve}=g_{ve}^{-1}\tilde g_{ve}$, we obtain $H_{ve} H_{ve}^\dagger= \mathbb{I}$, i.e. $H_{ve}\in \Su$.
}.

In the reference frame of the tetrahedron $t_e$, the original data defines 3D edge vectors $P_{e,ij}^{(4)} = \Lambda(g_{ve})^{-1} P_{ij}$, which relates to the bivectors $X_{ef}$ via \eqref{App:bivector}. Comparing with the procedure of $\Phi(\ell)$, the 4D edge vectors in the respective 4-simplex frames are related by $\tilde{P}_{ij} = \Lambda(x_v^{-1}) P_{ij}$. Thus, the 3D edge vectors in the reference frame transform as:
$$ \tilde{P}_{e,ij}^{(4)} = \Lambda(\tilde{g}_{ve})^{-1} \tilde{P}_{ij} = \Lambda(H_{ve}^{-1} g_{ve}^{-1} x_v) \Lambda(x_v^{-1}) P_{ij} = \Lambda(H_{ve})^{-1} \Lambda(g_{ve})^{-1} P_{ij} = \Lambda(H_{ve})^{-1} P_{e,ij}^{(4)}. $$
Therefore, the spatial bivectors transform under the $\mathrm{SU}(2)$ action as:
\be
\tilde{X}_{ef} = H_{ve}^{-1} X_{ef} H_{ve}. \label{app:XHXH}
\ee
Since both $\tilde{X}_{ef}$ and $X_{ef}$ are independent of $v$, we obtain for neighboring $\sig_v,\sig_{v'}$: $ H_{ve}^{-1} X_{ef} H_{ve} = H_{v'e}^{-1} X_{ef} H_{v'e} $. 
This relation holds for all four faces $\Delta_f \subset t_e$. Since the four bivectors $X_{ef}$ associated with a non-degenerate tetrahedron span the 3D space of spatial bivectors, the $\mathrm{SU}(2)$ element $H_{v'e} H_{ve}^{-1}$ must commute with all spatial rotations. The center of $\mathrm{SU}(2)$ is $\{\mathbb{I}, -\mathbb{I}\}$, so $H_{v'e} = \pm H_{ve}$. 

$\Psi$ and $\Phi$ are continuous, with $\Psi(\mathring s)=\mathring\ell$ and $\Phi(\mathring\ell)=\mathring s$ where $\mathring s$ denotes the background spinfoam data. So $\Phi\circ\Psi$ is continuous and fixes $\mathring s$. The small $U_{\mathcal S}$ makes $\tilde s$, a gauge-fixed representative, close to $\mathring s$. The original $s$ is close to $\mathring s$ simply because $s\in U_{\mathcal S}$. 

The pair $(x_v,H_{ve})$ is determined only up to $(x_v,H_{ve})\mapsto(-x_v,-H_{ve})$ for all $e\ni v$ \footnote{
	Flipping relative signs would cause $\tilde{g}_{ve}\to\pm \tilde{g}_{ve}$, which could not be both close to the background.
}. We can always choose the sign convention such that $x_v,H_{ve}$ are close to $\mathbb I$ rather than $-\mathbb I$ (since $\tilde s$ is close to $s$), then $H_{v'e}=\pm H_{ve}$ forces $H_{v'e}=H_{ve}\equiv H_e$ for every internal edge. Consequently
\be
\tilde g_{ve}=x_v^{-1}g_{ve}H_{ve},\qquad \tilde\xi_{ef}\propto_{\mathbb{C}} H_{ve}^{-1}\xi_{ef},\qquad \tilde z_{vf}\propto_{\mathbb{C}}(\tilde g_{ve}^{\dagger})^{-1}\tilde\xi_{ef}=x_v^{\dagger}(g_{ve}^\dagger)^{-1}\xi_{ef}\propto_{\mathbb{C}}x_v^{\dagger}z_{vf},\label{combineG25}
\ee
with $H_{ve}=H_e$ for internal edges, where we used $(H^\dagger)^{-1}=H$ for $H\in\mathrm{SU}(2)$. Eq.\eqref{combineG25} combines a vertex gauge transformation, an edge gauge transformation and a spinor rescaling. For a boundary edge $e$ there is no neighboring 4-simplex, and $H_{ve}$ is not constrained by the above argument. In this case \eqref{app:XHXH} gives $\tilde\xi_{eb}=e^{i\beta_{eb}}H_{ve}^{-1}\xi_{eb}$ for some phases $\beta_{eb}$,
so that $(\tilde g_{ve},\tilde\xi_{eb})$ and $(g_{ve},\xi_{eb})$ are related by a change of frame and of phase convention of the boundary data; moreover $\tilde k_b=k_b$, since both are determined by $\ell$ through Heron's formula.

The above shows  $\Phi(\Psi(s))=s$ in $\mathcal{S}$: $\Phi(\ell)$ recovers the original spinfoam critical-point data $(\eta_h,  g_{ve},  z_{vf}, k_b, \xi_{eb})$. Therefore $\Phi$ is surjective.

\vspace{1em}

Finally, we prove the injectivity of $\Phi$. Assume $\ell^{(1)}, \ell^{(2)} \in \mathcal{R}$ with $\Phi(\ell^{(1)}) = \Phi(\ell^{(2)})$ in $\mathcal{S}$. We want to prove $\ell^{(1)} = \ell^{(2)}$: 
For $\ell\in\mathcal{R}$, the data $\Phi(\ell)$ give, in the frame of $t_e$, the area vectors $\fa_f(\ell)\,\kappa_{ef}(v)\,\pi(Z_{vef})=-s_{ve}\,\fa_f(\ell)\,\vec n_{ef}$, where $s_{ve}=\pm1$ and $\vec n_{ef}$ is the outward face normal of the tetrahedron $t_e$ of the geometry $\ell$. By the uniqueness in Minkowski's theorem, the tetrahedron reconstructed from $\Phi(\ell)$ is congruent to $t_e$. Therefore $\Psi(\Phi(\ell))=\ell$, where $\Psi$ is well defined on classes in $\mathcal{S}$ (see above). If $\Phi(\ell^{(1)})=\Phi(\ell^{(2)})$ in $\mathcal{S}$, then $\ell^{(1)}=\Psi(\Phi(\ell^{(1)}))=\Psi(\Phi(\ell^{(2)}))=\ell^{(2)}$, since every segment of $\mathcal{K}^*$ is an edge of some tetrahedron. So $\Phi$ is injective.

\section{Deficit angle and dihedral angle}\label{Deficit angle and dihedral angle}

In this appendix, we demonstrate the correspondence between $E_h, E_b$ and the geometric quantities of deficit angle $\eps_h$ and dihedral angle $\Theta_b$.
\be
E_h= e^{i\g \eps_h/2},\qquad E_b= e^{i\g \Theta_b/2}.\label{app:angleCorrespond}
\ee
The relations are valid at the critical point discussed in the last section.

Inside each 4-simplex $\sig_v$, equation \eqref{parallelxiB19} can be written as
\be
g_{ve'}^{-1}g_{ve} \lt(\xi_{ef},J\xi_{ef}\rt)=\lt(\xi_{e'f},J\xi_{e'f}\rt) 
\begin{pmatrix}
	e^{\psi_{f}(v)+i\phi_f(v)} & 0 \\
	0 & e^{-\psi_{f}(v)-i\phi_f(v)} 
	\end{pmatrix},\quad e^{\psi_{f}(v)}=\frac{\|Z_{vef}\|}{\|Z_{ve'f}\|},\quad \phi_f(v)=\a^f_{ev}-\a^f_{e'v}.\label{app:ggxixi}
\ee
The boost angle $\psi_f(v)$ is related to the 4D dihedral angle $\Theta_f(v)$ between the tetrahedra $t_e$ and $t_{e'}$ (which share the triangle $\Delta_f$ in $\sigma_v$) by the relation $\psi_f(v) = \mu(v) \Theta_f(v)/2$ \cite{semiclassical,HZ} \footnote{
	Here $e$ denotes the edge through which the oriented boundary $\partial f$ enters $v$ and $e'$ the edge through which it leaves $v$, so that $\kappa_{ef}(v)=-1$ and $\kappa_{e'f}(v)=+1$. By \eqref{app:ggxixi}, the null vectors $L_e=g_{ve}\xi_{ef}\,\xi_{ef}^\dagger g_{ve}^\dagger$ and $L_{e'}=g_{ve'}\xi_{e'f}\,\xi_{e'f}^\dagger g_{ve'}^\dagger$ (4-vectors in the representation $X=X^I\sigma_I$) satisfy $L_e=e^{2\psi_f(v)}L_{e'}$. By \eqref{DefBivectorB18} and \eqref{App:bivectorSign}, $\xi_{ef}\xi_{ef}^\dagger$ is the null vector $\frac12\big(\hat t-\kappa_{ef}(v)s_{ve}\hat n_{ef}\big)$, hence $L_e=\frac{s_{ve}}{2}\big(N_{ve}-\kappa_{ef}(v)\,\nu_{ef}\big)$, and similarly for $L_{e'}$. Both lie in the timelike plane orthogonal to $\Delta_f$. Computing their ratio in this plane gives $\psi_f(v)=-|\Theta_f(v)|/2$ if the wedge at $\Delta_f$ is thin ($s_{ve}\neq s_{ve'}$) and $\psi_f(v)=+|\Theta_f(v)|/2$ if it is thick ($s_{ve}=s_{ve'}$), where $\cosh|\Theta_f(v)|=|N_{ve}\cdot N_{ve'}|$. With the convention of \cite{semiclassical}, in which the dihedral angle is positive for a thin wedge and negative for a thick wedge, both cases read $\psi_f(v)=-\Theta_f(v)/2$. The result does not depend on the orientation of $\partial f$. The opposite sign in \eqref{App:bivectorSign} gives the parity-related critical point, with $\psi_f(v)=+\Theta_f(v)/2$.
}
The sign $\mu(v) = \pm 1$ encodes the orientation of the simplex, and changes sign under the transformation $g_{ve} \to (g_{ve}^{\dagger})^{-1}$. For the background configuration, we choose $\mu(v) = -1$ for all $v$, and this choice remains valid under small perturbations.

Let us now consider an internal triangle $\Delta_h$, which is shared by several 4-simplices $\sigma_v$ with $v \in \partial h$. The associated loop holonomy is defined as $G_{h}(e) = \overrightarrow{\prod}_{v \in \partial h}\, g_{ve'}^{-1} g_{ve}$, where the ordered product follows the orientation of the loop $\partial h$ and $e$ denotes the starting edge. Using equation \eqref{app:ggxixi}, we find that at the critical point, $G_h(e)$ relates to the deficit angle $\eps_h$ by the following relation:
\be
G_h(e)&=&\lt(\xi_{eh},J\xi_{eh}\rt) 
\begin{pmatrix}
	e^{-\eps_{h}/2+i\theta_h/2} & 0 \\
	0 & e^{\eps_{h}/2-i\theta_h/2} 
	\end{pmatrix}\lt(\xi_{eh},J\xi_{eh}\rt)^{-1},\qquad \eps_{h}=\sum_{v\in \partial h}\Theta_h(v),\qquad \frac{\theta_{h}}{2}=\sum_{v\in \partial h}\phi_h(v)\ \nonumber\\
	&=&e^{(-\eps_{h}+i\theta_h)X_{eh}},\qquad X_{eh}=\frac{1}{2}\lt(\xi_{eh},J\xi_{eh}\rt) 
	\sig_3\lt(\xi_{eh},J\xi_{eh}\rt)^{-1}=\xi_{eh}\otimes \xi_{eh}^\dagger -\frac{1}{2}\mathbb{I}_{2\times 2}\ . \label{App:Ghe}
\ee
The bivector representation of $X_{ef}$ is given by \eqref{App:bivector}. Observe that the Lorentz transformation $\Lambda(g_{ve})$ acts as a parallel transport on the edge vectors: $P_{e,ij}^{(4)} = \Lambda(g_{ve})^{-1} P_{ij}$. When gluing 4-simplices with a shared tetrahedron $t_e$, the edge vectors $P_{e,ij}^{(4)}$ in each simplex match, so that, under the loop holonomy, the edge vector associated to the triangle $\Delta_f$ is mapped to itself:
\be
\L\lt(G_h(e)\rt)P_{e,ij}^{(4)}= P_{e,ij}^{(4)},\qquad  (i,j)\subset \Delta_f.
\ee
From Eq.\eqref{App:Ghe}, we see that $G_h(e)$ generates, on the triangle's plane, a rotation by angle $\theta_h$ in addition to the boost by the deficit angle. This implies 
\be
\theta_h = 0 \ \text{mod}\ 2\pi,\qquad e^{i\theta_h/2}=\pm1,\label{theta=0G5}
\ee
where this sign ambiguity reflects the fact that $\Slc$ is a double cover of the Lorentz group. 
Computing $E_h$ gives $e^{-i\theta_h/2}=E_h\,e^{i\gamma\sum_{v\in\partial h}\psi_h(v)}$. The background satisfies $\eps_h=0$ and is a real critical point, so $\mathring E_h=e^{-i\mathring\theta_h/2}=1$. It selects the lift $\mathring g_{ve}$. On the critical set $\cs\cong\mathcal{R}$ ($\cs$ is taken to be connected), $e^{i\theta_h/2}$ is a continuous function of the data with values in $\{\pm1\}$. It is therefore constant on $\mathcal{S}$, so $e^{i\theta_h/2}=1$ for all perturbed critical points.

Eq.\eqref{theta=0G5} have assumed that the 4-manifold is time-oriented, i.e. that the loop holonomy $\Omega_h$ of the spin connection of the Regge geometry lies in $\mathrm{SO}^+(1,3)$, thus preserving the time orientation. Note that $\L(G_h)\in\mathrm{SO}^+(1,3)$ always, and that $\L(G_h)=\pm\Omega_h$ for uniformly oriented 4-simplices. However, critical points may also occur where $\Omega_h$ is not contained in $\mathrm{SO}^+(1,3)$; in such cases $\L(G_h)=-\Omega_h$ and one finds $\theta_h = \pi$ modulo $2\pi$, as discussed in \cite{hanPI}.

Therefore, at the critical point, we have the relation between $E_h$ (recall the definition of $E_h$ and \eqref{App:cefe}) and deficit angle:
\footnote{
	Eqs. \eqref{App:ZZrelationB11} and \eqref{relationXiandZ} give that $g_{ve}\xi_{ef}=\frac{\|Z_{vef}\|}{\|Z_{ve'f}\|}e^{i\phi_f(v)}g_{ve'}\xi_{e'f}$ where $\phi_f(v)=\a^f_{ev}-\a^f_{e'v}$. 
	Applying \eqref{relationXiandZ} to $E_h$, we have $\frac{\langle Z_{vef},Z_{v'ef}\rangle }{\|Z_{vef}\|\|Z_{v'ef}\|}= e^{i(\alpha^{f}_{ev}-\alpha^{f}_{ev'})}$.
}
\be
E_h=e^{-i\g\sum_{v\in\partial h}\psi_h(v)}e^{-i\sum_{v\in\partial h}\phi_h(v)}=e^{i\gamma \eps_h/2}\ .
\ee
The correspondence between $E_b$ and the dihedral angle $\Theta_b = \sum_{v\in\partial b}\Theta_b(v)$ follows analogously. 
The same computation applies to a boundary face $b$ and gives
\be
E_b=e^{-i\gamma\sum_{v\in\partial b}\psi_b(v)}\,e^{-i\sum_{v\in\partial b}\phi_b(v)}=e^{i\gamma\Theta_b/2+i\phi_b},\qquad \phi_b\equiv-\sum_{v\in\partial b}\phi_b(v).
\ee
Here no holonomy condition fixes $\phi_b$, because $\partial b$ is not closed.
Nevertheless, by adjusting the phase convention of the boundary spinor $\xi_{eb}$ via $\xi_{eb} \to \xi_{eb} e^{i\beta_{eb}}$, one finds that $E_b$ transforms as $E_b \to E_b e^{\pm i\beta_{eb}}$. Therefore, the phase $\phi_b$ in $E_b$ is not physical and can always be set to zero by a suitable choice of the boundary spinor phases.

\bibliography{muxin.bib}

\end{document}